\documentclass[preprint,12pt]{elsarticle}
\usepackage[a4paper, total={6.5in, 10in}]{geometry}
\usepackage[export]{adjustbox}

\journal{\texttt{arXiv}} 

\RequirePackage{fontenc,inputenc,subcaption,graphicx,sectsty,titlesec, caption, ragged2e}
\RequirePackage{amsthm,amsmath,amssymb,nccmath}
\RequirePackage{algorithm,paralist,multirow}
\RequirePackage[noend]{algpseudocode}
\RequirePackage{float}

\newcommand{\trans}{^{T}} 
\newcommand{\prm}{^{\prime}} 
\newcommand{\dprm}{^{\prime\prime}} 
\newcommand{\del}[2]{\frac{\partial #1}{\partial #2}} 
\newcommand{\deltwo}[2]{\frac{\partial^2 #1}{\partial #2^2}} 
\newcommand{\tI}{\text{I}} 
\newcommand\scalemath[2]{\scalebox{#1}{\mbox{\ensuremath{\displaystyle #2}}}}

\def\balpha{\mbox{\boldmath$\medmath{\alpha}$}}
\def\bbeta{\mbox{\boldmath$\medmath{\beta}$}}
\def\bgamma{\mbox{\boldmath$\medmath{\gamma}$}}
\def\bdelta{\mbox{\boldmath$\medmath{\delta}$}}

\def\bx{\mbox{\boldmath$x$}}

\newtheorem{mythe}{Theorem}

\begin{document}

\begin{frontmatter}

\title{Spatial risk estimation in Tweedie compound Poisson double generalized linear models}

\author[mainaddress]{Aritra Halder\corref{mycorrespondingauthor}}
\cortext[mycorrespondingauthor]{Corresponding author}
\ead{aritra.halder@uconn.edu}
\author[mainaddress]{Shariq Mohammed}
\author[mainaddress]{Kun Chen}
\author[mainaddress]{Dipak K. Dey}


\address[mainaddress]{Department of Statistics, University of Connecticut, Storrs, CT - 06269, USA.}

\centering
\begin{abstract} 
	\vspace*{.3in}
		\justify
		Tweedie exponential dispersion family constitutes a fairly rich sub-class of the celebrated exponential family. In particular, a member, compound Poisson gamma (CP-g) model has seen extensive use over the past decade for modeling mixed response featuring exact zeros with a continuous response from a gamma distribution. This paper proposes a framework to perform residual analysis on CP-g double generalized linear models for spatial uncertainty quantification. Approximations are introduced to proposed framework making the procedure scalable, without compromise in accuracy of estimation and model complexity; accompanied by sensitivity analysis to model mis-specification. Proposed framework is applied to modeling spatial uncertainty in insurance loss costs arising from automobile collision coverage. Scalability is demonstrated by choosing sizable spatial reference domains comprised of groups of states within the United States of America.
		
		\vspace*{.7in}
\end{abstract}


\end{frontmatter}


\section{Introduction}

Advanced geographic information systems (GIS) are increasingly being used to improve predictive accuracy in a variety of statistical systems. The idea is to incorporate previously untapped spatial information present inherently in recorded data to improve predictive performance of a statistical model. Spatial information can be recorded at many levels, geographical co-ordinates i.e., a longitude-latitude pair (point-referenced), a county/district at observation level, census tract information, a three/five-digit zipcode at observation level (areal-referenced). Such information is used to analyze variation in response across a granularity level of interest, for instance, at county/district level, or at zipcode level. Along with variation in the response across granularity levels, commonly recorded population covariates, like median age, number of teenage drivers etc. can be potential predictors. Apart from improving predictive accuracy, one may also be interested in identifying areas (at a desired level of granularity) that behave similarly. This encompasses the idea that neighboring regions show similarity in terms of response; it also includes the possibility of similar response surface characteristics manifesting in different (non-neighboring) locations, due to re-occurring dependence on covariate information. 
In such a scenario an \emph{ordinal} ranking can be established to identify boundaries derived from variation in response that are different from existing geographic borders. This is referred to as \emph{boundary analysis} in literature (\cite{banerjee2003directional}, \cite{lu2005bayesian}, and \cite{womble1951differential}). Another related concept which is relevant to context, is the idea of identifying areas that feature rapid changes with respect to a variable of interest. Identifying these zones can also be a part of boundary analysis, and is referred to as wombling (\cite{womble1951differential}). This is also known as ``barrier analysis" or ``edge detection" in studies of landscape topography, systematic biology, sociology, ecology, and public health.


Spatial information can be incorporated to an existing model in multiple ways. It is both desirable and advantageous to include such information without major changes to an existing modeling structure, by devising a methodology that utilizes already present implementation and builds on it.  Consequently, the same unaccounted for spatial information brings in extra variation, that was previously not quantified by the existing model; this excess variation can now be interpreted as \emph{risk} faced by the response. Depending on the nature of response, qualitative characterizations of this risk may vary, for instance if we are looking at number of road accidents on interstate highways, adverse characterizations of risk would follow. Naturally, the ordinal nature of rankings can then be of considerable interest to an investigator looking to put in place categorically different measures for spatial clusters identified solely on their ``riskiness". Considering the bigger picture, this results in a \emph{nested model}, that consists of structurally dependent and independent components, where proper specification of the structured component aids in explaining excess variation in response.

In practice, existing modeling implementation encountered most commonly are generalized linear models (GLMs) (\cite{nelder1972generalized}, \cite{mccullagh2018generalized}) , which provide a very flexible structure for modeling different types of responses. While constructing GLMs the response is modeled to follow a probability distribution, in that regard \emph{exponential families} of distributions provide a very general class of choices. The idea of GLMs could be extended to a more general class of models, namely \emph{dispersion models} (\cite{jorgensen1987exponential}, \cite{jorgensen1997theory}). The primary reason behind having these elaborate families is to relax unnecessary, restrictive assumptions of normality on the response, in case significant signs of non-normality are evident. GLMs employ a technique called analysis of deviance (which is a generalization of analysis of variance); for the time being if deviance is interpreted only as a measure of distance between realizations and a location parameter, then based on general functional forms and properties of deviance, dispersion models are divided into broad sub-classes. In this paper we are interested in exploring spatial risk estimation in GLMs for particularly one of those classes, called the Tweedie exponential dispersion models (EDMs) (\cite{jorgensen1997theory}, \cite{tweedie1984index}, \cite{dunn2001tweedie}).

Tweedie EDMs (as we shall describe in section (\ref{sec::DGLM-twmod}), table (\ref{tab::tweedie-pars}) and eq. (\ref{eq::tw-mod-dens-0}), (\ref{eq::tw-mod-dens-1})) have an additional index parameter, $p$ which classifies different sub-classes of distributions within the family. Commonly known distributions, like Poisson, gamma and inverse Gaussian are special cases of Tweedie EDMs (see table (\ref{tab::tweedie-edms}), \cite{dunn2018generalized}). Poisson distributions are obtained if index parameter, $p=1$, where as inverse Gaussian distributions fall in the positive stable class with $p=3$. Poisson distribution is commonly used to model count data, whereas gamma and inverse Gaussian distributions find their use in modeling positive continuous data. However when $p \in (1,2)$, this results in \emph{compound Poisson-gamma} distributions which are used in modeling positive continuous data with exact zeros. Applications include modeling weights of fish species in a single sample (trawl), appearance of exact zeros occurring if a particular species is not caught, otherwise a positive (continuous) weight is recorded (\cite{foster2013poisson}). The concept of catch per-unit effort (CPUE) in connection to relative fish stock and abundance is plagued by the zero-catch problem, which is dealt with using these distributions (\cite{shono2008application}). Other examples include rainfall and amount which can be modeled using such mixtures, naturally exact zeros occur in case there is no rainfall, otherwise a positive amount is recorded (\cite{dunn2003precipitation}). Simultaneous modeling of occurrence and size/amount of insurance claims also use compound Poisson-gamma distributions; here exact zeros occur in case of records that show no accidents during a chosen policy-period, whereas a positive claim size is recorded in case of an accident (\cite{jorgensen1994fitting}, \cite{smyth2002fitting}). While modeling insurance claims along with modeling the mean (\cite{jorgensen1994fitting}), dispersion modeling was also considered (\cite{smyth2002fitting}) to result in simultaneous GLMs for mean and dispersion, which are termed as double generalized linear models (DGLMs). Exact zeros in political contributions or donations (in US dollars) are also modeled similarly (\cite{lauderdale2012compound}). Forest degradation which involves sampling of biomass loss produces continuous data with a large number of exact zeros (no disturbance, implying no loss) which are again analyzed using such distributions (\cite{dons2016indirect}).

In all of the above examples existing alternatives include removing exact zeros, or adding a small constant 

\begin{table}[ht]
	\centering
	\caption{The Tweedie family of distributions for varying values of the index parameter, $p$, with respective supports $S$ and parameter spaces, $\mathit{\Omega}$.}\label{tab::tweedie-edms}
		\begin{tabular}{|l|c|cc|c|}
		\hline
		Tweedie EDMs & $p$ & $S$ & $\mathit{\Omega}$ & Examples\\
		\hline
		Extreme stable & $p<0$ & $\mathbb{R}$ & $\mathbb{R}^{+}$& \\
		Normal & $p=0$ & $\mathbb{R}$ & $\mathbb{R}$& --\\
		No EDMs exist & $0<p<1$ & -- & -- & --\\
		Discrete & $p=1$ & $\mathbb{N}\cup \{0\}$ & $\mathbb{R}^{+}$ & Poisson \\
		Poisson-gamma & $1<p<2$ & $\mathbb{R}^{+}\cup\{0\}$ & $\mathbb{R}^{+}$ & --\\
		Gamma & $p=2$ & $\mathbb{R}^{+}$ & $\mathbb{R}^{+}$& --\\
		Positive stable & $p>2$ & $\mathbb{R}^{+}$ & $\mathbb{R}^{+}$ & inverse Gaussian\\
		\hline
	\end{tabular}
\end{table}

\noindent to zeros such that ``$\log(0)$" problems are avoided when fitting GLMs. Apart from applications in varied fields of study, notable developments in methodology include variable selection procedures involving grouped elastic net being developed for Tweedie compound Poisson-gamma models (\cite{qian2016tweedie}), likelihood based Bayesian approaches, that are well-suited alternatives to quasi-likelihood methods in terms of inference for Tweedie compound Poisson \emph{mixed} models have been explored (\cite{zhang2013likelihood}). Machine learning algorithms, like gradient boosting have been used in conjunction with compound Poisson-gamma models to achieve better performance while predicting insurance premiums (\cite{yang2018insurance}). 

\begin{figure}[ht]
	\centering
	\begin{subfigure}{.33\textwidth}
		\centering
		\includegraphics[width=1\linewidth , height=1\linewidth]{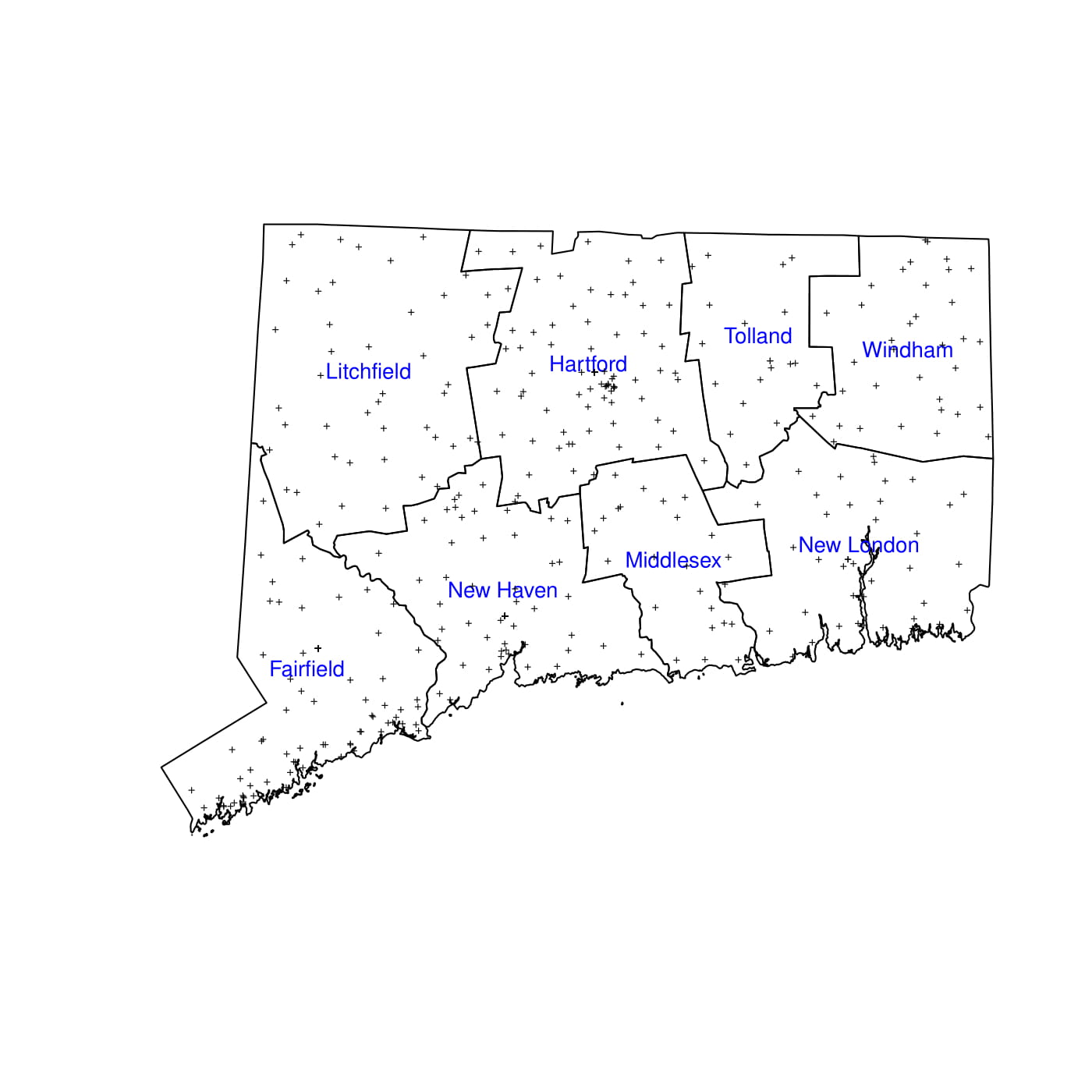}
		\caption{}
	\end{subfigure}
	\begin{subfigure}{.33\textwidth}
		\centering
		\includegraphics[width=1\linewidth , height=1\linewidth]{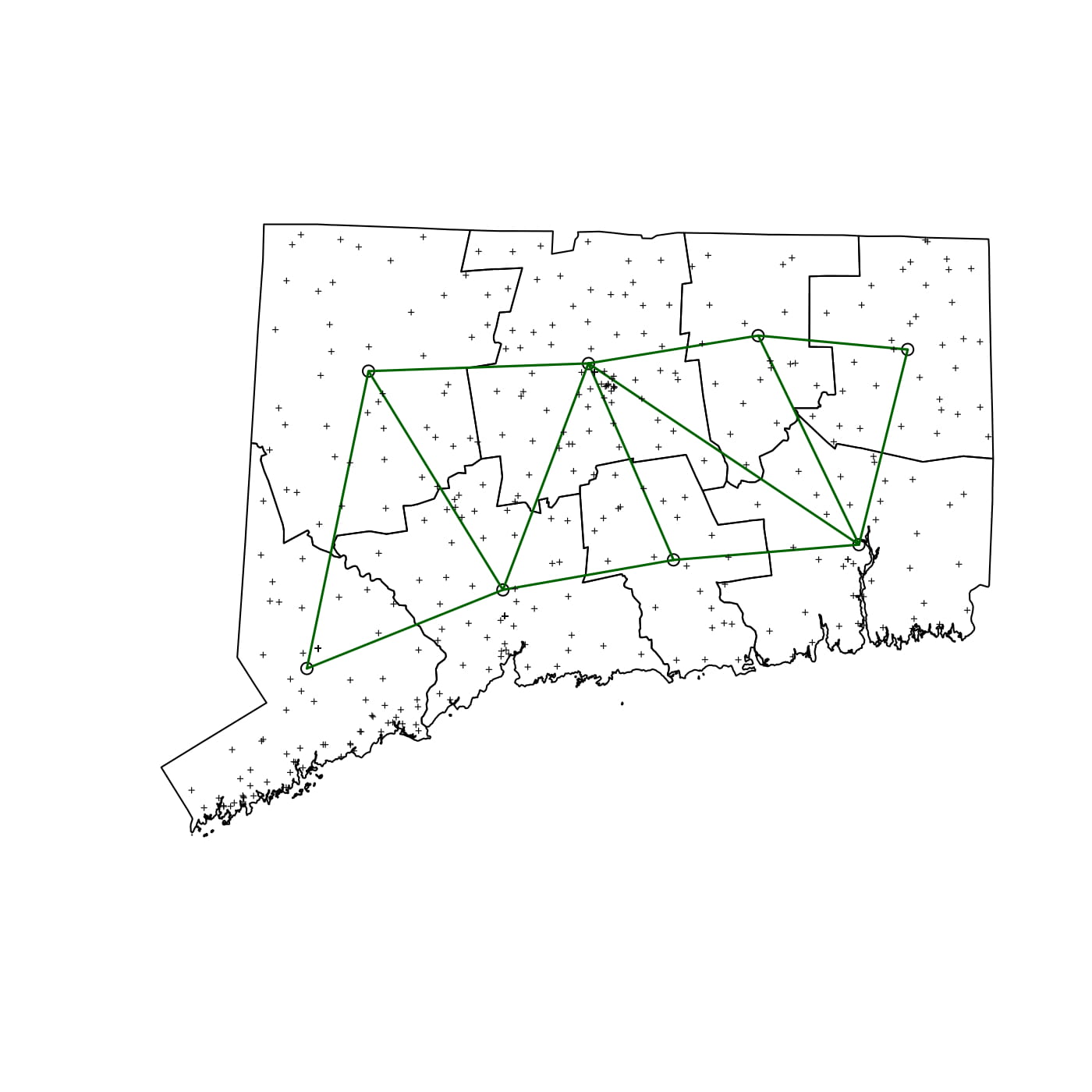}
		\caption{}
	\end{subfigure}%
	\begin{subfigure}{.33\textwidth}
		\centering
		\includegraphics[width=1.1\linewidth , height=1\linewidth]{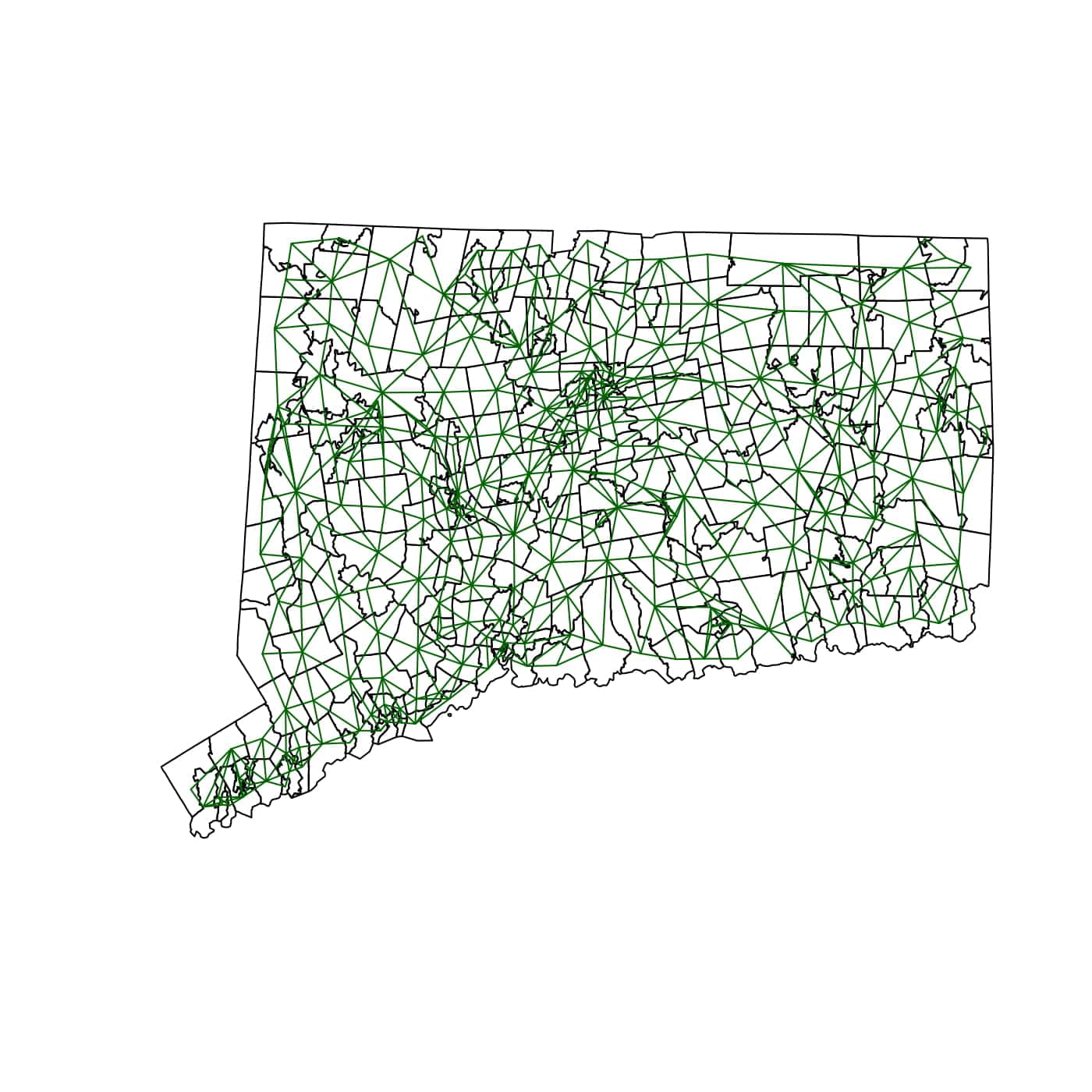}
		\caption{}
	\end{subfigure}
		\caption{Figure showing the (a) 8 counties with zipcodes, (b) construction of adjacency based on 8 counties, (c)  construction of adjacency based on 282 zip-codes for state of Connecticut.}
	\label{fig::ct-1}
\end{figure}

Another integral fact that is apparent from examples mentioned above, is existence of spatial information in terms of location, which may be unrecorded and hence associated variation being unaccounted for by implemented models. Presence of such information could help explorations into effects that adjacent regions have on the response. Throughout this paper we use a \emph{first-order adjacency} to account for spatial correlation among locations or regions. The concept of first-order adjacency simply states, if two regions share boundaries then we put an edge between their centroids, and call them neighbors. Naturally, adjacency information has dual representations, a graph (as shown in (b) and (c) in fig. \ref{fig::ct-1}) or a matrix, the order being number of locations, $L$. We can also see from figures (b) and (c) that depending on a chosen level of granularity, an adjacency graph can have small (county level) to sufficiently large number (zipcode level) of edges.  Adjacency matrices have been used in multiple situations to capture spatial correlation, some of the earliest and celebrated applications being simultaneous and conditional autoregressive areal models (SAR and CAR respectively) (\cite{whittle1954stationary}, \cite{besag1974spatial}). They are also employed extensively in developing hierarchical Bayesian spatial areal models (\cite{besag1991bayesian}). Formally stating, if we denote an adjacency matrix by $A=(A_{i_1i_2})$ of the order $L \times L$ then,

\begin{align}\label{eq::neighbor}
A_{i_1i_2}=\begin{cases}
1	& \text{if } i_2 \in \mathcal{N}(i_1)\\
0 & \text{otherwise}
\end{cases},
\end{align}
where $\mathcal{N}(i_1)$ denotes the set containing neighbors (i.e., locations that share a boundary) of location $i_1$. In the ensuing discussion we shall only look at adjacency matrices generated by locations that share boundaries, or are first-order neighbors and not locations that are ``neighbors" of neighbors (i.e second/higher order).

In section \ref{sec::DGLM-twmod} we consider revisiting the formulation, development and necessary details regarding double GLMs (DGLMs) for Tweedie compound Poisson-gamma distributions. Reviewing the choice of an appropriate likelihood function and further characteristics of the Tweedie family of distributions are also discussed in this section. We consider incorporating spatial information as an un-observable fixed effect into the formulated DGLM, where \emph{penalized estimation} of the same fixed effect leads to solving an optimization problem.  Section \ref{sec::comp} states and discusses necessary conditions for obtaining a solution to this problem. Consequently, the approach we describe essentially fits a DGLM having a penalized spatial fixed effect, where nature of the penalty chosen promotes structural shrinkage.  Simulations that demonstrate the efficacy of our approach over existing alternatives are shown in section \ref{sec::sim}. The penalty we propose can be interpreted as an extension of the ridge penalty (\cite{friedman2001elements}), naturally making it a baseline for comparing performance. Section \ref{sec::cs} describes the nature and results of our real data application, showcasing performance of our approach for characteristically different response surfaces in selective states or, groups of states in USA. Finally, sections \ref{sec::conc}, \ref{sec::disc-future} and Appendix include conclusions comments about further developments and required proofs, tables and additional figures respectively.

\section{Characterizations and DGLMs  for compound Poisson distributions}\label{sec::DGLM-twmod} 

We start this section with some notation, that will be used throughout the ensuing discussion. Let $y_{ij}, \mu_{ij} \in \mathbb{R}, \phi_{ij} \in \mathbb{R}^{+}$ denote responses, mean, dispersion respectively and, $x_{ij}\trans \in \mathbb{R}^{m_1}$, $z_{ij}\trans  \in \mathbb{R}^{m_2}$ denote observed covariate vectors (which can be adjusted to include an intercept, i.e. the first entry is 1) for $j$-th observation at $i$-th location, where $i=1,\ldots,L$ and $j=1,\ldots, n_i$; with $\sum_{i=1}^{L}n_i=N$ being total number of observations (notations without subscripts are used to describe model formulations).

\subsection{Probability density characterizations}

 Probability distributions for EDMs, where the response, $y$ can be discrete or continuous depending on the problem at hand can be expressed as,
 
\begin{align}\label{eq::edm-dens}
f(y;\theta,\phi) = a(y,\phi) \exp\Bigg\{\frac{y\theta-\kappa(\theta)}{\phi}\Bigg\}.
\end{align}
In eq. (\ref{eq::edm-dens}) $\theta$ is called the canonical parameter, $\kappa(\theta)$ is a known function called the cumulant function, $\phi$ is the dispersion parameter and $a(y,\phi)$ is a normalizing constant that ensures (\ref{eq::edm-dens}) is a probability function. Eq. (\ref{eq::edm-dens}) is called the canonical form for EDM densities, with other parametrizations being possible. As we can see from table (\ref{tab::tweedie-pars}) that $a(y,\phi)$ may not always have a closed form (in case of compound Poisson-gamma). For EDMs we have some well-known relations, $E(y)=\mu=\kappa\prm(\theta)$ and $\text{Var}(y)=\phi\kappa\dprm(\theta)$ (\cite{nelder1972generalized}, \cite{mccullagh2018generalized}, \cite{dunn2018generalized}). Due to the relationship/map between $\theta$ and $\mu$, $\kappa\dprm(\theta)$ can also be expressed as a function of $\mu$, which is denoted by \emph{variance function} $V(\mu)$; which uniquely corresponds to an exponential dispersion model (\cite{jorgensen1987exponential}, \cite{dunn2018generalized} pgs. 217, \cite{barndorff2014information}). We focus on EDMs with a power variance function $\phi\mu^p$, where $p \in (1,2)$, table (\ref{tab::tweedie-pars}) shows necessary details of probability density (or mass, if discrete) functions for other members of the Tweedie family. In particular, the probability density function for a compound Poisson-gamma distribution, where $p \in (1,2)$ can be expressed as,

\begin{align}\label{eq::tw-mod-dens-0}
\scalemath{0.95}{f(y;\mu,\phi, p) = a(y,\phi,p) \exp\Bigg\{-\frac{2}{\phi}\int_{y}^{\mu}\frac{y-u}{V(u)}\mathrm{d}u\Bigg\} =a(y,\phi,p) \exp\Bigg\{-\frac{2}{\phi}\int_{y}^{\mu}\frac{y-u}{u^p}\mathrm{d}u\Bigg\}},
\end{align}
where $d(y,\mu)=-2\int_{y}^{\mu}\frac{y-u}{V(u)}\mathrm{d}u$ is defined as the deviance, i.e. a measure of discrepancy between observation, $y$ and its expected value, $\mu$. An alternative characterization for a known $p \in (1,2)$, random variable $Y$ follows a compound Poisson-gamma distribution (\cite{jorgensen1987exponential}) if,

\begin{align}\label{eq::tw-mod-dens-1}
\small
Y=\begin{cases} 0 & M=0\\
\sum\limits_{i=1}^{M}C_i  & M>0 \end{cases}&,& M \sim \text{Poisson}(\xi) &,& C_i \stackrel{iid}{\sim} \text{Gamma}(\eta,\zeta),
\end{align}

\begin{table}[ht]
	\centering
	\caption{Common members of the Tweedie family of distributions with their index parameter ($p$), variance function ($V(\mu)$), cumulant function ($\kappa(\theta)$), canonical parameter ($\theta$), dispersion ($\phi$), deviance ($d(y,\mu)$), normalizing constant ($a(y,\phi)$), support ($S$), and respective parameter spaces for mean ($\mathit{\Omega}$) and the natural parameter ($\Theta$).}\label{tab::tweedie-pars}
	\resizebox{\linewidth}{!}{
		\begin{tabular}{|l|cccccccccc|}
			\hline
			&&&&&&&&&&\\
			Tweedie EDMs & $p$ & $V(\mu)$ & $\kappa(\theta)$ & $\theta$ & $\phi$ & $d(y,\mu)$ & $a(y,\phi)$ & $S$ & $\mathit{\Omega}$ & $\Theta$ \\
			\hline
			Normal & 0 & 1 & $\frac{\theta^2}{2}$ & $\mu$ & $\sigma^2$ & $(y-\mu)^2$ & $\exp\{-(y^2/\sigma^2+\log 2\pi)/2\}$ & $\mathbb{R}$ & $\mathbb{R}$ & $\mathbb{R}$\\
			Poisson & 1 & $\mu$ & $\exp(\theta)$ & $\log(\mu)$ & 1 & $2\Big\{y\log\frac{y}{\mu}-(y-\mu)\Big\}$ & $\frac{1}{y!}$ & $\mathbb{N}\cup\{0\}$ & $\mathbb{R}^{+}$ & $\mathbb{R}$\\
			Poisson-gamma & $(1,2)$ & $\mu^p$ & $\frac{\{(1-p)\theta\}^{(2-p)/(1-p)}}{2-p}$ & $\frac{\mu^{1-p}}{1-p}$ & $\phi$ & $2\Big\{\frac{\max(y,0)^{2-p}}{(1-p)(2-p)}-\frac{y\mu^{1-p}}{1-p}+\frac{\mu^{2-p}}{2-p}\Big\}$ & -- & $\mathbb{R}^{+}\cup\{0\}$ & $\mathbb{R}^{+}$ & $\mathbb{R}^{-}$\\
			Gamma & 2 & $\mu^2$ & $-\log(-\theta)$ & $-\frac{1}{\mu}$ & $\phi$ & $2\Big\{-\log\frac{y}{\mu}+\frac{y-\mu}{\mu}\Big\}$ & $\frac{\phi^{-1/\phi} y^{1/\phi-1}}{\Gamma(1/\phi)}$ & $\mathbb{R}^{+}$ & $\mathbb{R}^{+}$ & $\mathbb{R}$\\
			\hline
		\end{tabular}
	}
\end{table}
 \noindent where, $M$ is independent of $C_i$. Above definition also demonstrates the fact that at $M=0$, $Y=0$ with some  non-zero probability, followed by $M>0$ resulting in a sum of gamma random variables producing a skewed continuous distribution on $\mathbb{R}^{+}$. It can be shown that the two characterization are equivalent by deriving and equating cumulant generating functions for densities in eq. (\ref{eq::tw-mod-dens-0}) and (\ref{eq::tw-mod-dens-1}) (\cite{smyth2002fitting}, \cite{zhang2013likelihood}). From table (\ref{tab::tweedie-pars}) and eq. (\ref{eq::tw-mod-dens-1}) it is evident that $a(y,\phi,p)$ needs to be approximated for obtaining a closed form of the density while characterizing compound Poisson-gamma densities as EDMs. Analogously, evaluating the marginal density of $Y$ in alternative characterization shown in eq. (\ref{eq::tw-mod-dens-1}) results in an infinite sum representation of $a(y,\phi,p)$ which can be approximated in multiple ways (\cite{dunn2005series}, \cite{dunn2008evaluation}).

\subsection{Double generalized linear models}

DGLMs were considered as a further generalization to GLMs for exponential families (\cite{pregibon1984}), where the dispersion parameter $\phi$ was no longer required to be a constant across observations. This resulted in simultaneous GLMs where mean $\mu$ and dispersion $\phi$ both varied across observations as described through the general quasi-likelihood model,

\begin{align}\label{eq::DGLM}
g_1(\mu) = x\trans\bbeta &,& \mathrm{var}(y) = \phi V(\mu) &,& g_2(\phi) = z\trans\bgamma.
\end{align}
Here $g_1(\cdot)$, $g_2(\cdot)$ are monotonic link functions, $\bbeta \in \mathbb{R}^{m_1}$, $\bgamma \in \mathbb{R}^{m_2}$ are model coefficients and $x\trans \in \mathbb{R}^{m_1}$ and $z\trans \in \mathbb{R}^{m_2}$ are predictor vectors  for mean and dispersion GLMs respectively. Based on how we want to explain the response using predictors or covariates, the type of characterization along with choice of likelihood approximation adopted varies. For instance, in likelihood-based variable selection approaches (\cite{qian2016tweedie}), where $\phi$ is assumed to be constant resulting in a GLM for mean, any desired approximation of $a(y,\phi,p), ~ 1<p<2$ can be accommodated into the estimation procedure. Whereas in likelihood based methods like (\cite{zhang2013likelihood}), the type of approximation used affects the parameter estimation significantly. Alternatively, an approach that allows for a variable dispersion $\phi$ (like in eq. (\ref{eq::DGLM})) would require a particular approximation of $a(y,\phi,p)$, producing extended quasi-likelihoods (EQLs) for compound Poisson models (\cite{smyth2002fitting}, \cite{nelder1987extended}, \cite{nelder1992likelihood}). We will be working with DGLMs, using marginal likelihood approximations involving Fourier inversion of the characteristic function (\cite{dunn2008evaluation}) and joint likelihood for $(Y,M)$ (as defined in eq. (\ref{eq::tw-mod-dens-1})). The joint likelihood for $(y,m)$ is,

\begin{subequations}
	\begin{align}\label{eq::tw-mod-dens-2-1}
	\alpha=\frac{2-p}{p-1} &,& a(m,y,\phi,p) =\frac{1}{m!\Gamma(m\alpha)y}\Bigg\{\frac{y^\alpha\phi^{-(\alpha+1)}}{(p-1)^{\alpha}(2-p)}\Bigg\}^{m},
	\end{align}
	
	\begin{align}\label{eq::tw-mod-dens-2-2}
	f(y,m;\mu,\phi,p) = a(m,y,\phi,p)\exp\Bigg\{\frac{1}{\phi}t(y,\mu,p)\Bigg\} &,& t(y,\mu,p) = y\frac{\mu^{1-p}}{1-p}-\frac{\mu^{2-p}}{2-p}.
	\end{align}
\end{subequations}
However, if $m=y=0$, $f(0,0;\mu,\phi,p)=\exp\{-\frac{1}{\phi}t(0,\mu,p)\}=\exp\big\{-\frac{1}{\phi}\frac{\mu^{2-p}}{2-p}\big\}$, which completes the specification (\cite{jorgensen1994fitting}, \cite{smyth2002fitting}). In this paper we extend DGLM in eq. (\ref{eq::DGLM}) to include an additional fixed effect,

\begin{align}
	g_1(\mu) = x\trans\bbeta+r\trans\balpha &,& \mathrm{var}(y) = \phi V(\mu) &,& g_2(\phi) = z\trans\bgamma,
\end{align}
where $\balpha \in \mathbb{R}^{L\times 1}$ is the un-observable fixed effect. Particularly we interpret $\alpha_i$ as a spatial effect corresponding to location $i$, hence $r$ is a ${L\times 1}$ vector of 0's with 1 in exactly one entry indicating the corresponding index for a location. 

In case of an existing implementation of a DGLM, the quantities $o^{(1)} = x\trans\widehat{\bbeta}$ and $o^{(2)} = z\trans\widehat{\bgamma}$  and $p$ are known. Hence the \emph{negative log-likelihood} is given by,

\begin{align}\label{eq::tw-offset-neg-log-lik}
\scalemath{0.95}{
		\ell(\balpha) = -\sum\limits_{i=1}^{L}\sum\limits_{j=1}^{n_i}\frac{1}{g_2^{-1}\big(o^{(2)}_{ij}\big)}t\Big(y_{ij}, g_1^{-1}\big(o^{(1)}_{ij}+r_{ij}\trans\balpha\big),p\Big)+c\Big(m_{ij},y_{ij},g_2^{-1}\big(o^{(2)}_{ij}\big),p\Big)I(y_{ij}>0),
}
\end{align}
where $c(\cdot)$ is a known function (\cite{dunn2005series} pg. 6) and $I(\cdot)$ stands for the indicator function. We shall assume that $p$ is known from the existing implementation of DGLM. For most implementations link functions, $g_1(\cdot)$ and $g_2(\cdot)$ are both assumed to be logarithmic, and the covariate vectors/matrices $x\trans$ and $z\trans$ need not necessarily be the same (\cite{smyth2002fitting}). For a GLM, the \emph{canonical link} function is defined as $g(\cdot)=(\kappa\prm)^{-1}(\cdot)$ such that $E(y)=\kappa\prm(\theta)=g^{-1}(\theta)$, where $^{\prime}$ denotes the first derivative of a function. It can be readily seen from eqs. (\ref{eq::tw-mod-dens-0}), (\ref{eq::tw-mod-dens-2-2}) and  table (\ref{tab::tweedie-pars}) that a logarithmic function is \emph{not} the canonical link for compound Poisson GLMs. 

On further inspection, it can be seen that cross-derivatives of negative log-likelihood for the density $f$ in eq. (\ref{eq::tw-mod-dens-2-2}) have zero expectations, 

\begin{align}\label{eq::prop-cross-0}
\scalemath{0.9}{E\Bigg(-\frac{\partial^2}{\partial \phi\partial \mu}\log f\Bigg)=E\Bigg(\frac{1}{\phi^2}\frac{y-\mu}{\mu^p}\Bigg)=0} &,& \scalemath{0.9}{E\Bigg(-\frac{\partial^2}{\partial p\partial \mu}\log f\Bigg)=E\Bigg(-\frac{\log(\mu)}{\phi}\frac{y-\mu}{\mu^p}\Bigg)=0}.
\end{align}
As a result, off-diagonal elements for $\mu$ in the Fisher's information matrix of $(\mu,\phi,p)$ are 0, implying $\mu$ is statistically orthogonal to $\phi$ and $p$, resulting in $\bgamma$, $p$ being independent of $\bbeta$, $\balpha$. This property insulates the estimation of $\balpha$ from inaccuracies that are associated with using likelihood approximations (EQLs or penalized EQLs) in the existing implementation of a DGLM (for details see \cite{smyth2002fitting}, pgs. 148 or, \cite{zhang2013likelihood} pgs. 747, this will also be demonstrated through a sensitivity analysis to $p$ in section \ref{sec::sim}).

\section{Algorithm and computation}\label{sec::comp}

\subsection{Graph Laplacian}\label{subsec::graph-lap}

Before getting into the optimization problem, we digress briefly to illustrate relevant and related concepts in graph theory. A graph is represented by $(V,E)$, where $V$ is a set of vertices and $E$ is a set of edges, or pairs $(i_1,i_2) \in V$. A graph is said to be un-directed if $(i_1,i_2) \in E \Leftrightarrow (i_2,i_1) \in E$ (for example, see figure below in eq. (\ref{eq-fig::graph-lap})); in the notation of eq. (\ref{eq::neighbor}) $i_2 \in \mathcal{N}(i_1) \Leftrightarrow i_1 \in \mathcal{N}(i_2)$. A diagonal matrix defined as, $D=(D_{i_1i_1})=\sum_{i_2}A_{i_1i_2}$ is defined as a degree matrix, where $A$ is an adjacency matrix as defined in eq. (\ref{eq::neighbor}). An example is shown below in figure and equation (\ref{eq-fig::graph-lap}).

\begin{align}\label{eq-fig::graph-lap}
  \includegraphics[width=0.31\linewidth, valign=c]{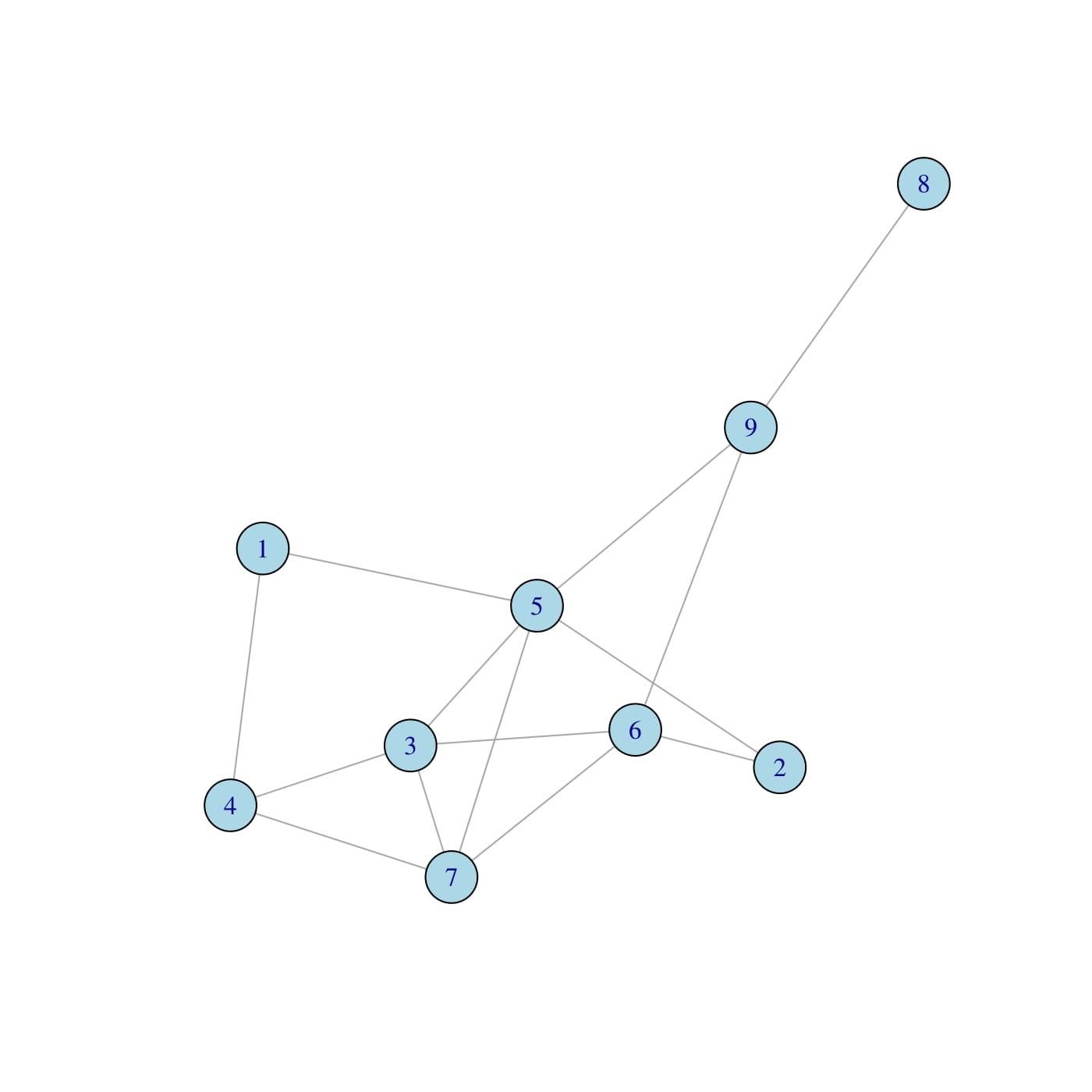} 
  && 
  A= \scalemath{0.7}{
  	\begin{pmatrix}
	  0 & 0 & 0 & 1 & 1 & 0 & 0 & 0 & 0 \\ 
	  0 & 0 & 0 & 0 & 1 & 1 & 0 & 0 & 0 \\ 
	  0 & 0 & 0 & 1 & 1 & 1 & 1 & 0 & 0 \\ 
	  1 & 0 & 1 & 0 & 0 & 0 & 1 & 0 & 0 \\ 
	  1 & 1 & 1 & 0 & 0 & 0 & 1 & 0 & 1 \\ 
	  0 & 1 & 1 & 0 & 0 & 0 & 1 & 0 & 1 \\ 
	  0 & 0 & 1 & 1 & 1 & 1 & 0 & 0 & 0 \\ 
	  0 & 0 & 0 & 0 & 0 & 0 & 0 & 0 & 1 \\ 
	  0 & 0 & 0 & 0 & 1 & 1 & 0 & 1 & 0 \\ 
  	\end{pmatrix}}%
   &,& 
  D=\scalemath{0.7}{
  \begin{pmatrix}
  2 & 0 & 0 & 0 & 0 & 0 & 0 & 0 & 0 \\ 
  0 & 2 & 0 & 0 & 0 & 0 & 0 & 0 & 0 \\ 
  0 & 0 & 4 & 0 & 0 & 0 & 0 & 0 & 0 \\ 
  0 & 0 & 0 & 3 & 0 & 0 & 0 & 0 & 0\\ 
  0 & 0 & 0 & 0 & 5 & 0 & 0 & 0 & 0\\ 
  0 & 0 & 0 & 0 & 0 & 4 & 0 & 0 & 0\\ 
  0 & 0 & 0 & 0 & 0 & 0 & 4 & 0 & 0\\ 
  0 & 0 & 0 & 0 & 0 & 0 & 0 & 1 & 0\\ 
  0 & 0 & 0 & 0 & 0 & 0 & 0 & 0 & 3\\ 
  \end{pmatrix}}%
\end{align}%
The \emph{Laplacian} for a graph is defined as $W=D-A$, which is an element-wise difference between degree and adjacency matrices. If we interpret spatial effect as functions, $\balpha: V \to \mathbb{R}$ then the graph Laplacian, $W$ can be equivalently defined as a linear operator,

\begin{align}\label{eq::graph-lap-1}
\balpha\trans W \balpha = \frac{1}{2}\sum\limits_{i_1=1}^{L}\sum\limits_{\scalemath{0.7}{i_2\in \mathcal{N}(i_1)}}(\alpha_{i_1}-\alpha_{i_2})^2 &,& \forall ~ \balpha \in \mathbb{R}^{L}.
\end{align}
As a result, regularization based on graph Laplacian matrices penalize change between adjacent vertices thereby introducing local smoothness (\cite{smola2003kernels}).  In this paper we work with un-directed graphs describing neighborhood structures corresponding to locations (for example, latitude-longitude pairs for counties or zipcodes, as shown in fig. (\ref{fig::ct-1})) on a map, associated adjacency, degree and Laplacian matrices.

\begin{figure}[H]
	\centering
	\begin{subfigure}{.33\textwidth}
		\centering
		\includegraphics[width=1\linewidth , height=1\linewidth]{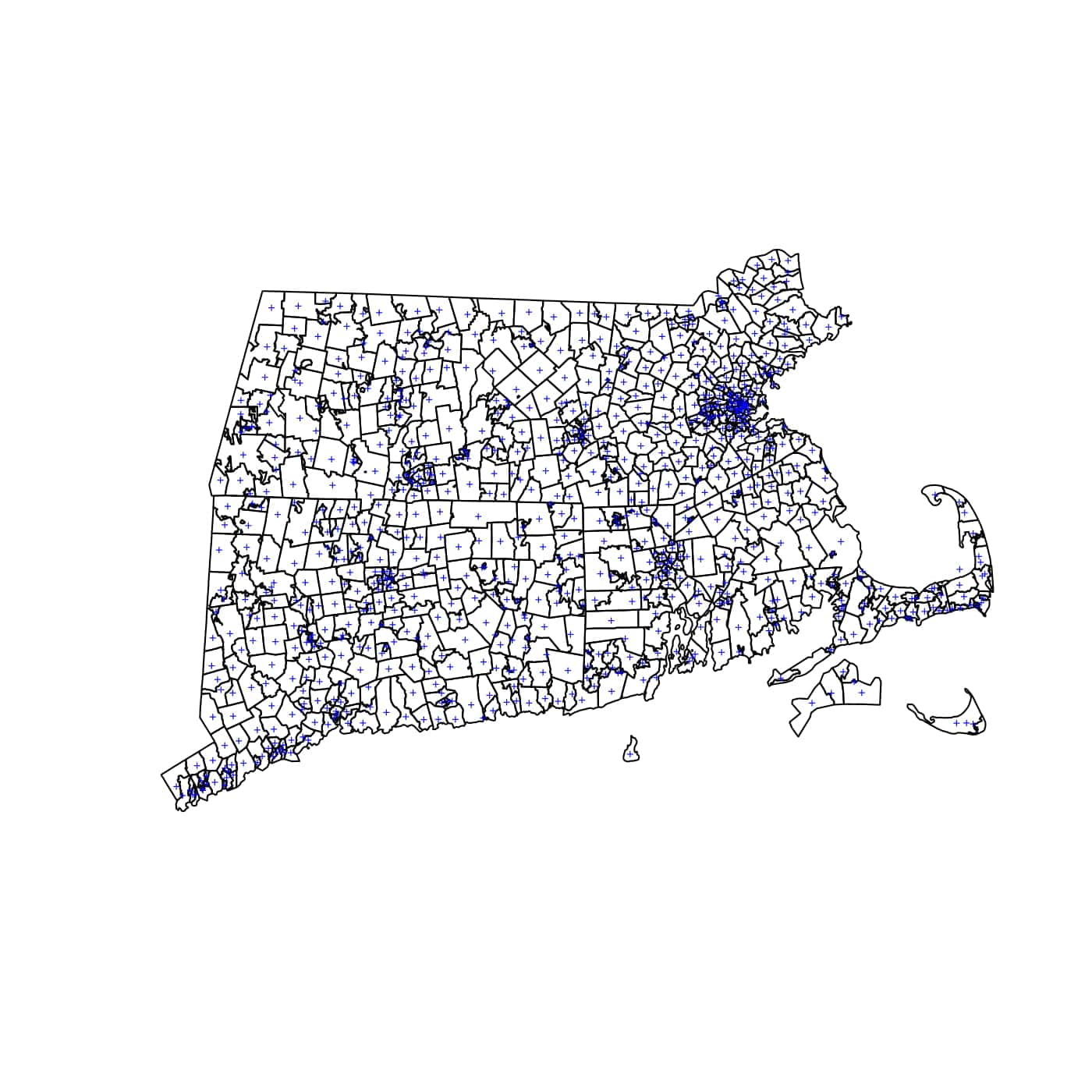}
		\caption{}
	\end{subfigure}
	\begin{subfigure}{.33\textwidth}
		\centering
		\includegraphics[width=1\linewidth , height=1\linewidth]{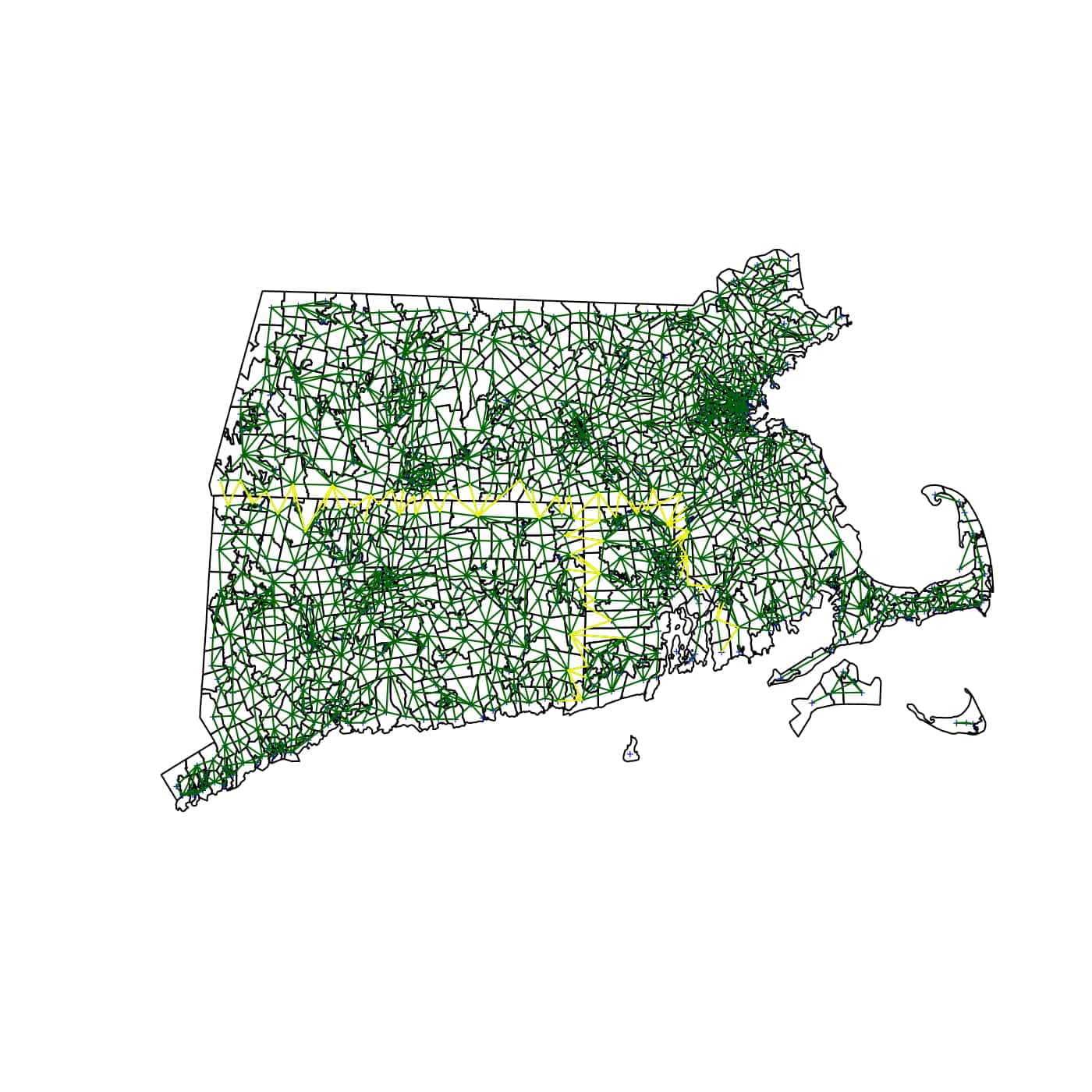}
		\caption{}
	\end{subfigure}%
	\begin{subfigure}{.33\textwidth}
		\centering
		\includegraphics[width=1.1\linewidth , height=1\linewidth]{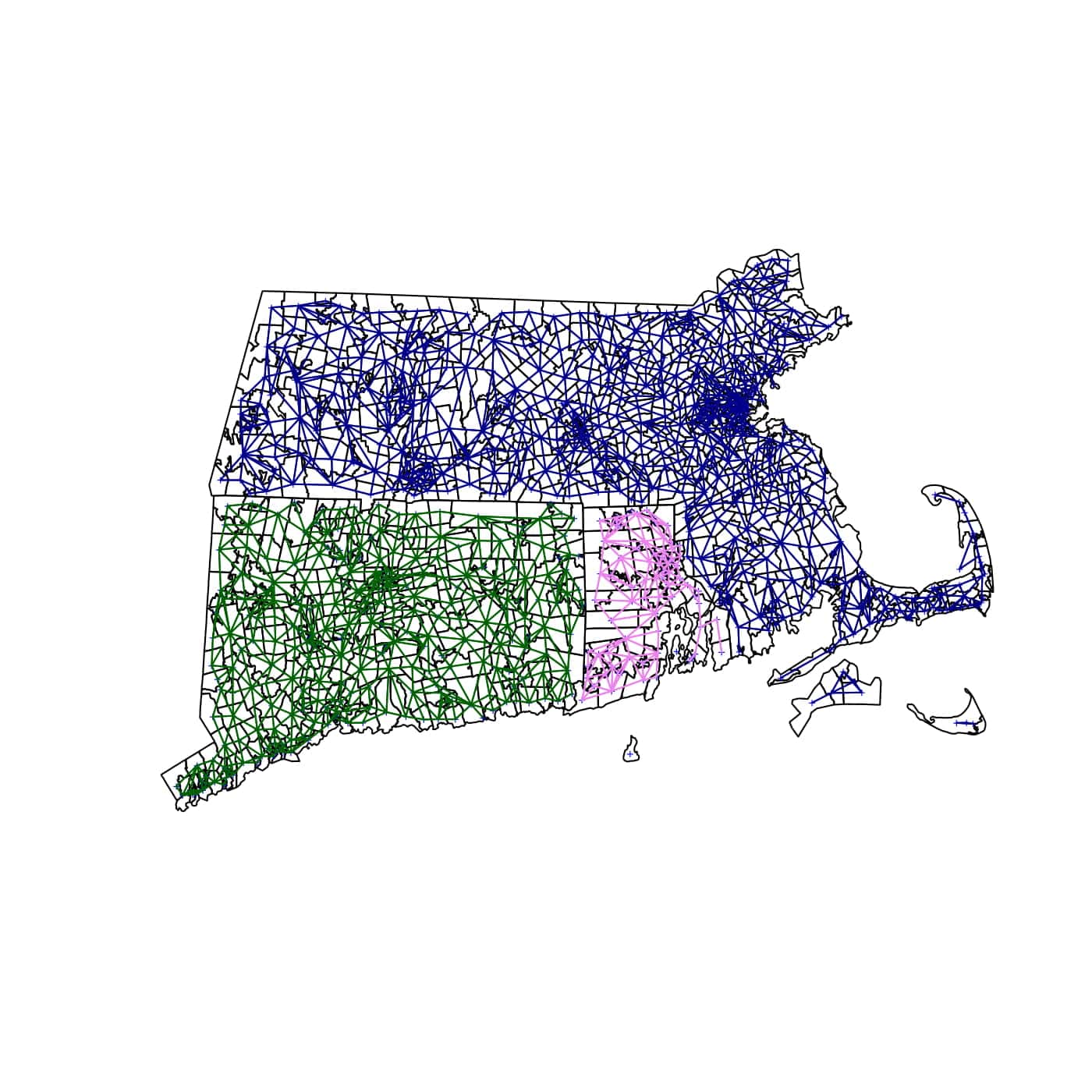}
		\caption{}
	\end{subfigure}
	\caption{Plots showing (a) 896 zipcodes for three states Connecticut (CT), Massachusetts (MA) and Rhode Island (RI), locations marked by \texttt{+} (b) the zipcode level adjacency matrix, with edges between zipcodes within any state colored \texttt{darkgreen}, and edges between zipcodes across states colored \texttt{yellow} (c) zipcode level adjacency graphs colored w.r.t states, after the boundary edges have been removed (adjacency graphs for CT, MA and RI are in \texttt{darkgreen}, \texttt{darkblue} and \texttt{violet} respectively).}
	\label{fig::3st-approx}
\end{figure}

Sparse matrices consist larger number of zero entries; the adjacency matrix, $A$ in eq. (\ref{eq-fig::graph-lap}) is an example of a sparse matrix. Working with a single state, adjacency matrices for neighborhood structures between zipcodes are comparatively less \emph{sparse} as opposed to adjacency matrices generated when considering multiple states together. If we consider proportion of zero entries as a measure of sparsity for adjacency matrices, then considering states of Connecticut (CT), Massachusetts (MA) ad Rhode Island (RI) in conjunction the adjacency matrix generated contains 99.4\% zeros; individually however, adjacency matrices for CT, MA and RI contain 98.1\%, 99.02\% and 94.1\% of zero entries. We use this sparsity to our advantage by introducing an \emph{approximation} to adjacency for matrices when considering multiple states in our analysis. Let us consider $S$ states in conjunction, if $A$ denotes the associated adjacency matrix, we approximate it by, $A_a$ defined as

\begin{align*}
A =A_a+A_\epsilon &, & A_a =  \begin{pmatrix}
A_1 & O & \cdots & O\\
O & A_2 & \cdots & O\\
\vdots & \vdots & \ddots & \vdots\\
O & O & \cdots & A_S 
\end{pmatrix}
&,& A_\epsilon = 
\begin{pmatrix}
O & \epsilon_{1,2} & \cdots &  \epsilon_{1,S}\\
\epsilon_{1,2} & O & \cdots & \epsilon_{2,S}\\
\vdots & \vdots & \ddots & \vdots\\
\epsilon_{1,S}  & \epsilon_{2,S}  & \cdots & O
\end{pmatrix},
\end{align*}
where $A_1, \ldots, A_S$ are the adjacency matrices for $S$ states respectively, and $\epsilon_{k_1,k_2}$, with $k_1, k_2 = 1,2,\ldots, S$, can be interpreted as the cross-state adjacency for pair $(k_1,k_2)$. Figure (\ref{fig::3st-approx}) illustrates this concept for $S=3$ states viz., CT, MA and RI. This approximation is carried forward to associated degree and Laplacian matrices. Consequently, resulting Laplacian matrices are block diagonal. The computational advantage of having block diagonal Laplacian matrices is immediately evident from nature of solution to the optimization problem described in the following.

\subsection{Optimization problem}\label{subsec::optim-prob}
Consider spatial effect, $\balpha = (\alpha_1, \alpha_2, \ldots, \alpha_L)\trans$, in this section we primarily focus on solving the minimization problem,

\begin{align}\label{eq::opt-prob}
\scalemath{0.95}{\widehat{\balpha} = \mathrm{arg}\min\limits_{\balpha} F(\balpha)} &,&  \scalemath{0.95}{F(\balpha) = \ell(\balpha) + P (\balpha; \lambda_1, \lambda_2)} &,& \scalemath{0.95}{P (\balpha; \lambda_1, \lambda_2) = \frac{1}{2}\big[\lambda_1\balpha\trans\balpha+\lambda_2\balpha\trans W\balpha\big],}
\end{align}
where $W$ is the graph Laplacian as defined in eq. (\ref{eq::graph-lap-1}), $\lambda_1, \lambda_2 >0$ are tuning parameters for the ridge and Laplacian regularization respectively. Under a logarithmic link the negative log-likelihood is,

\begin{align*}
\ell(\balpha) = \sum\limits_{i}\sum\limits_{\scalemath{0.7}{j}}\widehat{\phi}^{-1}_{ij}\Bigg[y_{ij}\frac{e^{-(p-1)(\widehat{\mu}_{ij}+r_{ij}\trans\scalemath{0.8}{\balpha})}}{p-1}+\frac{e^{(2-p)(\widehat{\mu}_{ij}+r_{ij}\trans\scalemath{0.8}{\balpha})}}{2-p}\Bigg] - c\Big(y_{ij},\widehat{\phi}_{ij},p\Big)I(y_{ij}>0),
\end{align*}
where $\widehat{\mu}_{ij}$ and $\widehat{\phi}_{ij}$ are fitted mean and dispersion respectively. Flexibility in structure and advantages of having $P(\balpha, \lambda_1,\lambda_2)$ as a penalty is immediately evident for instances where spatial clustering needs to be introduced. The ridge part penalizes magnitude of estimated spatial effects by regularizing $L^2$-norm, $||\balpha||^2_2=\langle \balpha, \balpha\rangle = \balpha\trans\balpha$, while penalty on the Laplacian promotes local neighborhood smoothing on vertices by regularizing induced semi-norm, $||\balpha||_W=\langle \balpha, W\balpha\rangle =\balpha\trans W\balpha$. Solution to optimization problem in eq. (\ref{eq::opt-prob}) is obtained using a majorization descent (MD) algorithm, which utilizes the majorization-minimization (MM) principle (for further details see \cite{lange2013mm}, \cite{lange2016mm}, \cite{wu2010mm}). Properties and details of proposed MD algorithm are shown below.

Let us denote the $L\times 1$ gradient vector and, $L\times L$ Hessian matrix for negative log-likelihood $\ell(\balpha)$, as $\nabla_1(\balpha)$ and $\nabla_2(\balpha)$ respectively, having following expressions,

\begin{subequations}
	\begin{align}\label{eq::grad-1}
	\nabla_{1,i}(\balpha) = \del{\ell(\balpha)}{\alpha_i}=\sum\limits_{\scalemath{0.7}{j}} \widehat{\phi}_{ij}^{-1}r\trans_{ij}\Big[-y_{ij}e^{-(p-1)\big(\widehat{\mu}_{ij}+r\trans_{ij}\scalemath{0.85}{\balpha}\big)}+e^{(2-p)\big(\widehat{\mu}_{ij}+r\trans_{ij}\scalemath{0.85}{\balpha}\big)}\Big],
	\end{align}
	
	\begin{align}\label{eq::grad-2}
	\scalemath{0.95}{\nabla_{2,ii}(\balpha) = \deltwo{\ell(\balpha)}{\alpha_i}=\sum\limits_{\scalemath{0.7}{j}}\widehat{\phi}_{ij}^{-1} r_{ij}\trans\Big[(p-1)y_{ij}e^{-(p-1)\big(\widehat{\mu}_{ij}+r_{ij}\trans\scalemath{0.85}{\balpha}\big)}+(2-p)e^{(2-p)\big(\widehat{\mu}_{ij}+r_{ij}\trans\scalemath{0.85}{\balpha}\big)}\Big]r_{ij},}
	\end{align}
\end{subequations}
where $i=1,\ldots,L$ and $\nabla_{2,i_1i_2}(\balpha)=0$ for all $i_1\ne i_2$. Hence, $\nabla_2(\balpha)$ is a diagonal matrix. If $\balpha^{(t)}$ is the updated estimate of spatial effect from $\balpha$, then we obtain $\balpha^{(t)}$ by solving

\begin{align}\label{eq::equiv-prob}
\scalemath{0.95}{\mathrm{arg}\min\limits_{\balpha^{(*)}} \ell(\balpha) +(\balpha^{(*)}-\balpha)\trans\nabla_1(\balpha)+\frac{1}{2}(\balpha^{(*)}-\balpha)\trans \big(\tI_L+\nabla_2(\balpha)\big)(\balpha^{(*)}-\balpha)+P(\balpha^{(*)};\lambda_1,\lambda_2),}
\end{align}
which admits a closed form solution. $\tI_L$ is the $L$-dimensional identity matrix. In fact, after some algebra it can be shown by the Karush-Kuhn-Tucker (KKT) conditions that,

\begin{align}\label{eq::mm-est}
	\balpha^{(t)} = \big[(\lambda_1+1)\tI_L+\lambda_2W+\nabla_2(\balpha)\big]^{-1}\big\{(\tI_L+\nabla_2(\balpha))\balpha-\nabla_1(\balpha)\big\}.
\end{align}
For the estimates $\balpha^{(t)}$ and $\balpha$ we have,

\begin{align}\label{eq::surr-fun}
	\ell(\balpha^{(t)}) \leq \ell(\balpha)+(\balpha^{(t)}-\balpha)\trans\nabla_1(\balpha)+\frac{1}{2}(\balpha^{(t)}-\balpha)\trans \big(\tI_L+\nabla_2(\balpha)\big)(\balpha^{(t)}-\balpha)=\mathcal{L}\big(\balpha^{(t)}\big|\balpha\big).
\end{align}
Above inequality in (\ref{eq::surr-fun}), follows from second order Taylor expansion. Therefore,

\begin{eqnarray}
F(\balpha^{(t)})-F(\balpha)&=& \ell(\balpha^{(t)})+\frac{1}{2}\bigg[\lambda_1||\balpha^{(t)}||^2_2+\lambda_2||\balpha^{(t)}||^2_W\bigg]-\ell(\balpha)-\frac{1}{2}\bigg[\lambda_1||\balpha||^2_2+\lambda_2||\balpha||^2_W\bigg],\nonumber\\
&\leq& \ell(\balpha)+(\balpha^{(t)}-\balpha)\trans\nabla_1(\balpha)+\frac{1}{2}(\balpha^{(t)}-\balpha)\trans \big(\tI_L+\nabla_2(\balpha)\big)(\balpha^{(t)}-\balpha)\nonumber\\
&&+\frac{1}{2}\bigg[\lambda_1||\balpha^{(t)}||^2_2+\lambda_2||\balpha^{(t)}||^2_W\bigg]-\frac{1}{2}\bigg[\lambda_1||\balpha||^2_2+\lambda_2||\balpha||^2_W\bigg]-\ell(\balpha),\nonumber\\
&\leq& 0.\nonumber
\end{eqnarray}
The first inequality follows from eq. (\ref{eq::surr-fun}) and the last equality follows from update in (\ref{eq::equiv-prob}) (for more detailed calculations see Appendix). The algorithm derived above is summarized as Algorithm \ref{algo::md-1}.

\begin{algorithm}[H]
	\caption{The MD algorithm for estimating penalized spatial effects from a fitted compound Poisson model.}\label{algo::md-1}
	\begin{enumerate}
		\item Fit a compound Poisson DGLM without spatial effects $\balpha$, to obtain
		\begin{itemize}
			\item fitted mean $\widehat{\mu}_{ij}$,
			\item fitted dispersion $\widehat{\phi}_{ij}$.
		\end{itemize} 
		\item Initialize $\balpha$.
		\item Repeat until $F(\balpha)$ converges,
		\begin{itemize}
			\item Compute $\nabla_1(\balpha)$ using eq. (\ref{eq::grad-1})
			\item Compute $\nabla_2(\balpha)$ using eq. (\ref{eq::grad-2})
			\item Compute $\balpha^{(t)}$ using eq. (\ref{eq::mm-est})
			\item Set $\balpha = \balpha^{(t)}$
		\end{itemize}
	\end{enumerate}
\end{algorithm}

\begin{mythe}\label{th::con-algo-1}
	For $\lambda_1>0$, the sequence $\{\balpha^{(t)}\}$ produced by Algorithm \ref{algo::md-1} satisfies,
	
	\begin{align*}
		F\big(\balpha^{(t)}\big)-F\big(\balpha^{(t+1)}\big) \geq \frac{1+\lambda_1}{2}||\balpha^{(t)}-\balpha^{(t+1)}||^2_2.
	\end{align*}
\end{mythe}
\noindent Theorem \ref{th::con-algo-1} shows that the objective function $F(\balpha)$ is guaranteed to decrease for all  $\lambda_1 > 0$. Proof for theorem \ref{th::con-algo-1} is postponed to Appendix for sake of brevity. 

In Algorithm \ref{algo::md-1}, a \emph{convergence criteria} can be selected based on either the objective function, $F(\balpha)$ or iterative estimates $\balpha^{(t)}$, $\balpha$, i.e. for an arbitrarily small quantity $\epsilon$, repeat until $F(\balpha)-F(\balpha^{(t)})<\epsilon$, or equivalently $||\balpha^{(t)}-\balpha||^2_2 < 2\epsilon/(\lambda_1+1)$. An intercept $\alpha_0$ can be included in the model, as is common practice, it is not penalized. Its estimate can be obtained by direct minimization of the negative log-likelihood at each step. An intercept is interpreted as an overall average spatial effect for all $L$ locations. 

The estimated spatial effects with no penalty (un-penalized) and the ridge penalty, which are used as baselines for comparison, can be obtained by following a similar algorithm, only change being $\lambda_1=\lambda_2=0$ and $\lambda_2=0$ respectively. Therefore, if $\balpha^{(t)}_0$ and $\balpha^{(t)}_r$ denote their respective solutions, we have

\begin{align*}
	 \balpha^{(t)}_0 =\balpha- \big[\tI_L+\nabla_2(\balpha)\big]^{-1}\nabla_1(\balpha)&,& \balpha^{(t)}_r = \big[(\lambda_1+1)\tI_L+\nabla_2(\balpha)\big]^{-1}\big\{(\tI_L+\nabla_2(\balpha))\balpha-\nabla_1(\balpha)\big\}.
\end{align*}

When considering estimation of spatial effects for $S$ states together using algorithm (\ref{algo::md-1}),  the increased dimension of $W$ affects computational complexity adversely. One can alternatively suggest running the algorithm in parallel for individual states, resulting in $2S$ tuning parameters, which is undesirable. In that regard, advantage of the suggested approximation to Laplacian matrices discussed in section (\ref{subsec::graph-lap}) is immediately evident when applied to solution in eq. (\ref{eq::mm-est}). It results in,

\begin{align}\label{eq::mm-est-approx}
	\balpha^{(t)} \approx \big[(\lambda_1+1)\tI_L+\lambda_2W_a+\nabla_2(\balpha)\big]^{-1}\big\{(\tI_L+\nabla_2(\balpha))\balpha-\nabla_1(\balpha)\big\},
\end{align}
where $W_a$ is block diagonal matrix consisting of $S$ blocks, $W_{k}$, $k=1,2,\ldots, S$. Each block $W_{k}$ is the \emph{exact} Laplacian for state $k$ and $L =\sum_{k}L_k$, $L_k$ being the number of locations in state $k$. Therefore, matrix inverse in eq. (\ref{eq::mm-est-approx}) can be computed in $O\big(\sum_{k}L_k^3\big)$ operations (instead of $O(L^3)$), affecting scalability of algorithm (\ref{algo::md-1}) significantly by allowing sufficient scope for parallelization while keeping the number of tuning parameters fixed. It is important to note here that, theorem \ref{th::con-algo-1} still holds for approximate Laplacian, since $W_a$ is still positive semi-definite (p.s.d.) (see Appendix for proof). 
  
\section{Simulation}\label{sec::sim}

The aims of presented simulation study are,
\begin{enumerate}
	\item to assess performance of algorithm \ref{algo::md-1} under different spatial patterns explained and demonstrated in the ensuing discussion,
	\item to demonstrate and compare performance of algorithm \ref{algo::md-1}
	\begin{itemize}
		\item[(i)] using exact solution in eq. (\ref{eq::mm-est}) and,
		\item[(ii)] using approximate solution in eq. (\ref{eq::mm-est-approx})
	\end{itemize}
	for multiple states.
	\item a sensitivity analysis for estimated spatial effects to the index parameter, $p$.
\end{enumerate}

In what follows it is important to note that, we use a state (or group of states) only as an example, results shown are in no way indicative of true responses in the region. Their sole purpose is to create an instance that can serve as a test case for the algorithm. We start by describing some error metrics that we will be using for all of the examples listed in this section. Let $\balpha_{(O)}$,  $\widehat{\balpha}_0$,  $\widehat{\balpha}_r$ and $\widehat{\balpha}$ denote true, un-penalized, ridge and the estimated spatial effect vectors from the algorithm \ref{algo::md-1} respectively. The response is simulated from a compound Poisson distribution, using likelihood approximations in \cite{dunn2005series} and \cite{dunn2008evaluation}, made available to use in \texttt{R}-package \texttt{tweedie}. We assume a logarithmic link for both GLMs. The index parameter $p=1.5$ is kept fixed, while the dispersion, $\phi$ and mean $\mu$ are allowed to vary across simulated response. For the entirety of this section simulated data will be split into training and validation sets, with tuning parameters $\lambda_1$, $\lambda_2$ being estimated using a five-fold leave-one-out cross-validation (LOOC) on the training data by minimizing deviance

\begin{align*}
d(y;\widehat{\mu}, \balpha) = \sum\limits_{i=1}^{L}\sum\limits_{\scalemath{0.7}{j\in \mathcal{N}(i)}}\frac{2}{\widehat{\phi}_{ij}}\Bigg[y_{ij}e^{-(\widehat{\mu}_{ij}+r_{ij}\trans\scalemath{0.8}{\balpha})/2}+e^{(\widehat{\mu}_{ij}+r_{ij}\trans\scalemath{0.8}{\balpha})/2}\Bigg],
\end{align*}
over a holdout set in each fold. As a measure of loss, we use error sum of squares (SSE) for estimated spatial effects, whereas prediction error over both training and validation set is measured using a ratio of deviances given by,

\begin{align}\label{eq::error-metrics}
SSE = \big|\big|\balpha_{(O)}-\balpha\big|\big|_2^2 &,& d_r\big(y; \widehat{\mu},\balpha_{(O)},\balpha\big) = \frac{d(y,\widehat{\mu},\balpha)}{d\big(y,\widehat{\mu},\balpha_{(O)}\big)},
\end{align}
respectively, where $\balpha$ can be any one of three estimates, $\widehat{\balpha}_0$,  $\widehat{\balpha}_r$ or $\widehat{\balpha}$. Lower $SSE$ and, $d_r(y; \widehat{\mu},\balpha_{(O)},\balpha)$ closer to 1 are desirable.

\begin{figure}[t]
	\centering
	\begin{subfigure}{.5\textwidth}
		\centering
		\includegraphics[width=1\linewidth , height=1\linewidth]{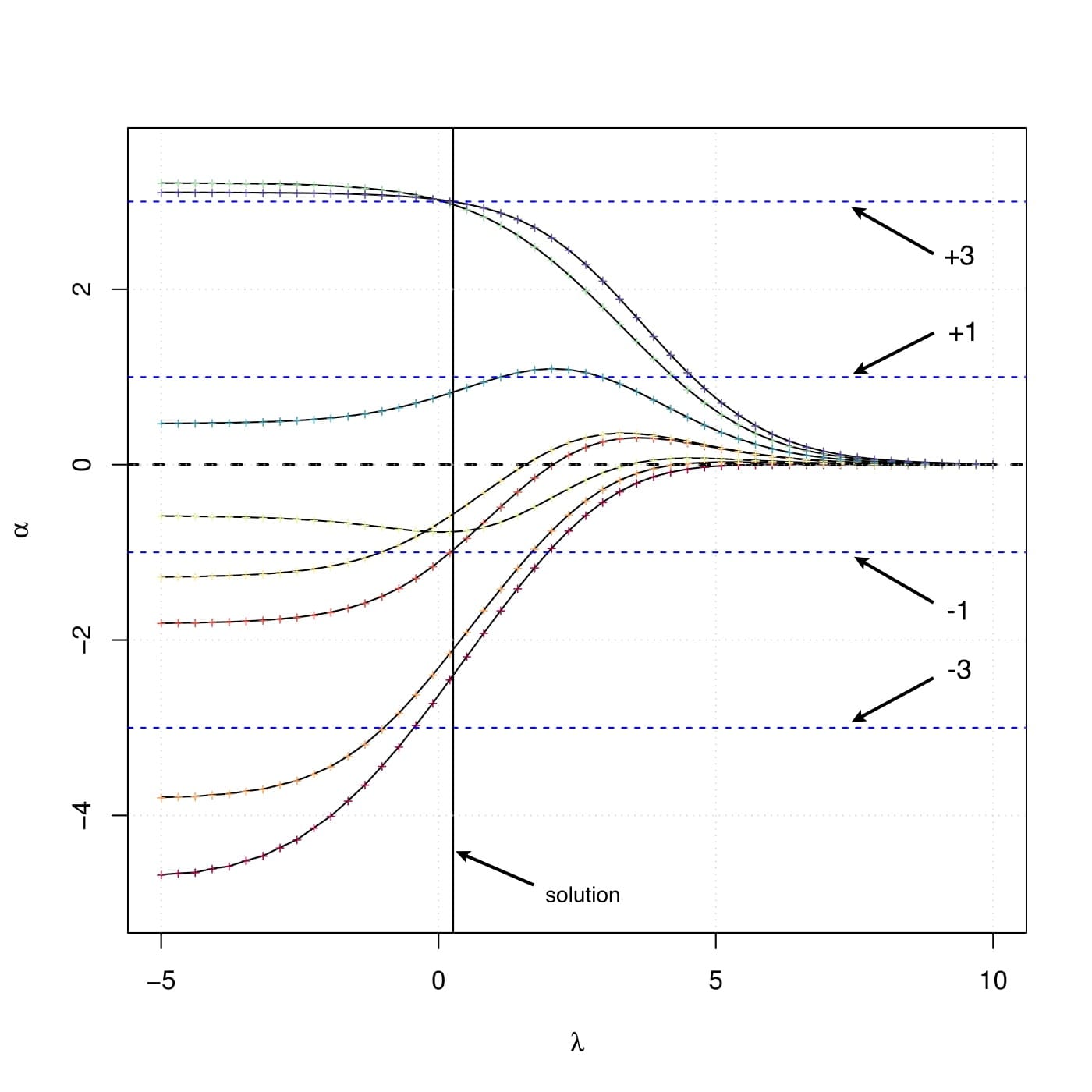}
		\caption{}
	\end{subfigure}%
	\begin{subfigure}{.5\textwidth}
		\centering
		\includegraphics[width=1\linewidth , height=1\linewidth]{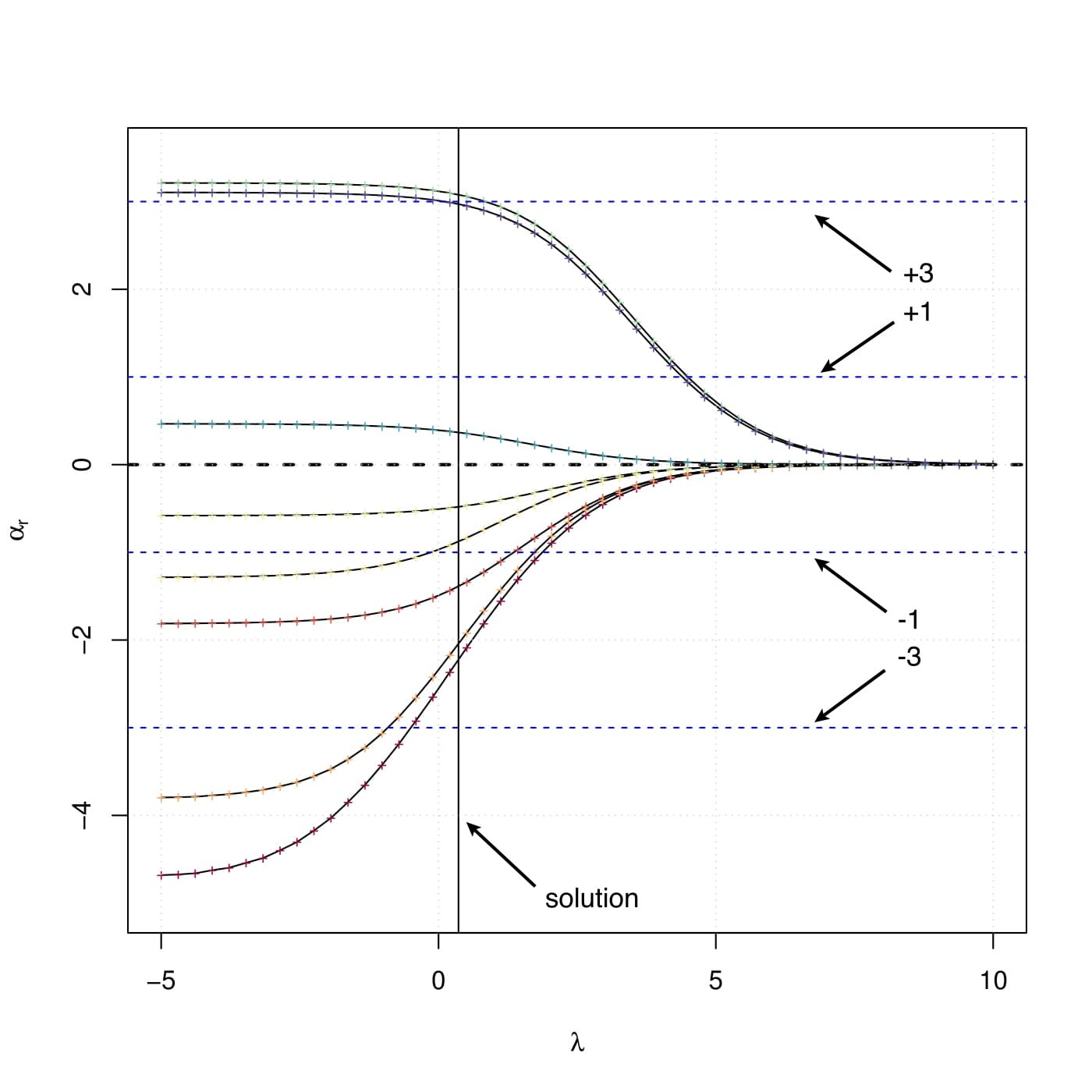}
		\caption{}
	\end{subfigure}%
	\caption{Figure showing solution path with $\lambda$ in logarithmic scale for (a) proposed solutions ($\balpha$) and, (b) ridge solutions ($\balpha_r$). Dashed blue horizontal lines indicate true values, vertical line indicates value of $\lambda$ at which minimum deviance was obtained.}
	\label{fig::solpath}
\end{figure}

Ridge estimates, ($\balpha_r$) are a natural baseline for the proposed estimates. To assess the difference between them it is not enough to just show solutions obtained for a single $\lambda$ or $\lambda_1, \lambda_2$. To make ease for comparison we use a variant of the penalty shown in eq. (\ref{eq::opt-prob}), $P(\balpha; \lambda)=\lambda(0.4||\balpha||_2+(1-0.4)||\balpha||_W)$. We vary $\lambda$ on a logarithmic scale in the range $[\lambda_l, \lambda_u]$, shown in figure \ref{fig::solpath}. We choose a small sample size, county level spatial effects for state of CT (8 counties). True values assigned viz., \{-3,-1,1,3\} are indicated in the figure. For large values of $\lambda$, the problem is un-penalized, i.e. $\balpha=\balpha_r=0$. Starting from a value of 0, estimates under both penalties are obtained by using a ``warm-start" strategy, that involves using the estimate, $\balpha^{(t)}$ as a starting value for next iteration $\balpha^{(t+1)}$. This warm-start strategy will be used when working with penalty, $P(\balpha;\lambda_1,\lambda_2)$ in eq. (\ref{eq::opt-prob}). In that scenario, $\lambda_1, \lambda_2 \in [\lambda_{1l}, \lambda_{1u}]\times [\lambda_{2l}, \lambda_{2u}] \subset \mathbb{R}^2$. Differences in solution paths are apparent from fig. (\ref{fig::solpath}). The value of $\lambda$ for which the path of an estimate first hits the true value , i.e. ``hitting-time" occurs much earlier in the proposed estimates. Furthermore, the value of $\lambda$ at which the solution is obtained under the two penalties, has estimates closer to true values in the case of proposed algorithm indicating lower $SSE$s between estimated and true spatial effects.

We evaluate the performance of algorithm \ref{algo::md-1} under four different sample sizes and proportions of zeros in simulated response. Chosen sample sizes for this simulation study vary from 10,000 -- 50,000, while proportions of zeros vary in the range 0.15--0.80. The estimated spatial effects are evaluated under each combination of settings to provide a detailed demonstration about efficacy of the proposed algorithm. Under each setting 100 replications are carried out, reported values of error metrics in tables shown are averages over all replications, accompanied by their respective standard deviations. We use four different spatial patterns as examples viz., \emph{block}, \emph{smooth}, \emph{hot-spot} and \emph{structured};  their construction and features will be explained in the following subsection. 

For the first part of this simulation we use the state of Connecticut as an instance, second part uses three states as instances viz., Connecticut, Massachusetts and Rhode Island (reason behind such a choice being their adjacent locations). There are 282, 537 and 77  zipcodes  in the respective state(s). Tables showing relevant results and details are postponed to Appendix B for maintaining continuity.

\begin{figure}[t]
	\centering
	\begin{subfigure}{.25\textwidth}
		\centering
		\includegraphics[width=1\linewidth , height=0.8\linewidth]{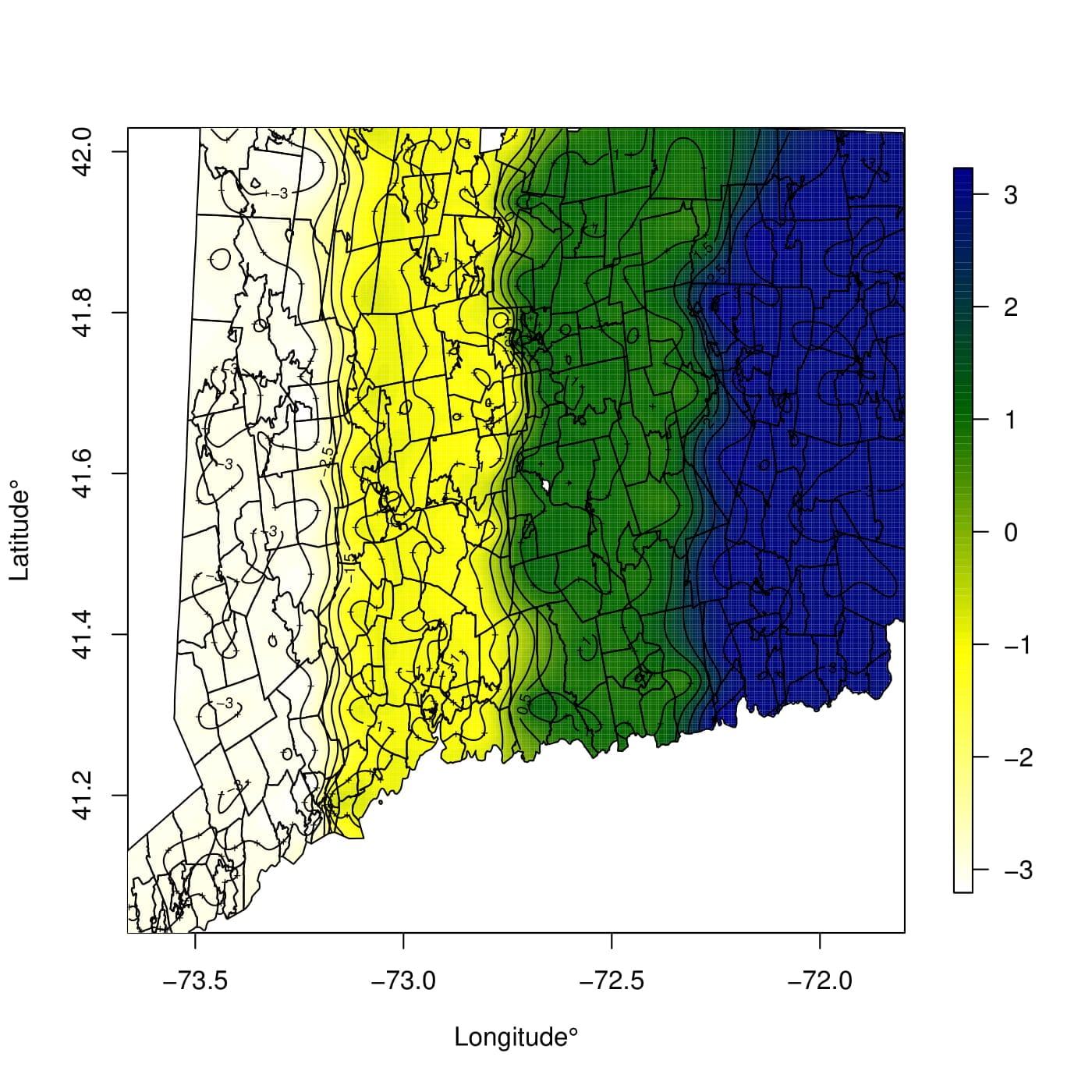}
	\end{subfigure}%
	\begin{subfigure}{.25\textwidth}
		\centering
		\includegraphics[width=1\linewidth , height=0.8\linewidth]{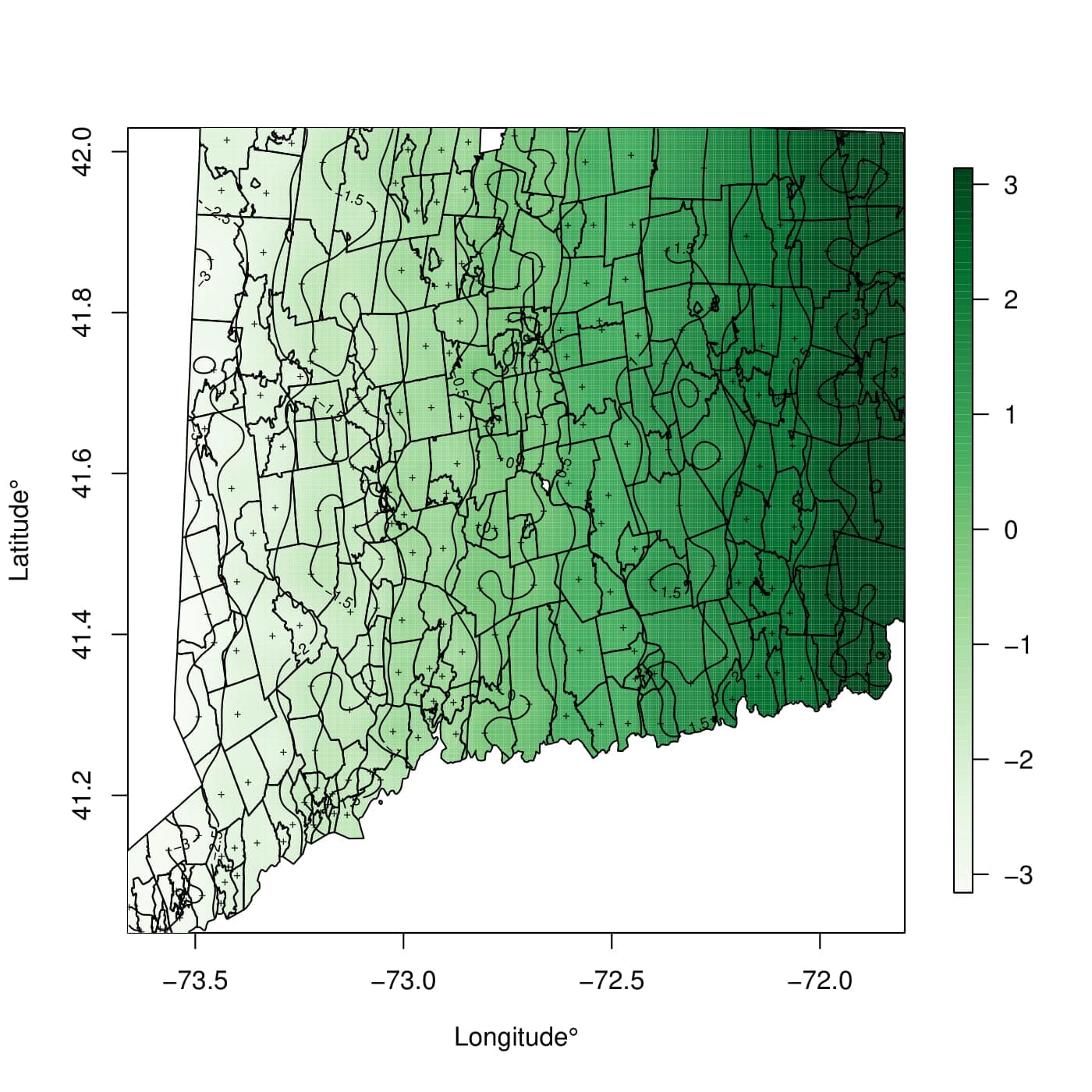}
	\end{subfigure}%
	\begin{subfigure}{.25\textwidth}
		\centering
		\includegraphics[width=1\linewidth , height=0.8\linewidth]{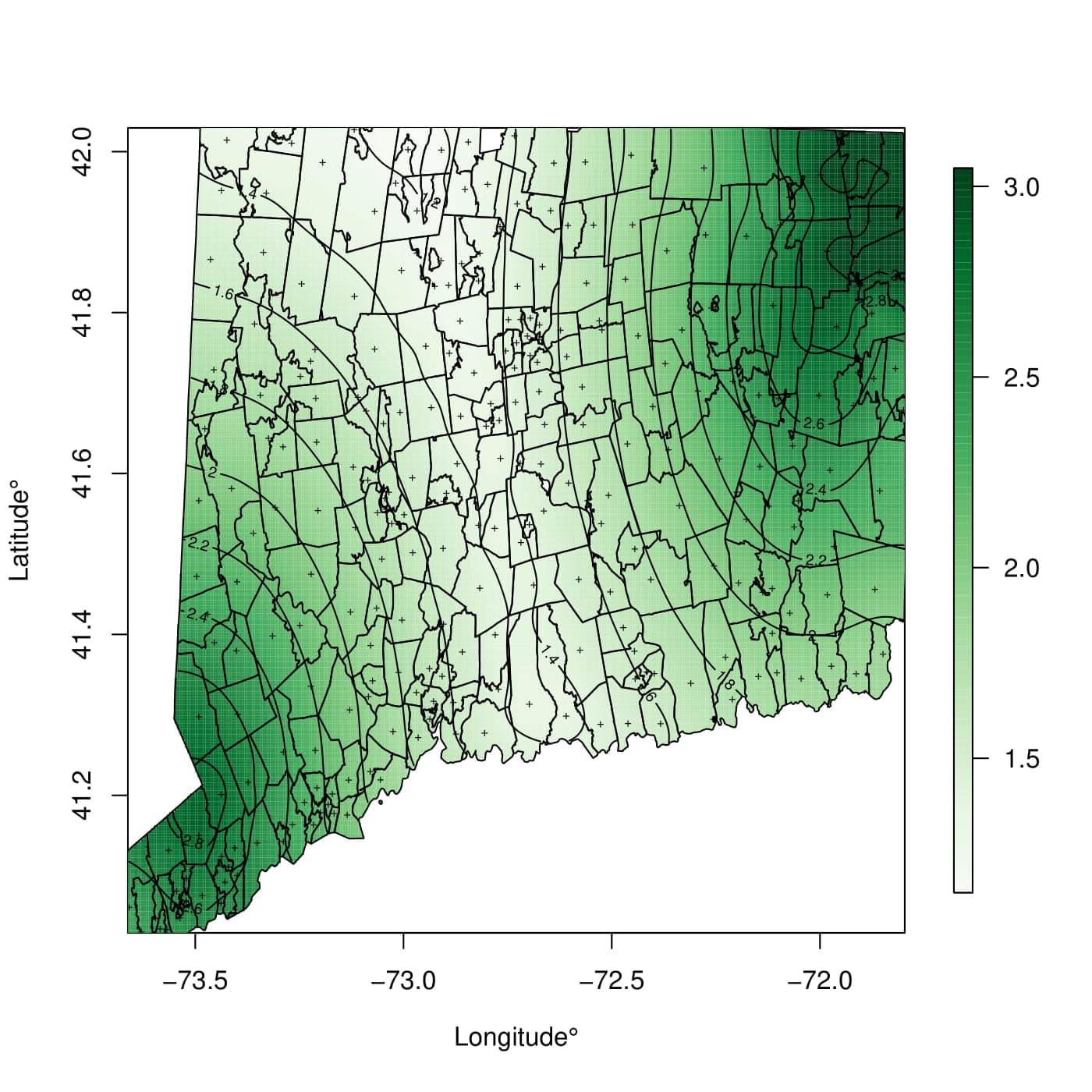}
	\end{subfigure}%
	\begin{subfigure}{.25\textwidth}
		\centering
		\includegraphics[width=1\linewidth , height=0.8\linewidth]{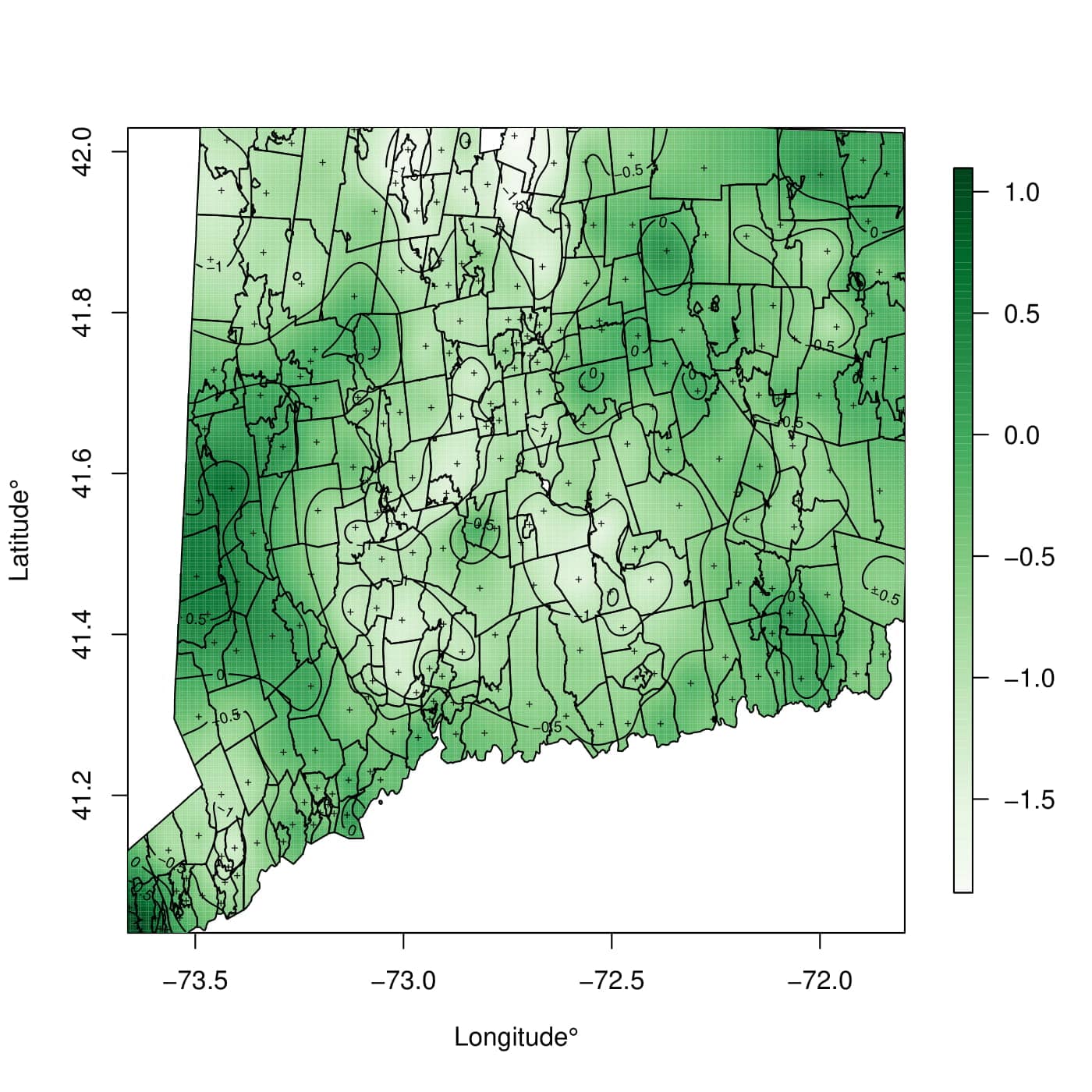}
	\end{subfigure}\\
	\begin{subfigure}{.25\textwidth}
		\centering
		\includegraphics[width=1\linewidth , height=0.8\linewidth]{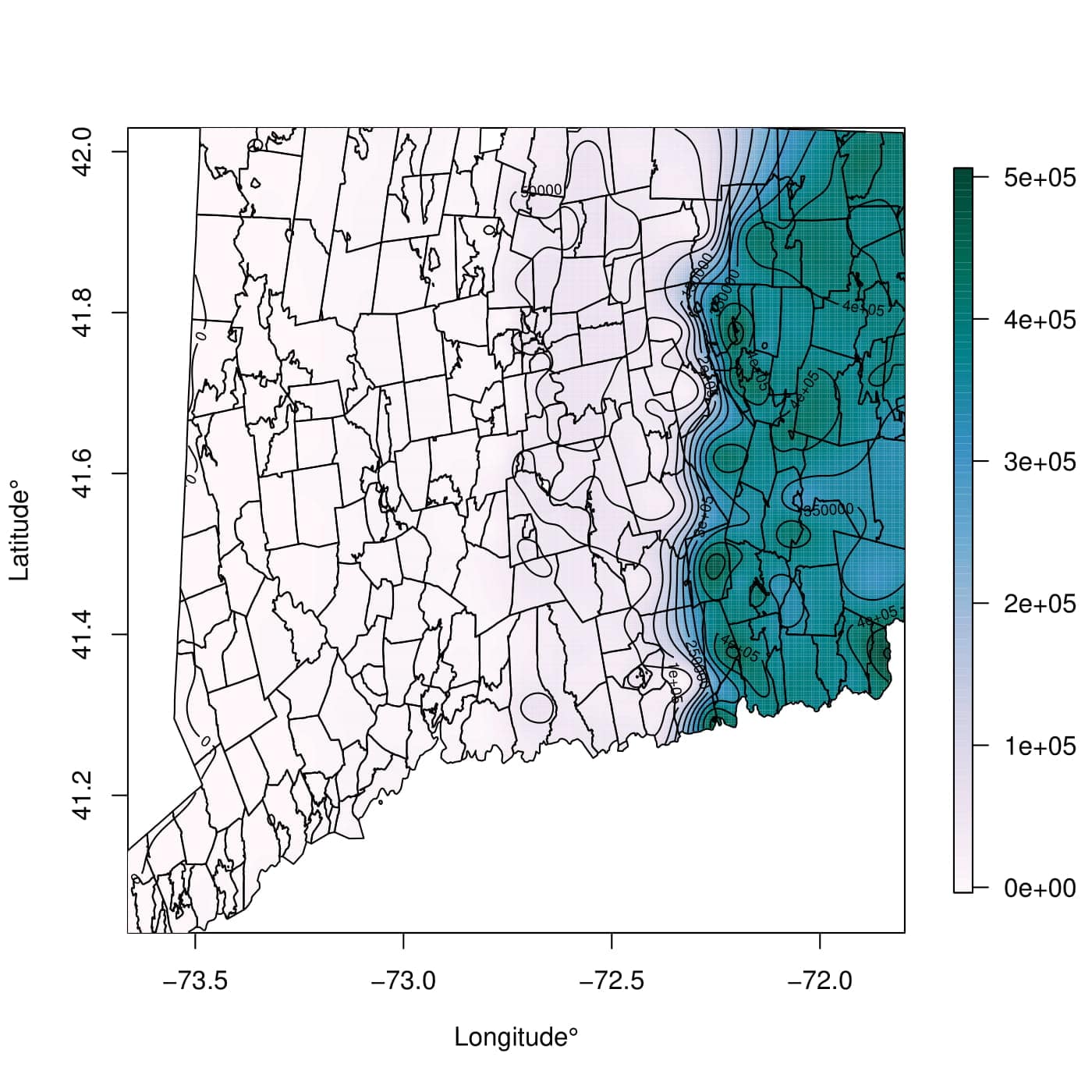}
		\caption{}
	\end{subfigure}%
	\begin{subfigure}{.25\textwidth}
		\centering
		\includegraphics[width=1\linewidth , height=0.8\linewidth]{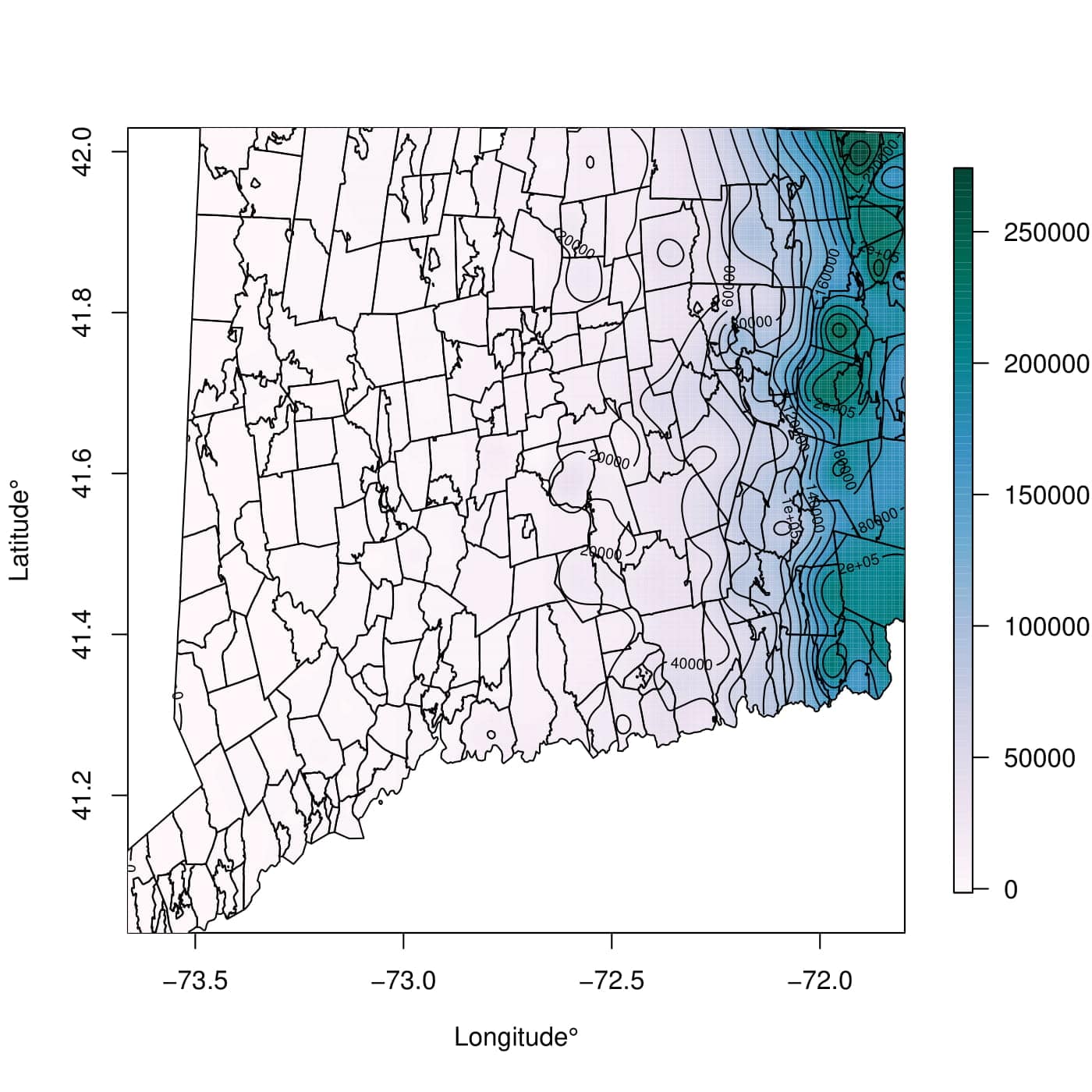}
		\caption{}
	\end{subfigure}%
	\begin{subfigure}{.25\textwidth}
		\centering
		\includegraphics[width=1\linewidth , height=0.8\linewidth]{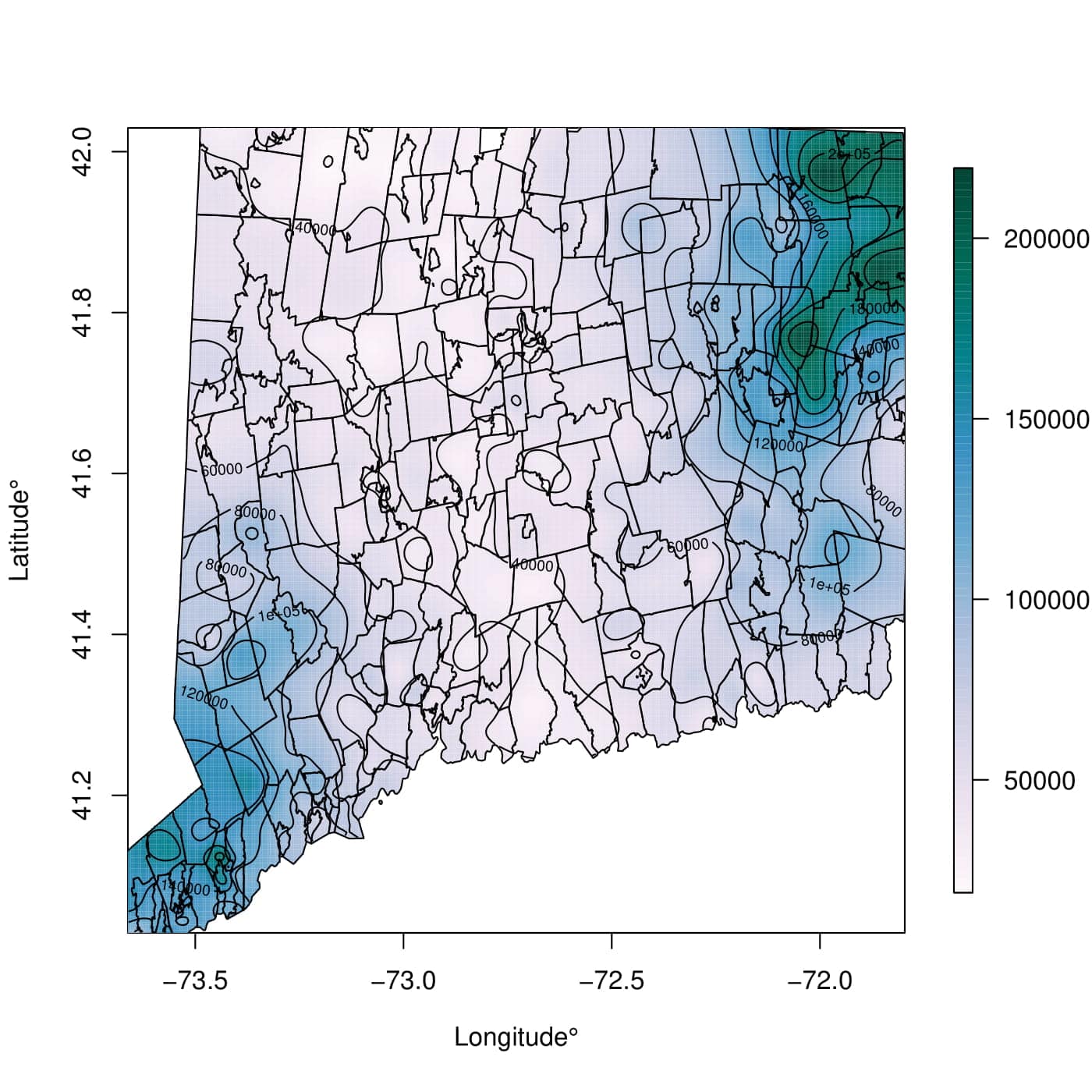}
		\caption{}
	\end{subfigure}%
	\begin{subfigure}{.25\textwidth}
		\centering
		\includegraphics[width=1\linewidth , height=0.8\linewidth]{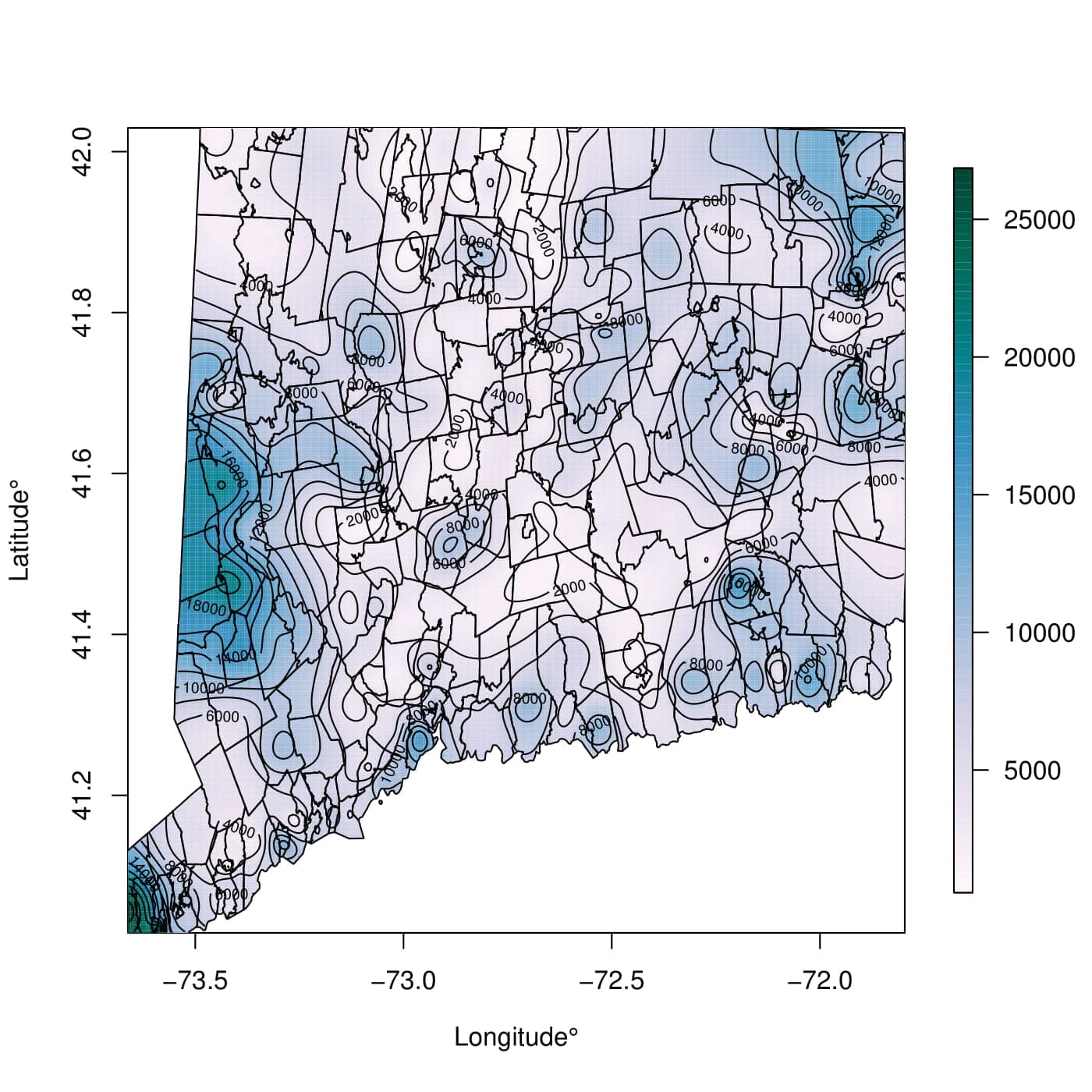}
		\caption{}
	\end{subfigure}
	\caption{Spatial plots showing simulated response (bottom row) for (a) block (b) smooth (c) hot-spot and (d) structured spatial patterns (top row), with legends alongside plots showing scales for true spatial effects and resulting simulated response.}
	\label{fig::ct-sim}
\end{figure}

\subsection{Examples of different spatial patterns}\label{subsec::sp-pattren}

In this subsection we describe a comparative study between un-penalized and penalized estimates discussed in section \ref{subsec::optim-prob} under different spatial patterns. In the ensuing examples, construction for each pattern is discussed in detail, followed by comments explaining results at the end. 

Under a logarithmic link and index parameter $p=1.5$, referring to table (\ref{tab::tweedie-pars}), relationship between the canonical parameter $\theta$ and mean $\mu$ for compound Poisson is given by $\theta=2\mu^{-1/2} \Rightarrow \mu=4/\theta^2$. In all examples shown below, we simulate, $\theta \sim N(-0.16,0.02^2) \Rightarrow \mu \in (83,400)$ approx., with $N(\cdot,\cdot)$ denoting a Gaussian distribution. To introduce larger number of zeros in simulated response under each settings described below we use the equivalent relationship between $(\mu, \phi, p)$ and $(\xi, \eta, \zeta)$ for characterizations in eqs. (\ref{eq::tw-mod-dens-0}) and (\ref{eq::tw-mod-dens-1}). We have,

\begin{align*}
	\mu=\xi\eta\zeta &,& \phi=\frac{\xi^{1-p}\cdot(\eta\zeta)^{2-p}}{2-p}.
\end{align*}
Hence for a fixed mean and index parameter $\mu$, $p$ with $p \in (1,2)$, increasing dispersion $\phi$, would result in increased mean, $\eta\zeta$ of individual  gamma components and a decreased rate, $\xi$ for Poisson. A smaller rate for Poisson means higher probability of obtaining exact zeros which is our requirement. Table \ref{tab::sim-param-summary} shows respective reference distributions used for sampling dispersion offsets under each setting to obtain desired proportions of zeros in simulated response with, $U(\cdot,\cdot)$ denoting a uniform distribution.

\medskip

\noindent \emph{Example 1 \& 2: Block and Smooth}

\medskip

\noindent To obtain a block pattern with four blocks, the longitudinal range for a state is divided into four regions. For example, figure (\ref{fig::ct-sim}) column (a), shows the boundaries for four regions. Each region is assigned a fixed spatial effect, in this case $\{-3,-1,1,3\}$. Under a logarithmic link it evaluates to a multiplicative effect towards the mean of magnitudes, $\{e^{-3},e^{-1},e^{1},e^{3}\}$. As a result, means for regions with lower magnitudes of spatial effects are expected to have higher number of zeros. Increasing the overall dispersion $\phi$ introduces more zeros in the simulated response for individual regions, as can be seen from lower plot for column (a) in fig. (\ref{fig::ct-sim}). Finally, for each setting the response is simulated  from $y \sim \text{Tweedie}(\mu, \phi, p)$. Figure (\ref{fig::ct-sim}) column (a) shows one such instance. For each replication within a setting, $10\times 10$ grids were chosen for tuning $(\lambda_1, \lambda_2)$ in algorithm (\ref{algo::md-1}), where $(\lambda_1, \lambda_2) \in [-5,0]\times [       -3,2]$ in \emph{log-scale}, similarly for the ridge penalty a line search is conducted on $10\times 1$ vector for $\lambda_1 \in [-5,0]$ in \emph{log-scale}.

Smooth pattern is designed to be a more general version of block patterns having finer divisions for the spatial effect thereby, producing lesser discreteness across their values. Column (b) in figure (\ref{fig::ct-sim}) shows spatial effects smoothly varying in the range $[-3,3]$ over 10 regions and the resulting simulated response.  Grid and line searches for optimizing tuning parameters for ridge and algorithm (\ref{algo::md-1}) were conducted over $10\times 1$ line and $10\times 10$ grid, where $\lambda_1 \in [-5,2]$ and, $(\lambda_1, \lambda_2) \in [-5,2]\times [-5,5]$ in \emph{log-scale} respectively. Table \ref{tab::sim-sm} shows related results in detail.

\medskip

\noindent \emph{Example 3: Hot-spots}

\medskip

\noindent A hot spot is defined as location(s) that exhibit higher magnitudes of response, with response magnitudes tapering off with increasing distance from the hot-spot (for further details on hot spots and their detection see \cite{cressie1992statistics}). An example is shown in fig. (\ref{fig::ct-sim}) column (c) top plot. For the scope of this simulation we create two hot-spots in Connecticut viz. north-east and south-west corners, being assigned a spatial effect of 3, tapering off with increasing distance (euclidean), $l^{(2)}$ from them via an exponential kernel (i.e. $\exp(-\phi l^{(2)})$). Hence zipcodes located in the center (equidistant from both hot spots) are expected to have lower spatial effects. Therefore,  simulated response varies accordingly showing higher magnitudes in two hot spots. Associated grids and lines remain same as the  ``smooth" pattern. Table \ref{tab::sim-hot} shows relevant results for this pattern settings.

\medskip

\noindent \emph{Example 4: Structured}

\medskip

\noindent A structured pattern is created using a distance based covariance kernel to simulate fixed spatial effects from a zero mean multivariate Gaussian. Explicitly, $\balpha_{(O)} \sim N_L(\mathbf{0}, \Sigma)$, where $N_L(\cdot,\cdot)$ , $\mathbf{0}$ are the $L$-dimensional multivariate Gaussian and zero vector respectively, with $\Sigma=\sigma^2 \exp(-\phi l^{(2)})$ being specified using an exponentially structured covariance kernel for the scope of this simulation, with $\sigma^2, \phi=1$. $l^{(2)}$ is an $L\times L$ euclidean distance matrix with all  operations being entry-wise. Figure (\ref{fig::ct-sim}), column (d) (upper plot) shows an example of structured effect and its resulting simulated response. We simulate the spatial fixed effect \emph{once}, to be used across all different combinations of settings for sample sizes and proportion of zeros. Associated grids and lines remain same as the  ``smooth" and ``hot-spot"  patterns. Relevant results for different settings in this pattern are shown in Table \ref{tab::sim-struc}.

\medskip

\noindent \emph{Comments and Interpretation:}

\medskip

The results in table \ref{tab::sim-param-summary} are to be considered in conjunction with those in tables (\ref{tab::sim-bl}, \ref{tab::sim-sm}, \ref{tab::sim-hot}  and \ref{tab::sim-struc}). With baselines for comparison being ridge and un-penalized estimates, under all different spatial patterns, results \emph{on an average} can be summarized as follows,
\begin{itemize}
	\item [(a)] estimates from proposed algorithm have comparatively lower $SSE$s,
	\item[(b)] if proportion of zeros in simulated response is assumed to be a metric for signal to noise ratio then proposed estimates also show lower $SSE$s under very low signal to noise ratios (i.e. higher proportion of zeros in response),
	\item[(c)] under low sample sizes and signal to noise ratios, proposed estimates show low $SSE$s in comparison to estimates from the ridge penalty,
	\item[(d)] training and validation deviance ratios, $d_r\big(y; \widehat{\mu},\balpha_{(O)},\widehat{\balpha}\big)$ are closer to 1 when compared to corresponding un-penalized and ridge versions.
\end{itemize}

With regard to above conclusions, flexibility of proposed algorithm over ridge penalty is demonstrated in table \ref{tab::sim-param-summary} from values of regularization parameters obtained under each spatial pattern for the two penalties. Having a function of Laplacian as an additional penalty term primarily allows for superiority in performance under the presence of spatial variation in simulated response. This is evident particularly in simulation results from a hot-spot pattern shown in table (\ref{tab::sim-param-summary}). Regularization parameters for both ridge and ridge part of proposed penalty show the same values (i.e. no penalty, $\lambda_1=-5.000$),  accompanied by a large value of $\lambda_2=2.708$ for Laplacian part of the penalty; on referring to corresponding table \ref{tab::sim-hot} we note that estimates from algorithm \ref{algo::md-1} are superior to ridge estimates.

\medskip

\subsection{Comparison between algorithms featuring approximate and exact solutions}

\begin{figure}[t]
	\centering
	\begin{subfigure}{.5\textwidth}
		\centering
		\includegraphics[width=1\linewidth , height=0.75\linewidth]{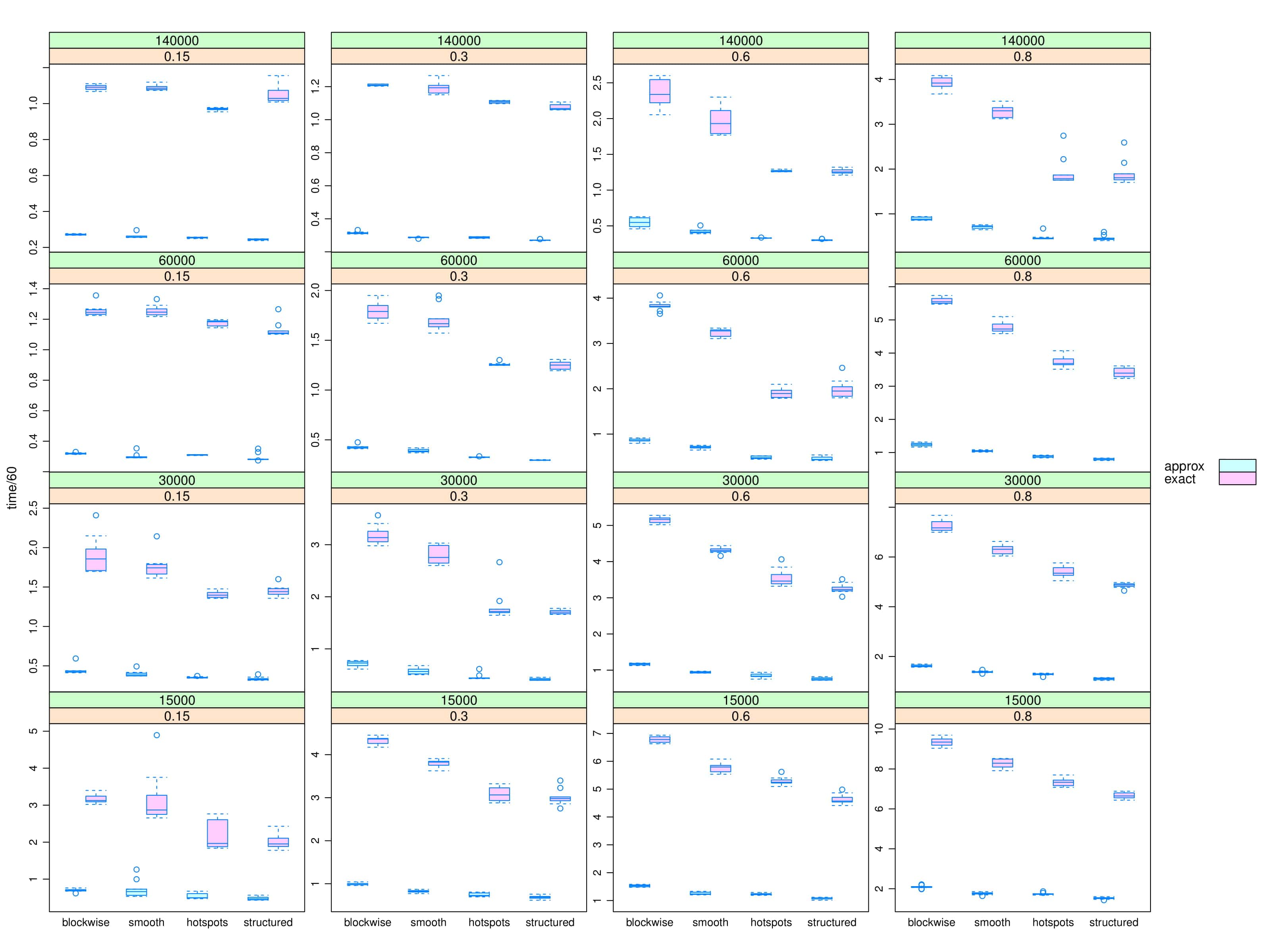}
		\caption{}
	\end{subfigure}%
	\begin{subfigure}{.5\textwidth}
		\centering
		\includegraphics[width=1\linewidth , height=0.75\linewidth]{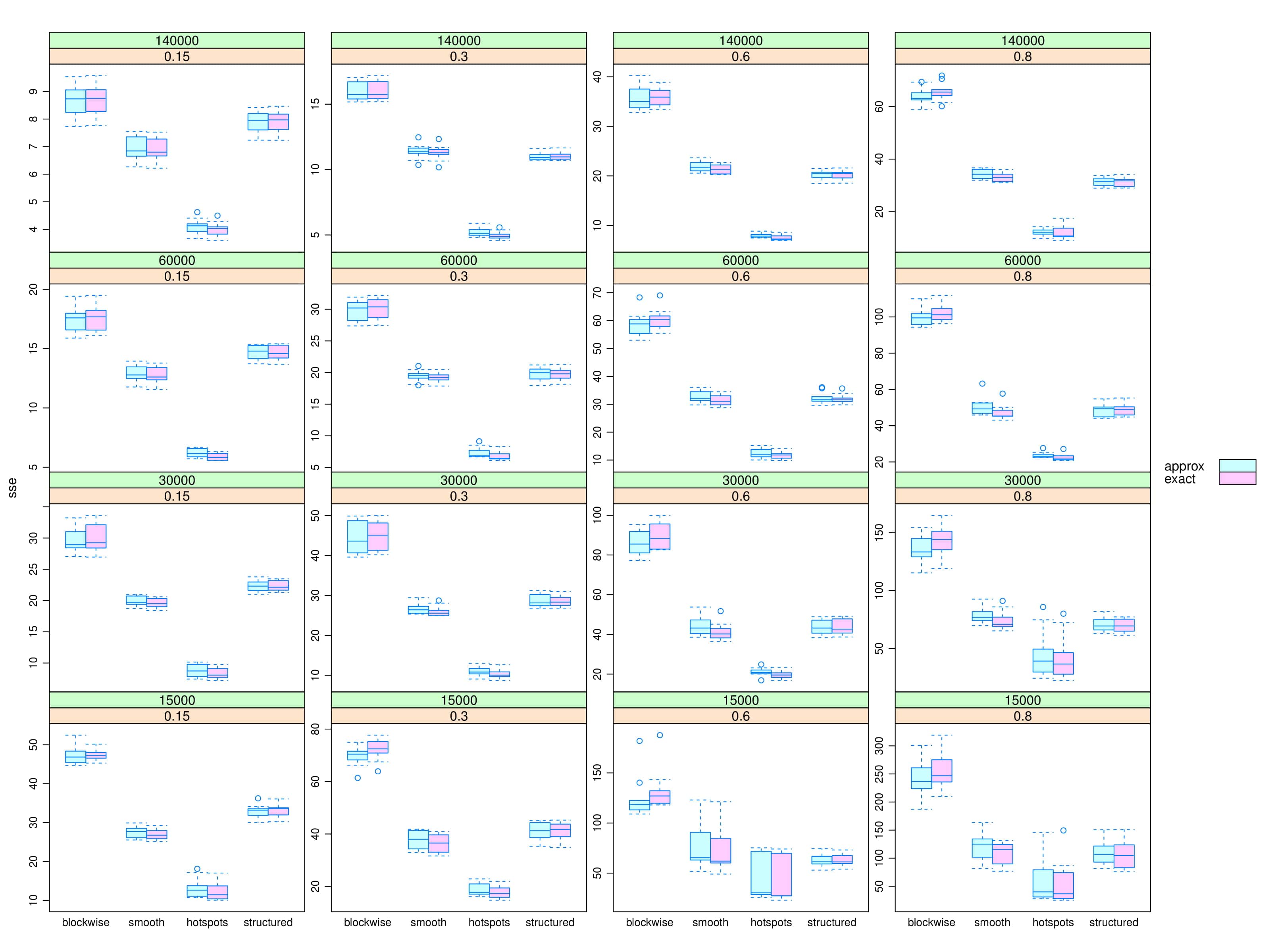}
		\caption{}
	\end{subfigure}%
	\caption{Figures showing (a) time taken in minutes (b) $SSE$ as defined in eq. (\ref{eq::error-metrics}), from 10 replications under each possible combination of sample size and proportion of zeros in simulated response for different spatial patterns. Each figure shows grouped box-plots for approximate and exact estimates respectively under different settings.}
	\label{fig::time-see-comp}
\end{figure}

This subsection primarily compares performance of two estimates, exact and approximate derived in eq. (\ref{eq::mm-est}) and (\ref{eq::mm-est-approx}) respectively. The metric used to compare them is a trade-off between error sum of squares and time taken to convergence under the same spatial patterns considered in section \ref{subsec::sp-pattren}. Aim being to demonstrate that proposed approximation in eq. (\ref{eq::mm-est-approx}) produces estimates with error sum of squares comparable to that of exact estimates in eq. (\ref{eq::mm-est}), but with significant improvement in time taken.  Figure (\ref{fig::time-see-comp}) summarizes results for the comparative study.

We considered three adjacent states in the southern New England group for this simulation viz., Connecticut, Massachusetts and Rhode Island. Sample sizes chosen are 15,000, 30,000, 60,000 and 140,000, with proportions of zeros in the simulated response varying in 0.15, 0.30, 0.60 and 0.80. The \emph{approximated} adjacency used, was a block diagonal matrix with blocks of the order, $282\times 282$, $537 \times 537$ and $77 \times 77$ respectively for individual states, as opposed to an exact adjacency matrix of the order $896\times 896$.  Associated tuning parameters were estimated over a $10\times 10$ grid, $[-5,5]\times [-5,5]$ in logarithm scale, for all different spatial patterns. 

As can be readily seen from fig. (\ref{fig::time-see-comp}b), approximate estimates show performance comparable to exact counterparts in terms of $SSE$; furthermore in-sample and out-sample deviance ratios (as defined in eq. (\ref{eq::error-metrics})) showed similar findings. Referring to fig. (\ref{fig::time-see-comp}a) the significant advantage in computational complexity that approximate estimates demonstrate makes it a worthy alternative when considering scalability of algorithm (\ref{algo::md-1}). In particular when we have discreetness in the spatial pattern, for ex. ``blockwise" patterns, referring to fig. (\ref{fig::time-see-comp}b) we can see that approximate estimates prevent unnecessary smoothing, thereby outperforming their exact counterparts (for more details regarding computation, see sec. \ref{sec::disc-future}).

\subsection{Sensitivity to choice of index parameter, $p$}

We elaborate on relevant consequences of properties discussed at the end of section (\ref{sec::DGLM-twmod}.2) (particularly in eq. (\ref{eq::prop-cross-0})), by looking further into sensitivity of estimates to the choice of index parameters. Working with offsets from a DGLM for the Tweedie family, where the index parameter, $p$ distinguishes between models, it is desirable and expected for our proposed approach to remain unaffected by erroneous model specification.

Explicitly we consider situations where true value of index parameter, $p \ne 1.5$, however a value of 1.5 was reported (or assumed since missing) for the index parameter. Under orthogonality of mean, $\mu$ to index and dispersion parameters, $p$, $\phi$ respectively for Tweedie models we expect spatial information in the mean to remain relatively unperturbed. Metrics used for this simulation are (i) $SSE$ and, (ii) $L^2$-norm (Euclidean)  for true against estimated effects. Figure (\ref{fig::norm-see-sens}) in Appendix C, shows results for the simulation study conducted. Previous simulation shows exact and approximate estimates demonstrating comparable results in terms of $SSE$, hence conclusions derived for exact can be extended to hold for approximate estimates. As a result, we limit the scope of this simulation to estimating exact spatial effects for state of CT under previously described settings for the index parameter. 

Additional simulation settings used consist of the same spatial patterns as described in examples from section (\ref{subsec::sp-pattren}), with sample sizes varying in 10,000,  20,000, 30,000 and 50,000, index parameters varying in 1.3, 1.4, \ldots, 1.9 and dispersion parameters are altered accordingly to result in simulated response having 30\% zeros under all combinations. Associated tuning parameters were estimated over a $10\times 10$ grid, $[-5,5]\times [-5,5]$ in logarithm scale.

Figure (\ref{fig::norm-see-sens}a) shows that using $p=1.5$ as a reference, significant departures are located from true effects for extreme values of $p$ (ex. $p=1.9$), which become relatively pronounced under low sample sizes and coarser spatial patterns (ex. block-wise). In fig. (\ref{fig::norm-see-sens}b) norms for estimated and true spatial effects display a similar behavior under all patterns considered. This leads us to conclusively state that under large sample sizes and relatively smooth spatial variations in response, estimates produced by algorithm (\ref{algo::md-1}) remain stable w.r.t. model misspecification.

\section{Case Studies}\label{sec::cs}

Case studies considered aim to demonstrate the application and performance of proposed algorithm with both, exact and approximate estimates in detecting residual spatial effect while modeling response, i.e. \emph{loss costs per unit insured} related to personal automobile insurance collision claims. Exact estimates using algorithm (\ref{algo::md-1}) are obtained for state of CT. While, for approximate estimates we consider two case studies featuring, (i) six states in New England (group of states in the east coast) consisting lower number of zipcodes compared to, (ii) three adjacent states in West Coast, having larger number of zipcodes, within the United States of America (USA). All of these are subsets of data obtained from a more comprehensive repository named Highway Loss Data Institute (HLDI) maintained by the independent non-profit, Insurance Institute for Highway Safety (IIHS) \cite{iihs} working towards reducing losses arising from motor vehicle crashes. We shall refer to this as the HLDI database.

 We briefly describe the HLDI database. It contains data at an individual level. The data contains covariates associated with the individual on,
\begin{itemize}
	\item accident and model year of the vehicle, ranging from 2000--2015 and 1981 -- 2016 respectively,
	\item risk of the policy having two levels ``S", ``N",
	\item age, gender, marital status and gender of partner, where missing values are denoted by 0 and ``U" respectively for age and rest of these predictors,
	\item number of claims, payments (i.e. loss cost), exposure (measured in policy years, eg. 0.5 indicates individual insured for half a year) and deductible limit (categorical with 8 categories).
	\item 5-digit zipcode indicating location, i.e. areally-referenced.
\end{itemize} 
 
 \noindent Derived predictors like age categories, vehicle age (accident--model year) in years can be obtained and used in the DGLM. 
 
 For all case studies we use implementations and approximations for compound Poisson DGLMs as in \cite{smyth2002fitting}. Policy exposures ($w_{ij}$) are used as weights for associated dispersion parameter $\phi_{ij}$ in the DGLM, i.e. as $\phi^{*}_{ij}=\phi_{ij}/w_{ij}$. Response variable used is defined as the ratio of payments to exposure. Fitted mean model consists of deductible, accident year, gender and marital status, while the dispersion model consists of a categorical version of age (consisting of 6 categories) in addition to predictors in the mean model. Index parameter used is $p=1.6$. We use all of the data to obtain fitted mean and dispersions for the DGLM, which we aim to use as offsets in the process of estimating zipcode-level spatial effects. Following which for each case study shown, we then randomly divide the data into training and validation subsets using a 60-40 \emph{stratified} split, with stratification at a zipcode level for state(s). Proposed algorithm along with ridge and un-penalized counterparts are fitted on the training sets. We replicate this procedure 20, 10 and 5  times for exact and approximate (two case studies) procedures respectively. While evaluating models we consider predictions in terms of loss costs or payments for all fitted models, after adjusting for exposure. This is done to avoid issues involving misalignment in out-sample predictions.
 
  It is important to note that in our proposed model, a spatial fixed effect for zip-codes is present in the mean model. 
  Consequently, we expect to see substantial improvement in zip-code level aggregated loss costs rather than at individual level. As error metrics we choose deviance at the observation level (as in eqs. \ref{eq::tw-mod-dens-2-1}, \ref{eq::tw-mod-dens-2-2}) and mean square error (between observed and predicted loss costs), $MSE=\sum_{i=1}^{L}\big(\sum_{j=1}^{n_i}w_{ij}(y_{ij}-\widehat{y}_{ij})\big)^2/L$, at the aggregated level for model comparison. Regarding estimated spatial effects in each replication, we compute associated means, standard deviations and 95\% quantiles for each zip-code.

\subsection{Connecticut}

In HLDI data, the state of Connecticut contains 22,337,318 records on 282 zipcodes, with 96.18\% of them \noindent being exact zeros. Number of claims observed range from 0--8 within given exposure periods. The range of observed losses were 0--190,487 (in US dollars) with mean loss of 167.202,  estimated losses, $\widehat{\mu}$ from the DGLM were,  0.002--4023.710 with a mean of 158.337. Estimated dispersions, $\widehat{\phi}$ were 267.045--1038.674.

\begin{figure}[t]
	\centering
	\begin{subfigure}{.33\textwidth}
		\centering
		\includegraphics[width=1\linewidth , height=0.9\linewidth]{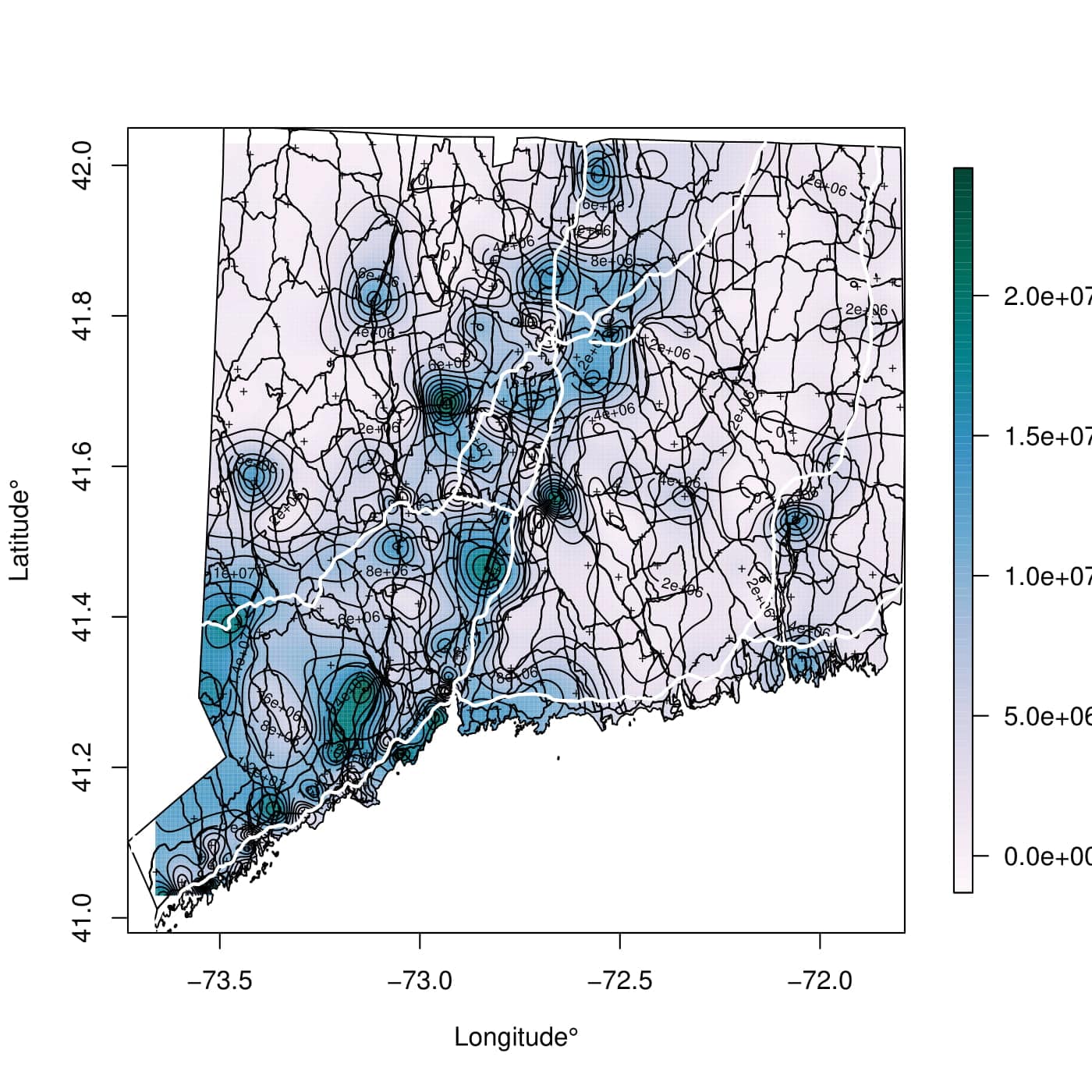}
		\caption{}
	\end{subfigure}~~%
	\begin{subfigure}{.33\textwidth}
		\centering
		\includegraphics[width=1\linewidth , height=0.9\linewidth]{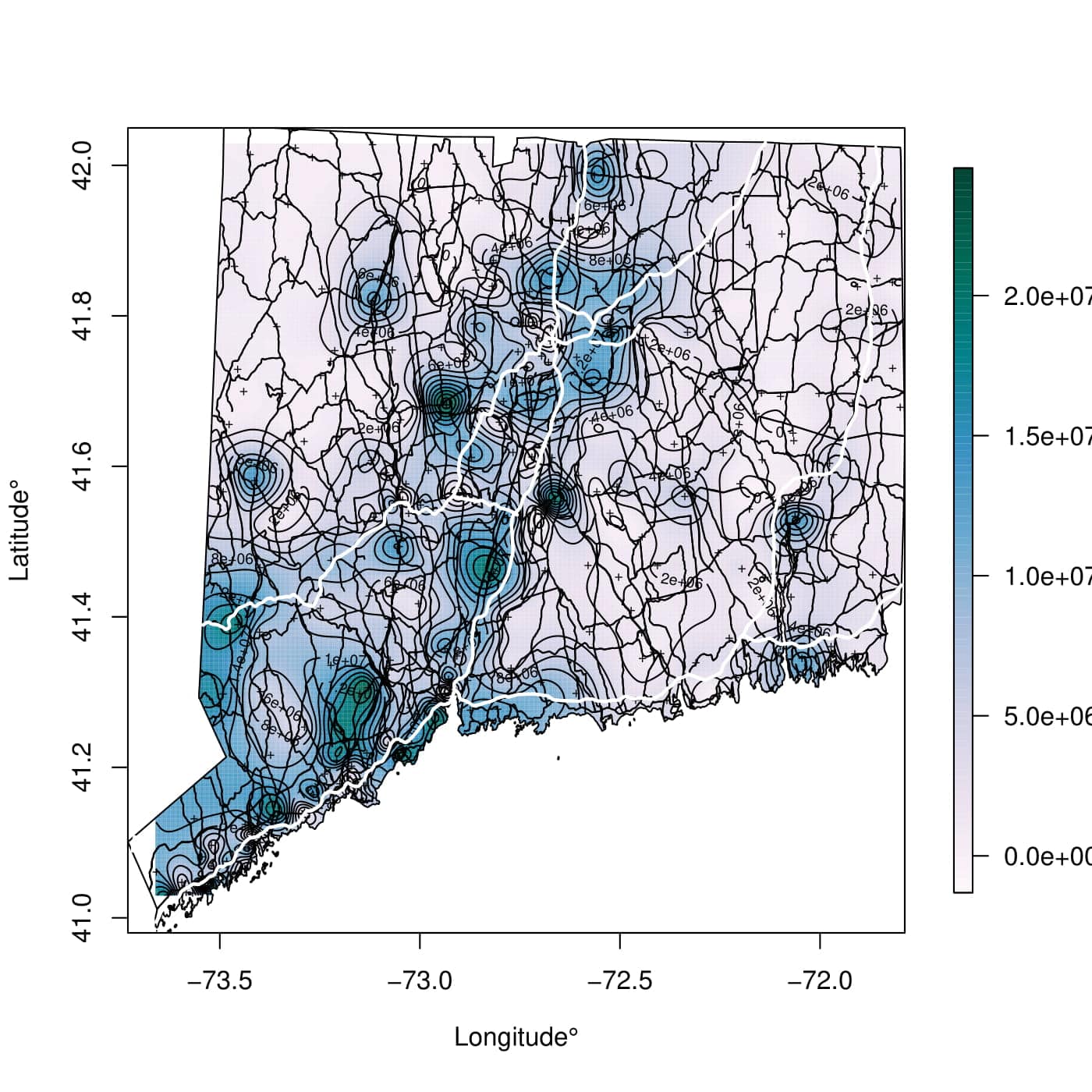}
		\caption{}
	\end{subfigure}~~%
	\begin{subfigure}{.33\textwidth}
		\centering
		\includegraphics[width=1\linewidth , height=0.9\linewidth]{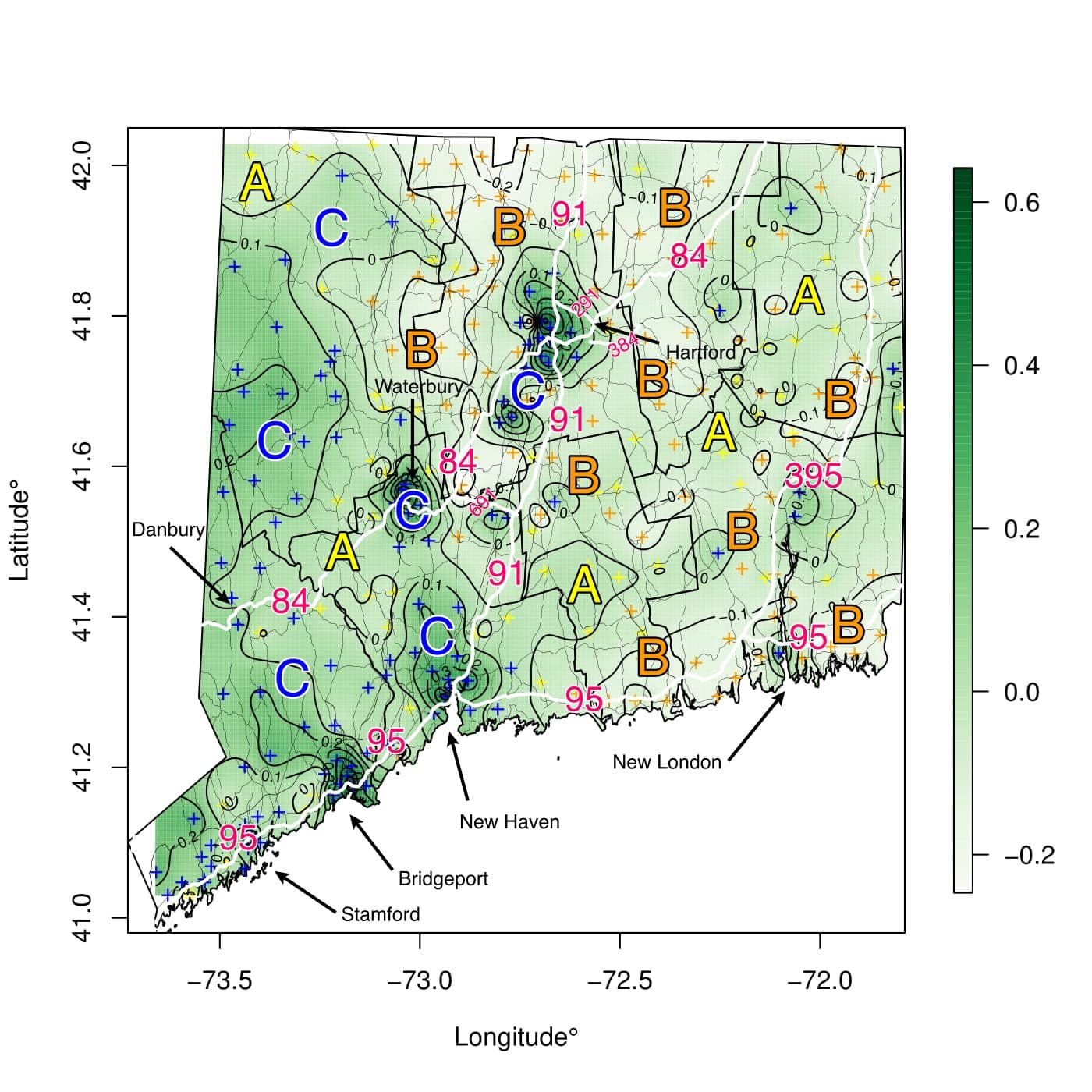}
		\caption{}
	\end{subfigure}%
	\caption{Spatial plots showing \emph{average} zip-code level out-sample (a) observed losses (b) predicted losses adjusted for spatial effects,  (c) estimated spatial effects from algorithm \ref{algo::md-1} for state of CT. (\emph{Note:} Plot (a), (b) and (c) shows primary interstate (I- 84, 384, 91, 95, 691, 291, 95, 395), secondary state highways and major cities in CT with bold (``white"), usual lines and arrows respectively. Also in plot (c), across replications, zipcodes whose approximate 95\% CIs contain zero,  both limits are negative and positive are color coded in ``yellow" (A), ``orange" (B) and ``blue" (C) respectively.)}
	\label{fig::ct-real}
\end{figure}

Estimates from the proposed algorithm (\ref{algo::md-1}), along with their ridge and un-penalized counterparts were obtained for 20 replications involving random training and validation splits. Out-sample prediction results for deviance and $MSE$ are shown in table \ref{tab::real-data-ct}. In both observation level and aggregated cases DGLM serves as the baseline. As expected, the improvement at aggregated level is much larger in comparison to the observation level, with an average of 0.33\% and 93.10\%  improvements for predictions adjusted with spatial effects from proposed algorithm respectively. Figure (\ref{fig::ct-real}a, \ref{fig::ct-real}b) show resulting zip-code level aggregated observed and predicted loss costs after adjustment with estimated spatial effect from proposed algorithm.

\begin{table}[b]
	\centering
	\caption{Table showing averaged out-sample results for CT, consisting of deviance and $MSE$ at the observation and zip-code  aggregated level (lower is better), accompanied by mean percentage improvements obtained with DGLM as a baseline for un-penalized, ridge and proposed estimates (GL). Standard deviations, first ($Q_1$) and third ($Q_3$) for \%ge improvements are shown in following rows.}\label{tab::real-data-ct}
	\resizebox{\linewidth}{!}{
		\begin{tabular}{@{\extracolsep{20pt}}l|cccccccc@{}}
			\hline
			\multirow{4}{2.5cm}{\centering Percentage Improvements}& \multicolumn{4}{c}{\multirow{2}{*}{Observation Level -- Deviance $(\times 10^5)$}} & \multicolumn{4}{c}{\multirow{2}{*}{Aggregated Level -- MSE $(\times 10^{10})$}}\\
			&&&&&&&&\\
			\cline{2-5}\cline{6-9}
			& \multirow{2}{*}{DGLM} & \multirow{2}{*}{GL} & \multirow{2}{*}{MLE} & \multirow{2}{*}{Ridge} & \multirow{2}{*}{DGLM} & \multirow{2}{*}{GL} & \multirow{2}{*}{MLE} & \multirow{2}{*}{Ridge}  \\
			&&&&&&&&\\
			\hline
			Values& 3.743265 & 3.730861 & 3.730994 & 3.730907 & 122.371867 & 8.437873 & 8.675707 & 8.541663 \\
			Mean (\%) & -- & 0.331350 & 0.327798 & 0.330115 & -- & 93.100513 & 92.904780 & 93.016643 \\
			Std. Dev & -- & (0.008448) & (0.009071) & (0.008366) & -- & (0.764774) & (0.820851) & (0.773495) \\
			$Q_1$ & -- & 0.319138 & 0.314905 & 0.318801 & -- & 91.514987 & 91.318403 & 91.554967 \\
			$Q_3$ & -- & 0.346402 & 0.344539 & 0.345565 & -- & 94.144875 & 93.966229 & 94.066700 \\
			\hline
		\end{tabular}
	}
\end{table}

Associated tuning parameters were estimated over a $50\times 50$ grid constructed within $[-5,5]\times [-5,5] \in \mathbb{R}^2$ in a logarithmic scale for all replications. Estimated tuning parameters $\lambda_1$, $\lambda_2$ (with standard deviations shown in brackets followed by an approximate 95\% interval) for algorithm (\ref{algo::md-1}) are (i) $\widehat{\lambda}_1=1.37~ (2.39)$,  $(0.01,6.95)$, (ii) $\widehat{\lambda}_2=19.44~ (2.58)$, $(15.72,23.65)$. For the ridge variant we have, $\widehat{\lambda}_1=37.25 ~ (4.04)$, $(32.12,43.62)$. The overall mean estimated spatial effect across all zipcodes and replications was 0.033 (0.167), i.e. a multiplicative effect of $e^{0.033}\approx 1.034$ to loss costs in the state on an average. Figure (\ref{fig::ct-real}c) shows a spatial plot of estimated spatial effects using algorithm (\ref{algo::md-1}). 

Across replications an approx. 95\% confidence interval (CI) is calculated for the spatial effect in each zip-code. Based on them containing zero and algebraic sign of their limits indicating whether the multiplicative effect is identity (if interval contains 0), decreasing (if both limits are negative) and increasing (both limits are positive) these effects are color coded into three categories which are shown in the spatial plot (\ref{fig::ct-real}c). For perspective, primary interstate highways and secondary roads passing through and within the state are shown as well. From the southwest corner until the intersection of primary interstates i.e. I-91 and 84 at Hartford, regions marked with ``C" having a higher multiplicative spatial effect, in comparison to northwest or the eastern part of CT, which are coded with ``A" or ``B" indicating the multiplicative factor is 1 or $<1$ respectively. Obtaining ordinal spatial risk ranking is rather straightforward, making this a demonstration of boundary analysis in CT, where primary interstate highways, their intersections and big cities serve as defining boundaries for spatial risk.

In comparison to ridge and un-penalized estimates, our proposed estimates show lower standard deviation (average of 7.22\% and 15.62\% respectively) for zipcodes that did not contain 0 in their approx. 95\% CI. The number of zipcodes with 0 belonging to their approx. CIs, less and more than 0  changed from $\{86, 94, 102\}$ in un-penalized and ridge, to $\{67,104, 111\}$ in proposed estimates.

\subsection{New England}

New England is a group of six states in the east coast of USA viz., Connecticut, Rhode Island, Massachusetts, Vermont, New Hampshire and Maine. HLDI data for New England consists of 72,118,625 records on 1831 zipcodes. The percentage of exact zeros in the response is 95.27\%. Losses (payments) ranged between 0--427,510 with an average loss of 184.23 (in US dollars). Estimated mean, $\widehat{\mu}$ from fitted DGLM ranged from 0.02--26,791.86, with an average of 184.37 (in US dollars) while, estimated dispersions $\widehat{\phi}$ were in the range, 267.04--1038.67.

\begin{table}[b]
	\centering
	\caption{Table showing averaged out-sample results for New England, consisting of deviance and $MSE$ at the observation and zip-code  aggregated level (lower is better), accompanied by mean percentage improvements obtained with DGLM as a baseline, compared to proposed \emph{approximate} estimates (GL). Standard deviations,  first ($Q_1$) and third ($Q_3$) quartile for \%ge improvements are shown in following rows.}\label{tab::real-data-ne}
	\resizebox{\linewidth}{!}{
		\begin{tabular}{@{\extracolsep{20pt}}l|cccc@{}|cc@{}}
			\hline
			\multirow{4}{2.5cm}{\centering Percentage Improvements}& \multicolumn{2}{c}{\multirow{2}{*}{Observation Level -- Deviance $(\times 10^6)$}} & \multicolumn{2}{c|}{\multirow{2}{*}{Aggregated Level -- MSE $(\times 10^{10})$}}&  \multicolumn{2}{c}{\multirow{2}{*}{Tuning Parameters}}\\
			&&&&&&\\
			\cline{2-3}\cline{4-5}\cline{6-7}
			& \multirow{2}{*}{DGLM} & \multirow{2}{*}{GL} &  \multirow{2}{*}{DGLM} & \multirow{2}{*}{GL} & \multirow{2}{*}{$\widehat{\lambda}_1$} & \multirow{2}{*}{$\widehat{\lambda}_2$}\\
			&&&&&&\\
			\hline
			Values& 2.965573 & 2.951098  & 93.567022 & 3.109582  &   &  \\
			Mean (\%) & -- & 0.488101 & -- & 96.676626 & \multirow{2}{*}{5.302584} & \multirow{2}{*}{26.466031}\\
			Std. Dev & -- & (0.004236)  & -- & (0.667377) & & \\
			$Q_1$ & -- & 0.483620  & -- & 95.233090  & \multirow{2}{*}{(0.938693)} & \multirow{2}{*}{(0.000000)}\\
			$Q_3$ & -- & 0.494925  & -- & 97.029910 & & \\
			\hline
		\end{tabular}
	}
\end{table}

Spatial effects are estimated using solution in  eq. (\ref{eq::mm-est-approx}) with algorithm (\ref{algo::md-1}). We have already observed in simulations comparing exact and approximates solutions, depending on smoothness of unobserved spatial effects performance of the approximate solution is comparable to its exact counterpart with substantial improvement in time complexity. This significantly improves scalability in case studies like the present one with multiple states involved. Tuning parameters are estimated over a $30\times 30$ grid within $[-5,5]\times [-5,5]$ in logarithmic scale. Average estimated tuning parameters from 10 replications are shown in table (\ref{tab::real-data-ne}) accompanied by percentage of improvement in deviance and $MSE$ at observation and zip-code level aggregation respectively. An improvement of 0.488\% and 96.677\% is seen at the observation and aggregated levels respectively. Average overall spatial effect estimated across all zipcodes is -0.077 (i.e. a multiplicative effect of $e^{-0.077}=0.926$), having a standard deviation of 0.167. Figure \ref{fig::ne-speff-real}, in Appendix C shows a spatially interpolated surface derived from estimated spatial effects. Letters ``A", ``B" and ``C" have the same interpretation as in earlier case study. On closer inspection of the figure following are readily apparent,

\begin{figure}[t]
	\centering
	\begin{subfigure}{.5\textwidth}
		\centering
		\includegraphics[width=1.1\linewidth , height=0.8\linewidth]{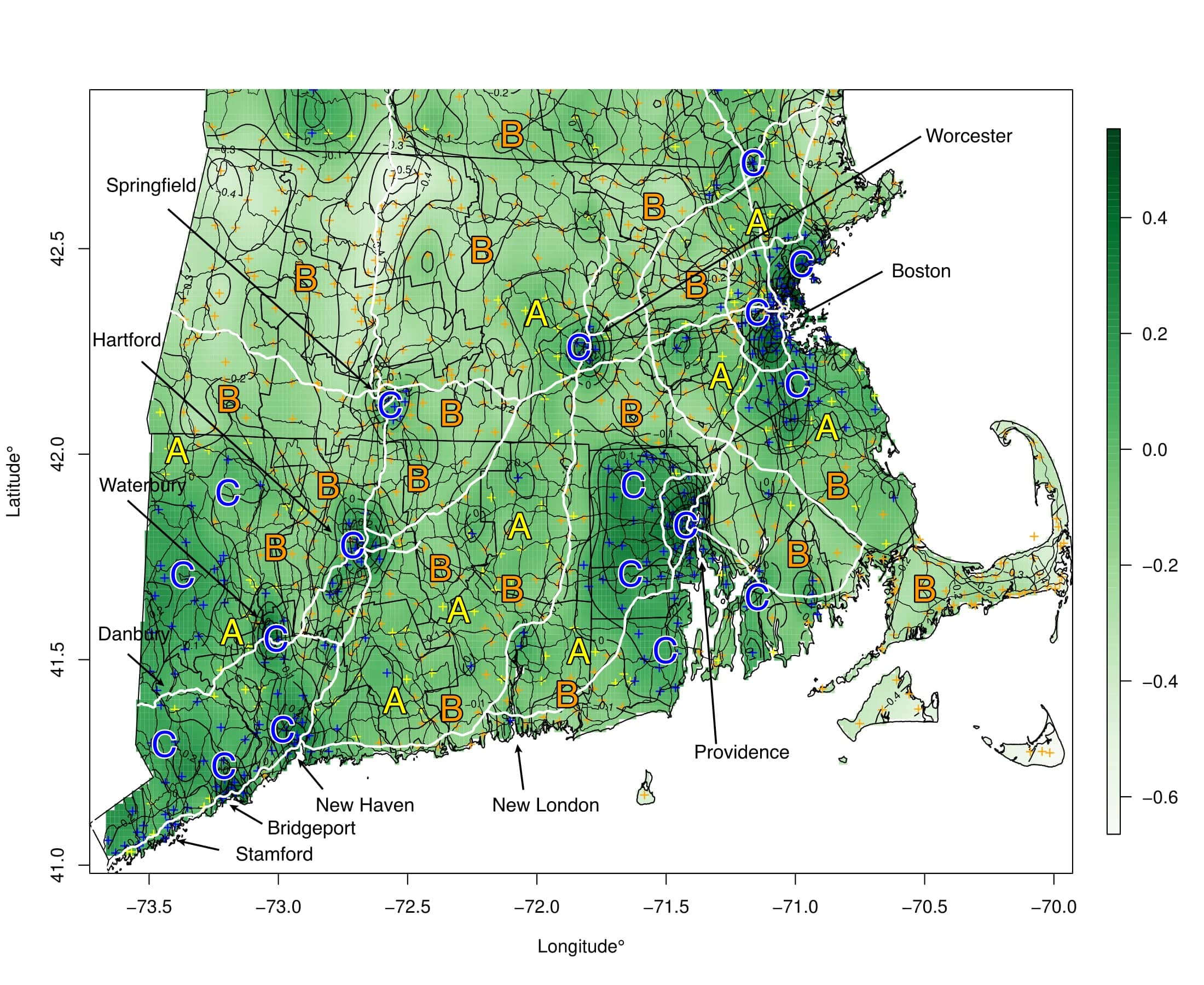}
		\caption{}
	\end{subfigure}~~~~%
	\begin{subfigure}{.5\textwidth}
		\centering
		\includegraphics[width=0.9\linewidth , height=0.8\linewidth]{plots/CTplots-1.jpg}
		\caption{}
	\end{subfigure}%
	\caption{Plots comparing estimated spatial effects obtained from the approximate algorithm for (a) states in southern New England (Connecticut, Rhode Island and Massachusetts) as an inset of fig. \ref{fig::ne-speff-real} and, exact estimates for (b) Connecticut, to demonstrate smoothness in estimated effects across state borders. Symbols and lines have the same description as in fig. \ref{fig::ct-real}.}
	\label{fig::ne-inset}
\end{figure}

\begin{itemize}
	\item regions in northern New England that are devoid of roads (consequently habitation) show sizable areas with negative spatial effects (marked by``B") that indicate a decreasing multiplicative effect,
	\item regions in southern New England show more frequently appearing regions with positive spatial effects (marked by``C") that indicate a increasing multiplicative effect.
	\item regions marked in ``A", i.e. neutral spatial regions act as \emph{separating boundaries} between regions marked by ``B" and ``C".
	\item Big cities and intersections of primary interstate highways (ex. I-95, 91, 84, 290 etc.) show positive spatial effects with large magnitudes. For instance, Suffolk (where state capital, Boston is a city) and Norfolk, MA are adjacent counties having spatial effect averaged (std. deviations) at the county level of 0.275 (0.280) and 0.015 (0.123) respectively. They are the only counties in MA having positive aggregated effects. They contain intersections of major interstate highways (viz., I-95, 90 and 93) and cities like Boston and Quincy. Boston and Quincy had aggregated (at city-level) effects (with std. deviations) of 0.242 (0.272) and 0.139 (0.096) respectively.
\end{itemize}

\begin{table}[t]
	\centering
	\caption{Table showing summarized aggregated (at city-level) spatial effects for cities shown in fig. \ref{fig::ne-speff-real}. Cities having one zipcode under them have no estimate of std. deviation, or quartiles ($Q_1,Q_3$)}\label{tab::ne-cities}
	\resizebox{\linewidth}{!}{
		\begin{tabular}{l|@{\extracolsep{8pt}}ccccccccc@{}}
			\hline
			\hline
			\multirow{6}{2cm}{\centering Estimated Spatial Effects}& \multicolumn{8}{c}{\multirow{2}{*}{State}} \\
			&&&&&&&\\
			\cline{2-9}
			& \multicolumn{2}{c}{\multirow{2}{*}{CT}} & \multirow{2}{*}{RI} & \multicolumn{2}{c}{\multirow{2}{*}{MA}} &\multirow{2}{*}{NH} &  \multirow{2}{*}{VT} & \multirow{2}{*}{ME} \\
			&&&&&&&\\
			\cline{2-3}\cline{5-6}
			& \multirow{2}{*}{Hartford} & \multirow{2}{*}{New Haven} & \multirow{2}{*}{Providence} & \multirow{2}{*}{Boston}  & \multirow{2}{*}{Worcester} & \multirow{2}{*}{Concord}  & \multirow{2}{*}{Montpelier} & \multirow{2}{*}{Augusta} \\
			&&&&&&&\\
			\hline
			Mean &  0.022375 &  0.068355 &  0.376642  &  0.242398 &  0.198258 & -0.137677 & -0.110704 & -0.055297  \\
			Std. Dev &  (0.090117) &  (0.160785) &  (0.241224)  &  (0.272502) &  (0.192553) & (0.037967) & -- & --  \\
			$Q_1$& -0.111165 & -0.138137 & -0.004110 & -0.099666 & -0.132779 & -0.163182 & -- & --  \\
			$Q_3$ &  0.121738 &  0.203204 &  0.604134 &  0.712228 &  0.400212 & -0.112172 &  -- & --  \\
			\hline
		\end{tabular}
	}	
\end{table}

\noindent Table \ref{tab::ne-cities} shows summaries for estimated spatial effects aggregated to the city level which support above findings.

 The only remaining concern is manifestation of unusual discreteness (ex. large differences in magnitude, or algebraic signs)  in estimated spatial effects across state borders. Consequently we compared zipcodes lying on the boundary for exact estimated effects in CT, and subset of CT from the approx. estimates, with respect to those in MA and RI. There are \{17, 22\} zipcodes on the CT-MA border and, \{9, 10\} zipcodes in CT-RI border respectively. Presence of Providence and, RI being a state with comparatively smaller area affects magnitudes of estimates on the CT-RI border (means of -0.057 and 0.091 respectively). However, this is not the case for CT-MA border, with Springfield being the only major city, estimates show differences of lower magnitude (means of -0.096 and -0.154 respectively). Figure \ref{fig::ne-inset} illustrates this observation visually.

\subsection{West Coast}

\begin{table}[b]
	\centering
	\caption{Table showing averaged out-sample results for West Coast, consisting of deviance and $MSE$ at the observation and zip-code  aggregated level (lower is better), accompanied by mean percentage improvements obtained with DGLM as a baseline, compared to proposed \emph{approximate} estimates (GL). Standard deviations,  first ($Q_1$) and third ($Q_3$) quartile for \%ge improvements are shown in following rows.}\label{tab::real-data-wc}
	\resizebox{\linewidth}{!}{
		\begin{tabular}{@{\extracolsep{20pt}}l|cccc@{}|cc@{}}
			\hline
			\multirow{4}{2.5cm}{\centering Percentage Improvements}& \multicolumn{2}{c}{\multirow{2}{*}{Observation Level -- Deviance $(\times 10^6)$}} & \multicolumn{2}{c|}{\multirow{2}{*}{Aggregated Level -- MSE $(\times 10^{11})$}}&  \multicolumn{2}{c}{\multirow{2}{*}{Tuning Parameters}}\\
			&&&&&&\\
			\cline{2-3}\cline{4-5}\cline{6-7}
			& \multirow{2}{*}{DGLM} & \multirow{2}{*}{GL} &  \multirow{2}{*}{DGLM} & \multirow{2}{*}{GL} & \multirow{2}{*}{$\widehat{\lambda}_1$} & \multirow{2}{*}{$\widehat{\lambda}_2$}\\
			&&&&&&\\
			\hline
			Values& 7.318123 & 7.280574  & 26.64058 & 1.183930  &   &  \\
			Mean (\%) & -- & 0.513094 & -- & 95.557069 & \multirow{2}{*}{4.719600} & \multirow{2}{*}{37.363330}\\
			Std. Dev & -- & (0.003630)  & -- & (0.245279) & & \\
			$Q_1$ & -- & 0.508669  & -- & 95.270866  & \multirow{2}{*}{(0.000000)} & \multirow{2}{*}{(0.000000)}\\
			$Q_3$ & -- & 0.517819  & -- & 95.754971 & & \\
			\hline
		\end{tabular}
	}
\end{table}

Three states namely, California (CA), Oregon (OR) and Washington (WA) compose the west coast. In HLDI, west coast consists of 235,329,963 records on 2775 zipcodes. This test case is considered as a demonstration of performance for proposed estimates under large number of zipcodes and sample size. The response consisted of 96.27\% zeros.  Losses range from 0--507,631 (in US dollars). The exposure in policy years range from 0.003 -- 14.825 with a mean of 0.612. Mean estimates, $\widehat{\mu}$ from fitted DGLM range from 0.054 -- 5,194.105, with estimated dispersion $\widehat{\phi}$ in the range 247.229 -- 1,294.303. The mean loss (payment) is 165.980, with an estimated mean loss of 165.721 US dollars from fitted DGLM. 
\begin{table}[t]
	\centering
	\caption{Table showing summarized aggregated (at city-level) spatial effects for cities shown in fig. \ref{fig::wc-real}. Standard deviations, first and third quartiles ($Q_1$,$Q_3$) are shown for multiple zipcodes under a city.}\label{tab::wc-cities}
	\resizebox{\linewidth}{!}{
		\begin{tabular}{l|@{\extracolsep{8pt}}cccccccc@{}}
			\hline
			\hline
			\multirow{6}{2cm}{\centering Estimated Spatial Effects}& \multicolumn{7}{c}{\multirow{2}{*}{State}} \\
			&&&&&&\\
			\cline{2-8}
			& \multicolumn{1}{c}{\multirow{2}{*}{WA}} & \multirow{2}{*}{OR} & \multicolumn{5}{c}{\multirow{2}{*}{CA}}\\
			&&&&&&\\
			\cline{4-8}
			& \multirow{2}{*}{Seattle} & \multirow{2}{*}{Portland} & \multirow{2}{*}{Sacramento} & \multirow{2}{*}{San Francisco}  & \multirow{2}{*}{San Jose} & \multirow{2}{*}{Los Angeles}  & \multirow{2}{*}{San Diego} \\
			&&&&&&\\
			\hline
			Mean & -0.044461 & -0.106216 &  0.156509 &  0.246519 &  0.119130 &  0.376755 & -0.026721 \\
			Std. Dev. &  (0.085933) &  (0.079656) &  (0.094374) &  (0.122083) &  (0.081754) &  (0.102605) &  (0.098419) \\
			$Q_1$ & -0.153033 & -0.209162 &  0.014194 &  0.106287 &  0.009060 &  0.187888 & -0.149399 \\
			$Q_3$ &  0.147396 &  0.039022 &  0.291680 &  0.517545 &  0.305338 &  0.569905 &  0.165796 \\
			\hline
		\end{tabular}
	}	
\end{table}

We considered 5 replications, with tuning parameters estimated over a $30 \times 30$ grid on $[-5,5]\times [-5,5]$ in log-scale. Results obtained are shown in table \ref{tab::real-data-wc}. Average (std. deviation) out-sample improvements of 0.513\% (0.004) and 95.557\% (0.245) at the observation and aggregated zip-code level were observed respectively over 5 replications. Estimated spatial effects had a mean (std. deviation) of -0.057 (0.222) ($\equiv$ multiplicative effect of $e^{-0.057}\approx 0.945$ to observed losses). Predicted losses (payments) after being adjusted with spatial effects ranged from 0.003--8,806.819 (in US dollars), with an estimated mean of 166.869 US dollars.

\begin{figure}[H]
	\centering
	\begin{subfigure}{1\textwidth}
		\centering
		\includegraphics[width=0.85\linewidth , height=1.3\linewidth]{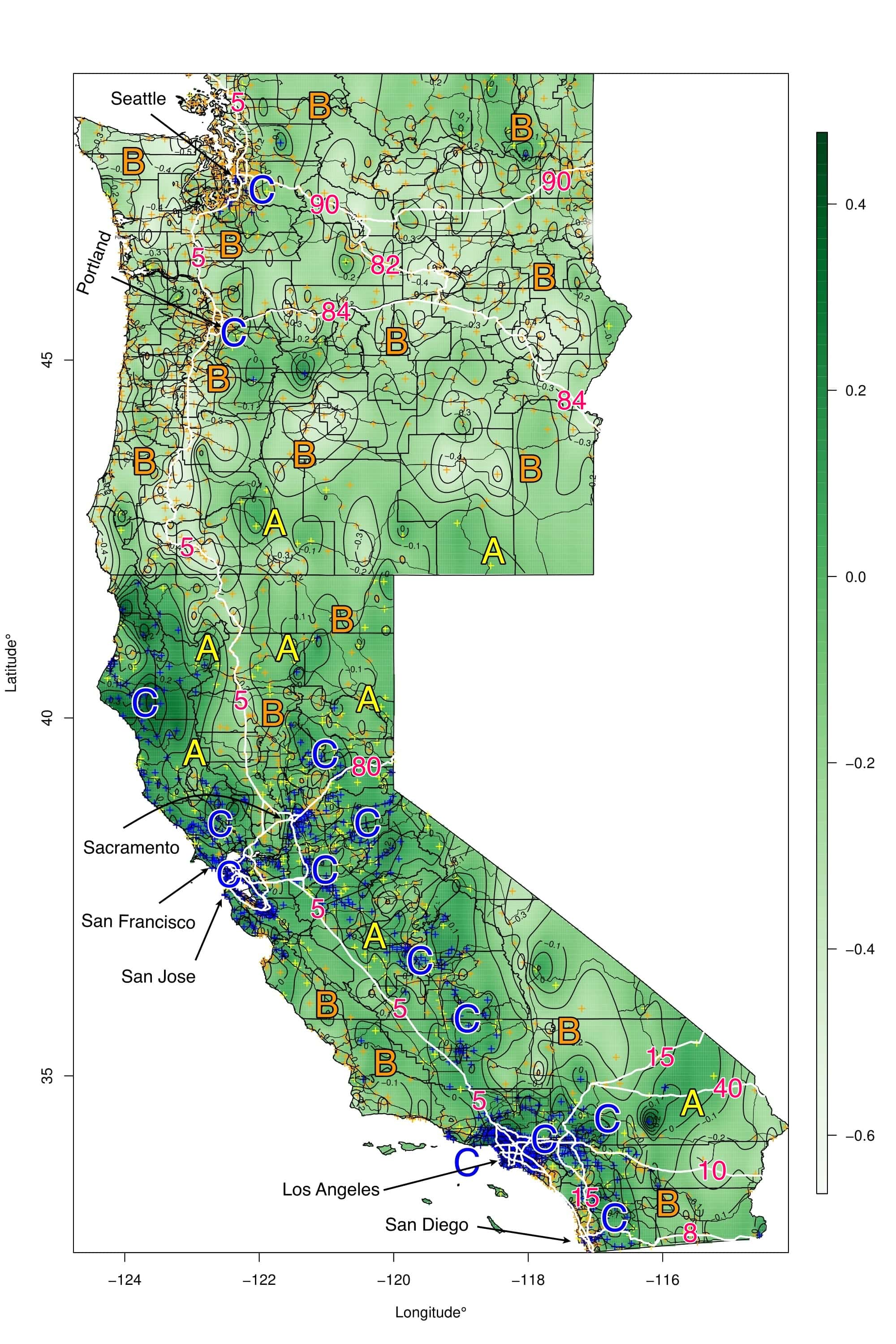}
		\caption{}
	\end{subfigure}
	\caption{Plot showing estimated spatial effects for states in West Coast (with county borders). It includes all primary (marked in bold ``white") and secondary (marked in ``black" lines) roads. Zipcodes (2775 in total) are color coded,  ones with 0 in their approx. CIs are ``yellow" (marked in ``A"), ones below zero are ``orange" (sizable regions marked with ``B"), and ones above are ``blue" (regions marked with ``C"). Some major cities are marked with arrows.}
	\label{fig::wc-real}
\end{figure}

Regarding boundary analysis, referring to figure \ref{fig::wc-real} we observe big cities and major intersections of primary interstates show positive spatial effects, for instance the cities shown in table \ref{tab::wc-cities}. However positive spatial effects (marked with ``C") occur more frequently across replication in California along interstate I-5, however Oregon and Washington show spatial effects that are negative (marked by ``B") or including zero in ther approx. 95\% CIs. Major cities like San Francisco, Los Angeles also form wombling boundaries i.e. show rapid changes in estimated spatial effects.

\section{Conclusion}\label{sec::conc}

Motivated by lack of methods for spatial estimation that scale well in terms of both processing and memory load, for data involving increasingly large number of locations, we developed exact and approximate (blockwise-exact) methods for penalized estimation of un-observable (fixed) spatial effects, while using predictions from a compound Poisson DGLM as offsets. Both exact and approximate algorithms utilize the majorization descent property which assures convergence. Using offsets paired with a zip-code level adjacency allows for significant improvement in out-sample performance when aggregated at the zip-code (areal) level over the DGLM. 

We presented detailed simulations accompanied by motivating case studies to illustrate the performance and features, including scalability of our proposed algorithm under different spatial settings. Methods shown can be applied to any existing GLM implementation after incorporating necessary changes in the link and deviance. With the increasing applicability of compound Poisson models and availability of geo-tagged areal data we feel that proposed algorithms will be of use in analyzing spatial variability in both individual and sizable groups of states.

\section{Discussion and future work}\label{sec::disc-future}

All computations shown here were performed on an \texttt{Intel (R) Xeon (R) Gold 6150 CPU @ 2.70GHz} with 128GB of RAM. Case studies featuring,
\begin{itemize}
	\item the state of Connecticut took an average (std. deviation) of 0.684 (0.053) minutes for each of 20 replications,
	\item New England took an average (std. deviation) of 4.618 (0.095) minutes for each of 10 replications
	\item West Coast took an average (std. deviation) of 2.087 (0.037) hours for each of 5 replications.
\end{itemize} 
 To give brief details regarding scale, HLDI data for different state(s) took disk spaces of 395.52MB, 0.96 and 3.48GB respectively for the three case studies considered. All spatial plots accompanied by computation shown was done using \texttt{R} \cite{rcite}. The library \texttt{MBA} \cite{rmbacite}, was used to spatially interpolate estimates and produce the wombling surface shown in plots using multilevel B-splines. Standard libraries like \texttt{sp} (\cite{rsp1}, \cite{rsp2}) and \texttt{maptools} (\cite{rmap}) were helpful in obtaining shapefiles necessary for plots.

There is scope for considerable future work. We aim to establish an efficient unified framework for estimating the DGLM parameters simultaneously with spatial effects that can incorporate variable selection methods like elastic net (for ex. in \cite{qian2016tweedie}) . Occurrence of exact zeros, as we have seen is primarily controlled by the dispersion model, thus hence having a spatial effect there would be another future goal.

\section*{Acknowledgments}

The authors would like to thank Brien Aronov,  whose valuable input regarding real world applications, data and necessary computational tools facilitated this research to a great extent.

\newpage

\section*{Appendix}\label{sec::appendix}

\subsection*{Appendix A: Proofs}

\emph{Proof of Theorem \ref{th::con-algo-1}:}

\medskip

\noindent $\mathcal{L}\big(\balpha^{(t)}\big|\balpha\big)$ in eq. (\ref{eq::surr-fun}) is the majorizing function. For any $\bdelta \in \mathbb{R}^{L}$ we have,

\begin{eqnarray}
\scalemath{0.88}{\mathcal{L}\big(\balpha^{(t)}+\bdelta\big|\balpha\big)- \mathcal{L}\big(\balpha^{(t)}\big|\balpha\big)}&=&\scalemath{0.88}{\bigg[\ell(\balpha)+(\balpha^{(t)}+\bdelta-\balpha)\trans\nabla_1(\balpha)+\frac{1}{2}(\balpha^{(t)}+\bdelta-\balpha)\trans \big(\tI_L+\nabla_2(\balpha)\big) (\balpha^{(t)}+\bdelta-\balpha)\bigg]}\nonumber\\
&&\scalemath{0.88}{-\bigg[\ell(\balpha)+(\balpha^{(t)}-\balpha)\trans\nabla_1(\balpha)+\frac{1}{2}(\balpha^{(t)}-\balpha)\trans \big(\tI_L+\nabla_2(\balpha)\big) (\balpha^{(t)}-\balpha)\bigg]}.\nonumber
\end{eqnarray}
Substituting the solution from eq. (\ref{eq::mm-est}) we have, 

\begin{align*}
\scalemath{0.88}{\bdelta\trans\nabla_1(\balpha)+\bdelta\trans\big(\tI_L+\nabla_2(\balpha)\big)(\balpha^{(t)}-\balpha)=-\bdelta\trans\bigg[\lambda_1\tI_L+\lambda_2W\bigg]\balpha^{(t)}}.
\end{align*}
Hence,
\begin{align*}
\scalemath{0.88}{\mathcal{L}\big(\balpha^{(t)}+\delta\big|\balpha\big)- \mathcal{L}\big(\balpha^{(t)}\big|\balpha\big)=-\bdelta\trans\bigg[\lambda_1\tI_L+\lambda_2W\bigg]\balpha^{(t)}+\frac{1}{2}\bdelta\trans\big(\tI_L+\nabla_2(\balpha)\big)\bdelta}.
\end{align*}
Using $\bdelta=(\balpha-\balpha^{(t)})$, $\mathcal{L}\big(\balpha\big|\balpha\big)+P(\balpha;\lambda_1,\lambda_2)=F(\balpha)$ and (\ref{eq::surr-fun}) after some algebra we have,

\begin{align*}
F(\balpha)-F(\balpha^{(t)})	= \frac{1+\lambda_1}{2}||\balpha-\balpha^{(t)}||_2^2+\frac{1}{2}(\balpha-\balpha^{(t)})\trans\big(\nabla_2(\balpha)+\lambda_2W\big)(\balpha-\balpha^{(t)}).
\end{align*}
Noting that $\nabla_2(\balpha)$ and graph Laplacian $W$ are both p.s.d. matrices we have the result in theorem (\ref{th::con-algo-1}).

\bigskip

\noindent  \emph{Proof of positive semi-definiteness of $W_a$:}

\medskip

\noindent For $k=1,\ldots,S$ and arbitrary $\bx = (\bx_1,\ldots, \bx_S)\trans \in \mathbb{R}^{L}, \bx_k \in \mathbb{R}^{L_k}$ and $L=\sum_{k} L_k$, if $\bx\trans W_k\bx \geq 0$, i.e. each $W_k$ is assumed to be a p.s.d. matrix, then $\bx\trans W_a \bx = \sum_k \bx_k\trans W_k\bx_k \geq 0$, i.e $W_a$ is also p.s.d.

\newpage

\subsection*{Appendix B: Tables}

\begin{table}[H]
	\centering
	\caption{Table showing regularization parameters in log-scale for ridge (Ridge) and estimates from algorithm (\ref{algo::md-1}) (GL) averaged across \emph{all} replications and settings, with their respective standard deviations shown in brackets below. Also, reference distributions used for simulating dispersion $\phi$ to obtain differing proportions of zeros in simulated response are shown for each spatial pattern, with $U(\cdot,\cdot)$ denoting a uniform distribution.}\label{tab::sim-param-summary}
	\resizebox{\linewidth}{!}{
		\begin{tabular}{@{\extracolsep{8pt}}lccccccc@{}}
			\hline
			\hline
			\multirow{5}{*}{Spatial Pattern}&  \multicolumn{3}{c}{Regularization Parameters}  & \multicolumn{4}{c}{\multirow{3}{*}{Proportion of Zeros}}\\
			\cline{2-4}
			& \multirow{2}{2cm}{\hspace{0.5cm}Ridge} & \multicolumn{2}{c}{\multirow{2}{3cm}{\hspace{1.25cm}GL}} &  \\
			&&&&&&&\\
			\cline{3-4} \cline{5-8}
			&\multirow{2}{*}{$\lambda_1$} & \multirow{2}{*}{$\lambda_1$}& \multirow{2}{*}{$\lambda_2$}&\multirow{2}{*}{0.15}&\multirow{2}{*}{0.30}&\multirow{2}{*}{0.60}&\multirow{2}{*}{0.80}\\
			&&&&&&&\\
			\hline
			\multirow{2}{*}{Block}& \multirow{2}{1.5cm}{-1.111 (0.116)}& \multirow{2}{1.5cm}{-3.135 (0.050)} & \multirow{2}{1.5cm}{\hspace{0.15cm}0.002 (0.104)} & \multirow{2}{*}{$U(7,12)$} & \multirow{2}{*}{$U(12,30)$} & \multirow{2}{*}{$U(30,140)$} & \multirow{2}{*}{$U(140,400)$}\\
			& & & & & & & \\
			\hline
			\multirow{2}{*}{Smooth}& \multirow{2}{1.5cm}{-0.333 (0.194)}& \multirow{2}{1.5cm}{-4.270 (0.049)} & \multirow{2}{1.5cm}{\hspace{0.15cm}1.377 (0.088)} & \multirow{2}{*}{$U(7,11)$} & \multirow{2}{*}{$U(11,24)$} & \multirow{2}{*}{$U(24,100)$} & \multirow{2}{*}{$U(100,200)$}\\
			& & & & & & & \\
			\hline
			\multirow{2}{*}{Hot Spot}& \multirow{2}{1.5cm}{-5.000 (0.199)}& \multirow{2}{1.5cm}{-5.000 (0.000)} & \multirow{2}{1.5cm}{\hspace{0.15cm}2.708 (5.534)} & \multirow{2}{*}{$U(25,40)$} & \multirow{2}{*}{$U(40,70)$} & \multirow{2}{*}{$U(70,200)$} & \multirow{2}{*}{$U(200,500)$}\\
			& & & & & & & \\
			\hline
			\multirow{2}{*}{Structured}& \multirow{2}{1.5cm}{-0.333 (0.097)}& \multirow{2}{1.5cm}{-1.930 (0.090)} & \multirow{2}{1.5cm}{\hspace{0.15cm}0.774 (0.232)} & \multirow{2}{*}{$U(5,7)$} & \multirow{2}{*}{$U(6,14)$} & \multirow{2}{*}{$U(16,35)$} & \multirow{2}{*}{$U(40,80)$}\\
			& & & & & & & \\
			\hline
			\hline
		\end{tabular}
	}
\end{table}

\begin{table}[H]
	\centering
	\caption{Comparative results over 100 replications for the un-penalized (MLE), ridge (Ridge) and estimates from algorithm \ref{algo::md-1} (GL) for a \emph{block spatial pattern} using metrics described in section \ref{sec::sim}, are shown for varying proportions of zeros in simulated response and sample sizes. The respective standard deviations are shown in brackets below the value of an error metric.}\label{tab::sim-bl}
	\resizebox{\linewidth}{!}{
		\begin{tabular}{|c|c|@{\extracolsep{8pt}}ccccccccc|@{}}
			\hline
			\multirow{5}{*}{Sample Size}& \multirow{5}{*}{Prop. of zeros} & \multicolumn{3}{c}{\multirow{3}{*}{SSE}}  & \multicolumn{6}{c|}{\multirow{2}{*}{Deviance Ratio}} \\
			&&&&&&&&&&\\
			\cline{6-11}
			&  & & & &  \multicolumn{3}{c}{Training} &  \multicolumn{3}{c|}{Validation}\\ 
			\cline{3-5}\cline{6-8}\cline{9-11}
			&  & \multirow{2}{*}{MLE} & \multirow{2}{*}{Ridge} & \multirow{2}{*}{GL} & \multirow{2}{*}{MLE} & \multirow{2}{*}{Ridge} & \multirow{2}{*}{GL} & \multirow{2}{*}{MLE} & \multirow{2}{*}{Ridge} & \multirow{2}{*}{GL}\\
			&&&&&&&&&&\\
			\hline
			\multirow{8}{*}{10000} & \multirow{2}{*}{0.15}  & 20.04 & 16.42 & 11.13 & 0.9973 & 0.9975 & 0.9978 & 1.0029 & 1.0026 & 1.0021 \\ 
			& & (3.58) & (2.09) & (1.20) & (0.0002) & (0.0002) & (0.0003) & (0.0005) & (0.0005) & (0.0004) \\ 
			\cline{2-2}
			&  \multirow{2}{*}{0.30}  & 95.40 & 33.55 & 19.18 & 0.9940 & 0.9946 & 0.9959 & 1.0582 & 1.0057 & 1.0042 \\ 
			& & (94.56) & (3.54) & (1.94) & (0.0005) & (0.0005) & (0.0007) & (0.1460) & (0.0009) & (0.0008) \\ 
			\cline{2-2}
			&  \multirow{2}{*}{0.60} & 2252.97 & 98.83 & 42.69 & 0.9766 & 0.9846 & 0.9918 & 2.9615 & 1.0205 & 1.0106 \\ 
			& & (575.81) & (10.60) & (5.69) & (0.0019) & (0.0021) & (0.0027) & (1.0034) & (0.0044) & (0.0024) \\ 
			\cline{2-2}
			&  \multirow{2}{*}{0.80} & 7600.11 & 274.11 & 84.41 & 0.9216 & 0.9639 & 0.9859 & 10.2145 & 1.0725 & 1.0213 \\ 
			& & (720.21) & (22.39) & (13.56) & (0.0063) & (0.0084) & (0.0066) & (3.9764) & (0.0164) & (0.0063) \\ 
			\hline
			\multirow{8}{*}{20000} & \multirow{2}{*}{0.15} & 9.49 & 8.68 & 6.75 & 0.9986 & 0.9987 & 0.9988 & 1.0015 & 1.0014 & 1.0013 \\ 
			& & (1.21) & (0.96) & (0.71) & (0.0001) & (0.0001) & (0.0001) & (0.0003) & (0.0003) & (0.0002) \\ 
			\cline{2-2}
			& \multirow{2}{*}{0.30} & 24.95 & 17.91 & 11.92 & 0.9971 & 0.9973 & 0.9977 & 1.0169 & 1.0028 & 1.0023 \\ 
			& & (26.72) & (2.15) & (1.12) & (0.0002) & (0.0003) & (0.0003) & (0.1372) & (0.0004) & (0.0004) \\ 
			\cline{2-2}
			& \multirow{2}{*}{0.60} & 572.46 & 52.09 & 27.18 & 0.9890 & 0.9912 & 0.9942 & 1.5624 & 1.0102 & 1.0067 \\ 
			& & (283.87) & (5.66) & (2.98) & (0.0010) & (0.0010) & (0.0013) & (0.4604) & (0.0020) & (0.0014) \\ 
			\cline{2-2}
			& \multirow{2}{*}{0.80} & 4237.76 & 161.90 & 58.05 & 0.9585 & 0.9764 & 0.9890 & 5.4494 & 1.0361 & 1.0153 \\ 
			& & (595.69) & (14.27) & (8.43) & (0.0029) & (0.0035) & (0.0042) & (2.0800) & (0.0070) & (0.0037) \\ 
			\hline
			\multirow{8}{*}{30000} & \multirow{2}{*}{0.15} & 6.15 & 5.75 & 4.81 & 0.9991 & 0.9991 & 0.9992 & 1.0009 & 1.0009 & 1.0008 \\ 
			& & (0.86) & (0.72) & (0.54) & (0.0001) & (0.0001) & (0.0001) & (0.0002) & (0.0002) & (0.0002) \\ 
			\cline{2-2}
			& \multirow{2}{*}{0.30}  & 13.47 & 11.90 & 8.68 & 0.9981 & 0.9982 & 0.9984 & 1.0020 & 1.0019 & 1.0016 \\ 
			& & (1.85) & (1.25) & (0.89) & (0.0002) & (0.0002) & (0.0002) & (0.0003) & (0.0003) & (0.0003) \\ 
			\cline{2-2}
			& \multirow{2}{*}{0.60}  & 150.56 & 38.19 & 21.30 & 0.9927 & 0.9936 & 0.9954 & 1.1470 & 1.0069 & 1.0049 \\ 
			& & (137.63) & (3.69) & (2.16) & (0.0006) & (0.0006) & (0.0008) & (0.2768) & (0.0012) & (0.0009) \\ 
			\cline{2-2}
			& \multirow{2}{*}{0.80}  & 2748.21 & 110.68 & 45.42 & 0.9731 & 0.9831 & 0.9910 & 3.6249 & 1.0231 & 1.0114 \\ 
			& & (512.76) & (9.19) & (5.22) & (0.0022) & (0.0024) & (0.0031) & (1.2627) & (0.0053) & (0.0030) \\ 
			\hline
			\multirow{8}{*}{50000} & \multirow{2}{*}{0.15} & 3.57 & 3.43 & 3.04 & 0.9995 & 0.9995 & 0.9995 & 1.0006 & 1.0005 & 1.0005 \\ 
			& & (0.35) & (0.32) & (0.27) & (0.0000) & (0.0000) & (0.0000) & (0.0001) & (0.0001) & (0.0001) \\ 
			\cline{2-2}
			& \multirow{2}{*}{0.30} & 7.71 & 7.16 & 5.78 & 0.9988 & 0.9989 & 0.9990 & 1.0012 & 1.0012 & 1.0010 \\ 
			& & (1.02) & (0.81) & (0.64) & (0.0001) & (0.0001) & (0.0001) & (0.0002) & (0.0002) & (0.0002) \\ 
			\cline{2-2}
			&\multirow{2}{*}{0.60} & 36.91 & 24.78 & 15.20 & 0.9957 & 0.9961 & 0.9969 & 1.0121 & 1.0041 & 1.0032 \\ 
			& & (22.96) & (2.72) & (1.52) & (0.0003) & (0.0004) & (0.0004) & (0.0718) & (0.0007) & (0.0006) \\ 
			\cline{2-2}
			& \multirow{2}{*}{0.80}& 1221.82 & 69.85 & 33.67 & 0.9841 & 0.9882 & 0.9927 & 2.1542 & 1.0145 & 1.0086 \\ 
			& & (432.43) & (6.84) & (3.87) & (0.0014) & (0.0014) & (0.0020) & (0.6635) & (0.0029) & (0.0021) \\ 
			\hline
		\end{tabular}
	}
\end{table}

\vspace*{3cm}

\begin{table}[H]
	\centering
	\caption{Comparative results over 100 replications for the un-penalized (MLE), ridge (Ridge) and estimates from algorithm \ref{algo::md-1} (GL) for a \emph{smooth spatial pattern} using metrics described in section \ref{sec::sim}, are shown for varying proportions of zeros in simulated response and sample sizes. The respective standard deviations are shown in brackets below the value of an error metric.}\label{tab::sim-sm}
	\resizebox{\linewidth}{!}{
		\begin{tabular}{|c|c|@{\extracolsep{8pt}}ccccccccc|@{}}
			\hline
			\multirow{5}{*}{Sample Size}& \multirow{5}{*}{Prop. of zeros} & \multicolumn{3}{c}{\multirow{3}{*}{SSE}}  & \multicolumn{6}{c|}{\multirow{2}{*}{Deviance Ratio}} \\
			&&&&&&&&&&\\
			\cline{6-11}
			&  & & & &  \multicolumn{3}{c}{Training} &  \multicolumn{3}{c|}{Validation}\\ 
			\cline{3-5}\cline{6-8}\cline{9-11}
			&  & \multirow{2}{*}{MLE} & \multirow{2}{*}{Ridge} & \multirow{2}{*}{GL} & \multirow{2}{*}{MLE} & \multirow{2}{*}{Ridge} & \multirow{2}{*}{GL} & \multirow{2}{*}{MLE} & \multirow{2}{*}{Ridge} & \multirow{2}{*}{GL}\\
			&&&&&&&&&&\\
			\hline
			\multirow{8}{*}{10000} & \multirow{2}{*}{0.15}  & 18.33 & 13.92 & 8.23 & 0.9970 & 0.9972 & 0.9985 & 1.0088 & 1.0030 & 1.0020 \\ 
			&  & (25.39) & (1.61) & (0.71) & (0.0002) & (0.0003) & (0.0006) & (0.0564) & (0.0005) & (0.0004) \\ 
			\cline{2-2}
			& \multirow{2}{*}{0.30}& 40.73 & 25.80 & 11.97 & 0.9943 & 0.9951 & 0.9980 & 1.0302 & 1.0054 & 1.0030 \\ 
			& & (38.74) & (2.71) & (1.24) & (0.0005) & (0.0006) & (0.0009) & (0.1535) & (0.0011) & (0.0007) \\ 
			\cline{2-2}
			& \multirow{2}{*}{0.60}  & 821.07 & 69.87 & 22.49 & 0.9807 & 0.9848 & 0.9950 & 2.0941 & 1.0173 & 1.0058 \\ 
			& & (331.74) & (7.64) & (3.34) & (0.0017) & (0.0022) & (0.0031) & (0.9489) & (0.0033) & (0.0015) \\ 
			\cline{2-2}
			& \multirow{2}{*}{0.60} & 4425.71 & 149.59 & 29.43 & 0.9442 & 0.9673 & 0.9855 & 8.5876 & 1.0466 & 1.0097 \\ 
			& & (760.88) & (16.00) & (4.89) & (0.0046) & (0.0050) & (0.0026) & (2.7101) & (0.0100) & (0.0034) \\ 
			\hline
			\multirow{8}{*}{20000}  & \multirow{2}{*}{0.15}  & 7.58 & 7.06 & 5.12 & 0.9985 & 0.9986 & 0.9989 & 1.0016 & 1.0015 & 1.0012 \\ 
			& & (0.84) & (0.75) & (0.56) & (0.0001) & (0.0001) & (0.0003) & (0.0003) & (0.0003) & (0.0002) \\ 
			\cline{2-2}
			& \multirow{2}{*}{0.30}& 14.63 & 12.91 & 7.91 & 0.9972 & 0.9974 & 0.9986 & 1.0029 & 1.0027 & 1.0019 \\ 
			& & (2.31) & (1.43) & (0.71) & (0.0003) & (0.0003) & (0.0005) & (0.0005) & (0.0005) & (0.0003) \\ 
			\cline{2-2}
			& \multirow{2}{*}{0.60}& 127.21 & 38.58 & 15.40 & 0.9909 & 0.9923 & 0.9974 & 1.1263 & 1.0088 & 1.0040 \\ 
			& & (112.53) & (3.24) & (1.64) & (0.0006) & (0.0011) & (0.0014) & (0.2419) & (0.0014) & (0.0008) \\ 
			\cline{2-2}
			&\multirow{2}{*}{0.80} & 1636.48 & 87.83 & 25.20 & 0.9731 & 0.9806 & 0.9930 & 3.4035 & 1.0237 & 1.0072 \\ 
			& & (445.75) & (9.99) & (4.75) & (0.0022) & (0.0029) & (0.0040) & (1.1488) & (0.0039) & (0.0023) \\ 
			\hline
			\multirow{8}{*}{30000} &  \multirow{2}{*}{0.15} & 4.93 & 4.73 & 3.75 & 0.9990 & 0.9990 & 0.9992 & 1.0010 & 1.0010 & 1.0009 \\ 
			& & (0.67) & (0.55) & (0.36) & (0.0001) & (0.0001) & (0.0002) & (0.0002) & (0.0002) & (0.0002) \\ 
			\cline{2-2}
			&  \multirow{2}{*}{0.30} & 9.66 & 8.87 & 6.08 & 0.9981 & 0.9982 & 0.9987 & 1.0020 & 1.0019 & 1.0015 \\ 
			& & (1.09) & (0.85) & (0.55) & (0.0001) & (0.0001) & (0.0004) & (0.0003) & (0.0003) & (0.0003) \\ 
			\cline{2-2}
			& \multirow{2}{*}{0.60} & 41.61 & 27.19 & 12.53 & 0.9939 & 0.9948 & 0.9978 & 1.0126 & 1.0059 & 1.0031 \\ 
			& & (31.37) & (2.53) & (1.20) & (0.0005) & (0.0006) & (0.0010) & (0.0385) & (0.0011) & (0.0007) \\ 
			\cline{2-2}
			& \multirow{2}{*}{0.80} & 659.54 & 63.13 & 20.56 & 0.9828 & 0.9866 & 0.9953 & 1.9031 & 1.0155 & 1.0055 \\ 
			& & (309.16) & (7.70) & (3.15) & (0.0014) & (0.0021) & (0.0030) & (0.7373) & (0.0027) & (0.0014) \\ 
			\hline
			\multirow{8}{*}{50000} &  \multirow{2}{*}{0.15} & 2.88 & 2.80 & 2.41 & 0.9994 & 0.9994 & 0.9995 & 1.0006 & 1.0006 & 1.0005 \\ 
			& & (0.35) & (0.34) & (0.26) & (0.0001) & (0.0001) & (0.0001) & (0.0001) & (0.0001) & (0.0001) \\ 
			\cline{2-2}
			&  \multirow{2}{*}{0.30} & 5.43 & 5.20 & 4.01 & 0.9989 & 0.9989 & 0.9992 & 1.0011 & 1.0011 & 1.0009 \\ 
			& & (0.61) & (0.55) & (0.44) & (0.0001) & (0.0001) & (0.0002) & (0.0002) & (0.0002) & (0.0002) \\ 
			\cline{2-2}
			& \multirow{2}{*}{0.60} & 19.48 & 16.45 & 9.15 & 0.9964 & 0.9968 & 0.9982 & 1.0040 & 1.0036 & 1.0023 \\ 
			& & (2.76) & (1.63) & (0.79) & (0.0003) & (0.0004) & (0.0008) & (0.0007) & (0.0006) & (0.0005) \\ 
			\cline{2-2}
			& \multirow{2}{*}{0.80} & 135.68 & 41.28 & 15.93 & 0.9901 & 0.9917 & 0.9968 & 1.1472 & 1.0095 & 1.0043 \\ 
			& & (119.62) & (4.51) & (1.69) & (0.0009) & (0.0013) & (0.0018) & (0.2881) & (0.0018) & (0.0011) \\ 
			\hline
		\end{tabular}
	}
\end{table}

\vspace*{3cm}

\begin{table}[H]
	\centering
	\caption{Comparative results over 100 replications for the un-penalized (MLE), ridge (Ridge) and estimates from algorithm \ref{algo::md-1} (GL) for a \emph{hot-spot spatial pattern} using metrics described in section \ref{sec::sim}, are shown for varying proportions of zeros in simulated response and sample sizes. The respective standard deviations are shown in brackets below the value of an error metric.}\label{tab::sim-hot}
	\resizebox{\linewidth}{!}{
		\begin{tabular}{|c|c|@{\extracolsep{8pt}}ccccccccc|@{}}
			\hline
			\multirow{5}{*}{Sample Size}& \multirow{5}{*}{Prop. of zeros} & \multicolumn{3}{c}{\multirow{3}{*}{SSE}}  & \multicolumn{6}{c|}{\multirow{2}{*}{Deviance Ratio}} \\
			&&&&&&&&&&\\
			\cline{6-11}
			&  & & & &  \multicolumn{3}{c}{Training} &  \multicolumn{3}{c|}{Validation}\\ 
			\cline{3-5}\cline{6-8}\cline{9-11}
			&  & \multirow{2}{*}{MLE} & \multirow{2}{*}{Ridge} & \multirow{2}{*}{GL} & \multirow{2}{*}{MLE} & \multirow{2}{*}{Ridge} & \multirow{2}{*}{GL} & \multirow{2}{*}{MLE} & \multirow{2}{*}{Ridge} & \multirow{2}{*}{GL}\\
			&&&&&&&&&&\\
			\hline
			\multirow{8}{*}{10000} & \multirow{2}{*}{0.15} & 14.27 & 14.27 & 2.42 & 0.9944 & 0.9944 & 0.9980 & 1.0061 & 1.0061 & 1.0011\\
			&  & (1.44) & (1.44) & (0.49)& (0.0005) & (0.0005) & (0.0006) & (0.0011) & (0.0011) & (0.0004)\\
			\cline{2-2}
			& \multirow{2}{*}{0.30} & 24.62 & 24.67 & 3.76& 0.9906 & 0.9906 & 0.9967 & 1.0100 & 1.0100 & 1.0018\\
			&  & (2.61) & (2.61) & (0.90)& (0.0009) & (0.0009) & (0.0017) & (0.0019) & (0.0019) & (0.0007) \\
			\cline{2-2}
			& \multirow{2}{*}{0.60} & 75.68 & 73.94 & 5.53& 0.9774 & 0.9784 & 0.9925 & 1.0838 & 1.0311 & 1.0026\\
			&  & (35.23) & (8.99) & (0.99)& (0.0020) & (0.0020) & (0.0008) & (0.3862) & (0.0063) & (0.0013)\\
			\cline{2-2}
			& \multirow{2}{*}{0.80} & 1333.70 & 266.43 & 10.49& 0.9336 & 0.9578 & 0.9874 & 9.2482 & 1.1248 & 1.0048\\
			& & (477.50) & (27.07) & (3.50)& (0.0063) & (0.0088) & (0.0057) & (4.8585) & (0.0227) & (0.0032)\\
			\hline
			\multirow{8}{*}{20000} & \multirow{2}{*}{0.15} & 6.90 & 6.91 & 1.69 & 0.9972 & 0.9972 & 0.9985 & 1.0030 & 1.0030 & 1.0008\\
			&  & (0.66) & (0.66) & (0.22) & (0.0002) & (0.0002) & (0.0002) & (0.0005) & (0.0005) & (0.0003)\\
			\cline{2-2}
			& \multirow{2}{*}{0.30} & 11.87 & 11.87 & 2.16 & 0.9953 & 0.9953 & 0.9982 & 1.0049 & 1.0049 & 1.0010\\
			&  & (1.16) & (1.16) & (0.31) & (0.0004) & (0.0004) & (0.0002) & (0.0009) & (0.0009) & (0.0003) \\
			\cline{2-2}
			& \multirow{2}{*}{0.60} & 30.31 & 30.39 & 4.55 & 0.9887 & 0.9888 & 0.9956 & 1.0129 & 1.0129 & 1.0022\\
			&  & (3.33) & (3.36) & (0.90) & (0.0010) & (0.0010) & (0.0020) & (0.0022) & (0.0022) & (0.0008)\\
			\cline{2-2}
			& \multirow{2}{*}{0.80} & 142.30 & 105.68 & 6.05 & 0.9701 & 0.9722 & 0.9919 & 1.4604 & 1.0477 & 1.0032\\
			& & (85.35) & (14.43) & (1.27) & (0.0028) & (0.0028) & (0.0010) & (1.2308) & (0.0096) & (0.0015)\\
			\hline
			\multirow{8}{*}{30000} & \multirow{2}{*}{0.15} & 4.56 & 4.56 & 1.41 & 0.9981 & 0.9981 & 0.9988 & 1.0019 & 1.0019 & 1.0006\\
			&  & (0.35) & (0.35) & (0.13) & (0.0001) & (0.0001) & (0.0001) & (0.0003) & (0.0003) & (0.0002)\\
			\cline{2-2}
			& \multirow{2}{*}{0.30} & 7.68 & 7.69 & 1.77 & 0.9969 & 0.9969 & 0.9985 & 1.0032 & 1.0032 & 1.0008\\
			&  & (0.69) & (0.69) & (0.21) & (0.0003) & (0.0003) & (0.0002) & (0.0005) & (0.0005) & (0.0002) \\
			\cline{2-2}
			& \multirow{2}{*}{0.60} & 19.08 & 19.09 & 2.98 & 0.9926 & 0.9926 & 0.9975 & 1.0079 & 1.0079 & 1.0014\\
			&  & (1.84) & (1.85) & (0.69) & (0.0006) & (0.0006) & (0.0010) & (0.0014) & (0.0014) & (0.0006)\\
			\cline{2-2}
			& \multirow{2}{*}{0.80} & 62.30 & 61.54 & 5.24 & 0.9802 & 0.9806 & 0.9928 & 1.0813 & 1.0265 & 1.0025\\
			& & (25.68) & (8.42) & (0.92) & (0.0017) & (0.0017) & (0.0008) & (0.5582) & (0.0062) & (0.0010)\\
			\hline
			\multirow{8}{*}{50000} & \multirow{2}{*}{0.15} & 4.56 & 4.56 & 1.41 & 0.9981 & 0.9981 & 0.9988 & 1.0019 & 1.0019 & 1.0006\\
			&  & (0.25) & (0.25) & (0.14) & (0.0001) & (0.0001) & (0.0002) & (0.0002) & (0.0002) & (0.0001)\\
			\cline{2-2}
			& \multirow{2}{*}{0.30} & 2.70 & 2.70 & 1.09 & 0.9989 & 0.9989 & 0.9993 & 1.0011 & 1.0011 & 1.0005\\
			&  & (0.41) & (0.41) & (0.14) & (0.0002) & (0.0002) & (0.0001) & (0.0003) & (0.0003) & (0.0002) \\
			\cline{2-2}
			& \multirow{2}{*}{0.60} & 10.70 & 10.70 & 2.01 & 0.9957 & 0.9957 & 0.9982 & 1.0046 & 1.0046 & 1.0009\\
			&  & (0.89) & (0.89) & (0.29) & (0.0003) & (0.0003) & (0.0002) & (0.0008) & (0.0008) & (0.0003)\\
			\cline{2-2}
			& \multirow{2}{*}{0.80} & 30.65 & 30.69 & 4.32 & 0.9886 & 0.9886 & 0.9959 & 1.0128 & 1.0128 & 1.0019\\
			& & (3.20) & (3.21) & (0.76) & (0.0010) & (0.0010) & (0.0020) & (0.0023) & (0.0023) & (0.0008)\\
			\hline
		\end{tabular}
	}
\end{table}

\vspace*{3cm}

\begin{table}[H]
	\centering
	\caption{Comparative results over 100 replications for the un-penalized (MLE), ridge (Ridge) and estimates from algorithm \ref{algo::md-1} (GL) for a \emph{structured spatial pattern} using metrics described in section \ref{sec::sim}, are shown for varying proportions of zeros in simulated response and sample sizes. The respective standard deviations are shown in brackets below the value of an error metric.}\label{tab::sim-struc}
	\resizebox{\linewidth}{!}{
		\begin{tabular}{|c|c|@{\extracolsep{8pt}}ccccccccc|@{}}
			\hline
			\multirow{5}{*}{Sample Size}& \multirow{5}{*}{Prop. of zeros} & \multicolumn{3}{c}{\multirow{3}{*}{SSE}}  & \multicolumn{6}{c|}{\multirow{2}{*}{Deviance Ratio}} \\
			&&&&&&&&&&\\
			\cline{6-11}
			&  & & & &  \multicolumn{3}{c}{Training} &  \multicolumn{3}{c|}{Validation}\\ 
			\cline{3-5}\cline{6-8}\cline{9-11}
			&  & \multirow{2}{*}{MLE} & \multirow{2}{*}{Ridge} & \multirow{2}{*}{GL} & \multirow{2}{*}{MLE} & \multirow{2}{*}{Ridge} & \multirow{2}{*}{GL} & \multirow{2}{*}{MLE} & \multirow{2}{*}{Ridge} & \multirow{2}{*}{GL}\\
			&&&&&&&&&&\\
			\hline
			\multirow{8}{*}{10000} & \multirow{2}{*}{0.15} & 14.82 & 13.21 & 9.17 & 0.9945 & 0.9950 & 0.9964 & 1.0058 & 1.0052 & 1.0037 \\ 
			& & (1.57) & (1.21) & (0.90) & (0.0005) & (0.0005) & (0.0011) & (0.0011) & (0.0009) & (0.0008) \\ 
			\cline{2-2}
			& \multirow{2}{*}{0.30} & 25.08 & 20.69 & 12.19 & 0.9912 & 0.9924 & 0.9950 & 1.0098 & 1.0082 & 1.0051 \\ 
			&  & (2.49) & (1.80) & (1.08) & (0.0007) & (0.0007) & (0.0017) & (0.0017) & (0.0014) & (0.0011) \\ 
			\cline{2-2}
			& \multirow{2}{*}{0.60}& 100.39 & 47.84 & 19.90 & 0.9764 & 0.9844 & 0.9902 & 1.0768 & 1.0182 & 1.0081 \\ 
			& & (50.38) & (4.27) & (1.86) & (0.0019) & (0.0021) & (0.0031) & (0.1631) & (0.0034) & (0.0022) \\ 
			\cline{2-2}
			& \multirow{2}{*}{0.80}& 1187.51 & 98.75 & 28.30 & 0.9394 & 0.9660 & 0.9849 & 4.7115 & 1.0411 & 1.0120 \\ 
			& & (418.37) & (8.15) & (3.42) & (0.0050) & (0.0123) & (0.0044) & (2.4765) & (0.0069) & (0.0039) \\ 
			\hline
			\multirow{8}{*}{20000} & \multirow{2}{*}{0.15} & 7.26 & 6.86 & 5.50 & 0.9972 & 0.9974 & 0.9977 & 1.0029 & 1.0027 & 1.0022 \\ 
			& & (0.57) & (0.50) & (0.48) & (0.0002) & (0.0002) & (0.0003) & (0.0004) & (0.0004) & (0.0004) \\ 
			\cline{2-2}
			& \multirow{2}{*}{0.30} & 11.88 & 10.76 & 7.82 & 0.9956 & 0.9959 & 0.9968 & 1.0047 & 1.0042 & 1.0031 \\ 
			& & (1.16) & (0.96) & (0.81) & (0.0004) & (0.0004) & (0.0008) & (0.0008) & (0.0007) & (0.0006) \\ 
			\cline{2-2}
			& \multirow{2}{*}{0.60} & 33.45 & 26.21 & 14.41 & 0.9885 & 0.9905 & 0.9941 & 1.0131 & 1.0104 & 1.0058 \\ 
			& & (3.87) & (2.18) & (1.17) & (0.0010) & (0.0009) & (0.0023) & (0.0026) & (0.0020) & (0.0014) \\ 
			\cline{2-2}
			& \multirow{2}{*}{0.80}  & 161.38 & 55.14 & 21.28 & 0.9719 & 0.9830 & 0.9899 & 1.1974 & 1.0225 & 1.0094 \\ 
			& & (82.12) & (4.12) & (2.11) & (0.0026) & (0.0025) & (0.0035) & (0.3213) & (0.0040) & (0.0026) \\ 
			\hline
			\multirow{8}{*}{30000} & \multirow{2}{*}{0.15} & 4.73 & 4.56 & 3.87 & 0.9982 & 0.9982 & 0.9984 & 1.0019 & 1.0018 & 1.0015 \\ 
			& & (0.49) & (0.45) & (0.35) & (0.0002) & (0.0002) & (0.0002) & (0.0003) & (0.0003) & (0.0002) \\ 
			\cline{2-2}
			& \multirow{2}{*}{0.30} & 7.85 & 7.35 & 5.80 & 0.9970 & 0.9972 & 0.9976 & 1.0031 & 1.0029 & 1.0023 \\ 
			& & (0.77) & (0.63) & (0.52) & (0.0002) & (0.0002) & (0.0003) & (0.0005) & (0.0005) & (0.0004) \\ 
			\cline{2-2}
			&\multirow{2}{*}{0.60} & 21.15 & 17.98 & 11.18 & 0.9924 & 0.9933 & 0.9953 & 1.0084 & 1.0073 & 1.0048 \\ 
			& & (2.23) & (1.68) & (1.09) & (0.0007) & (0.0007) & (0.0014) & (0.0016) & (0.0013) & (0.0011) \\ 
			\cline{2-2}
			& \multirow{2}{*}{0.80} & 65.80 & 39.20 & 17.89 & 0.9815 & 0.9865 & 0.9918 & 1.0387 & 1.0156 & 1.0076 \\ 
			& & (32.03) & (3.05) & (1.76) & (0.0016) & (0.0014) & (0.0027) & (0.1055) & (0.0028) & (0.0020) \\ 
			\hline
			\multirow{8}{*}{50000} & \multirow{2}{*}{0.15} & 2.80 & 2.74 & 2.46 & 0.9989 & 0.9989 & 0.9990 & 1.0011 & 1.0011 & 1.0010 \\ 
			& & (0.24) & (0.24) & (0.22) & (0.0001) & (0.0001) & (0.0001) & (0.0002) & (0.0002) & (0.0002) \\ 
			\cline{2-2}
			& \multirow{2}{*}{0.30} & 4.41 & 4.25 & 3.67 & 0.9983 & 0.9984 & 0.9985 & 1.0017 & 1.0017 & 1.0015 \\ 
			& & (0.41) & (0.39) & (0.35) & (0.0001) & (0.0001) & (0.0002) & (0.0003) & (0.0003) & (0.0003) \\ 
			\cline{2-2}
			& \multirow{2}{*}{0.60} & 12.01 & 10.93 & 7.98 & 0.9955 & 0.9958 & 0.9967 & 1.0047 & 1.0043 & 1.0032 \\ 
			& & (1.06) & (0.85) & (0.76) & (0.0003) & (0.0003) & (0.0007) & (0.0008) & (0.0007) & (0.0006) \\ 
			\cline{2-2}
			& \multirow{2}{*}{0.80} & 31.25 & 24.54 & 13.80 & 0.9893 & 0.9911 & 0.9942 & 1.0121 & 1.0096 & 1.0056 \\ 
			& & (3.66) & (2.14) & (1.40) & (0.0009) & (0.0009) & (0.0019) & (0.0023) & (0.0018) & (0.0014) \\ 
			\hline
		\end{tabular}
	}
\end{table}

\newpage

\subsection*{Appendix C: Figures}

\vspace*{2cm}

\begin{figure}[H]
	\centering
	\begin{subfigure}{1\textwidth}
		\centering
		\includegraphics[width=1\linewidth , height=0.42\linewidth]{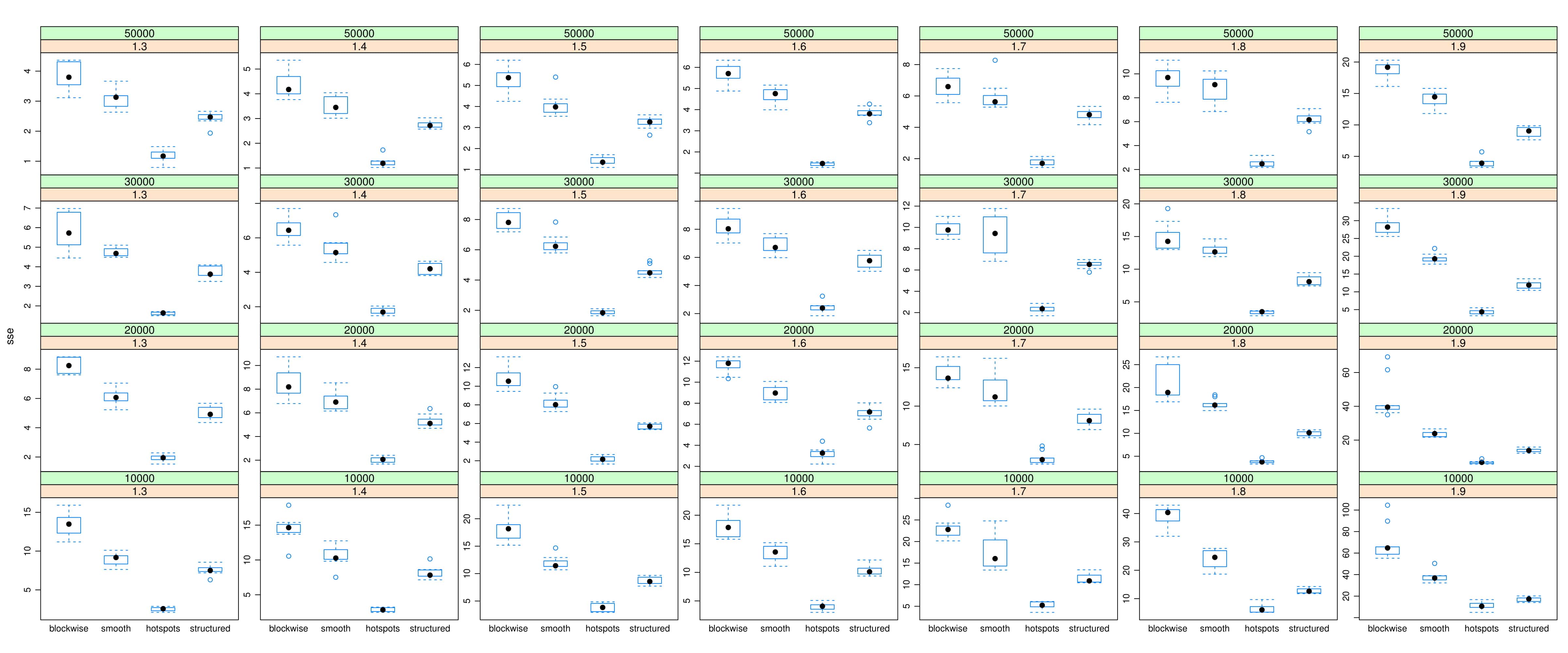}
		\caption{}
	\end{subfigure}\\
	\begin{subfigure}{1\textwidth}
		\centering
		\includegraphics[width=1\linewidth , height=0.42\linewidth]{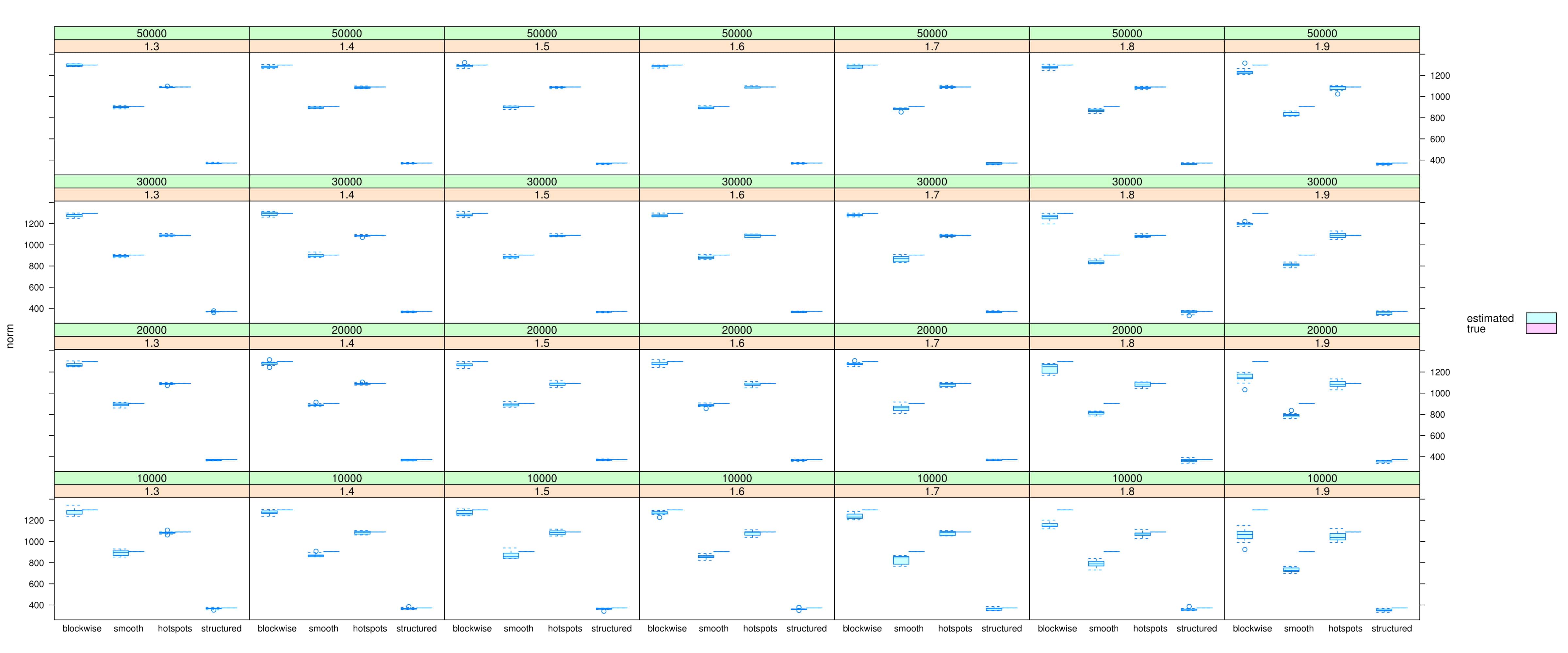}
		\caption{}
	\end{subfigure}%
	\caption{Figures showing (a) $SSE$s, (b) euclidean norms of estimated (using $p=1.5$) and true spatial effects for 282 zipcodes in CT, for different sample sizes and true values of the index parameter, $p$ under different spatial patterns. Results are obtained from 10 replications under each combination of sample and parameter settings. In both figures (a) and (b) the reference (where the index parameters match, i.e. simulated data and estimated effects have same index parameters) is the column with $p=1.5$, any significant departure is indicative of inferior performance. Figure (b) shows boxplots of norms for estimates under each pattern with the true norm shown as a horizontal line grouped along with it.}
	\label{fig::norm-see-sens}
\end{figure}

\vspace*{1cm}

\begin{figure}[H]
	\centering
	\begin{subfigure}{1\textwidth}
		\centering
		\includegraphics[width=1\linewidth , height=1.15\linewidth]{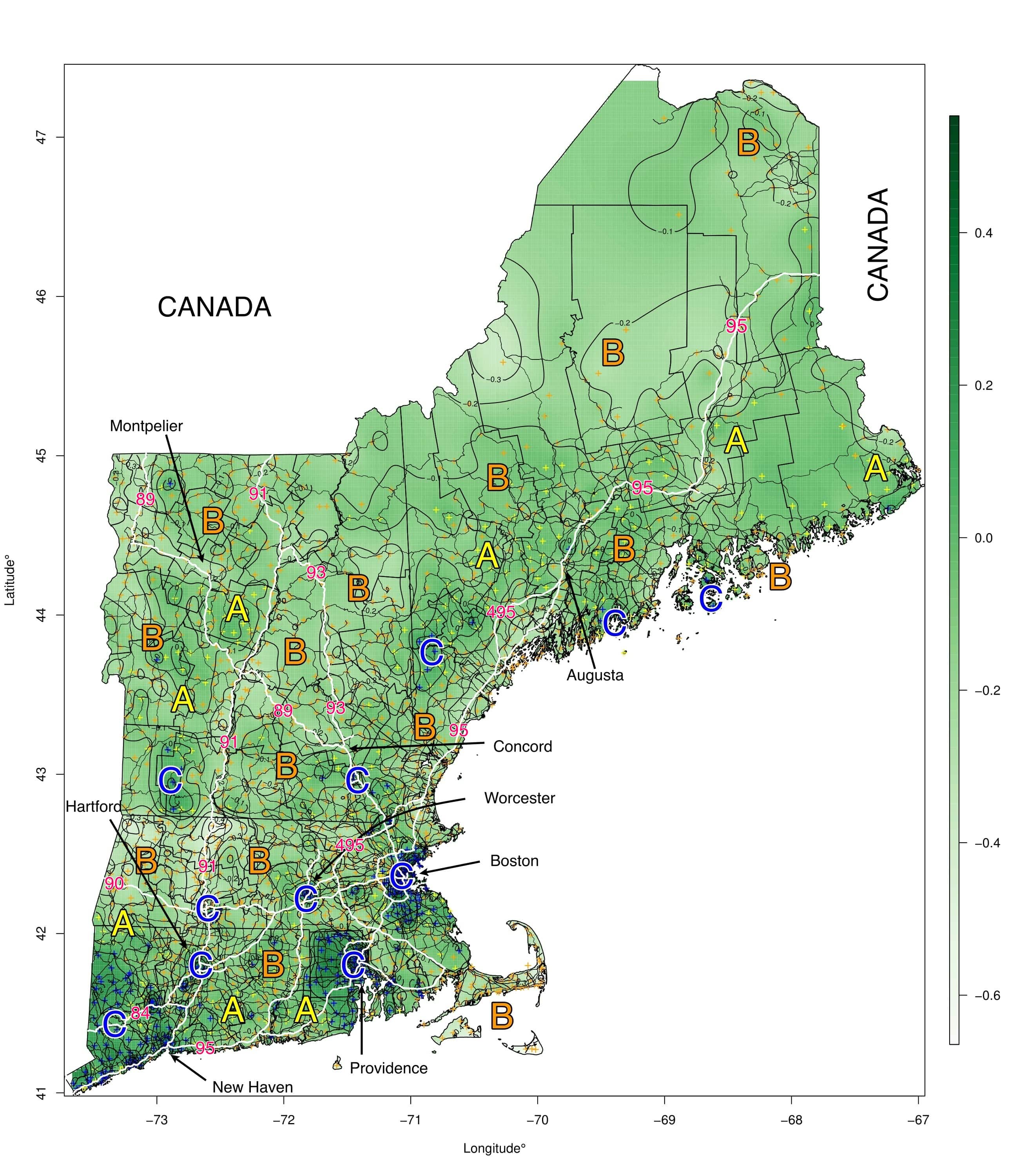}
	\end{subfigure}%
	\caption{Plot showing estimated spatial effects for states in New England (with county borders). It includes all primary (marked in bold ``white") and secondary (marked in ``black" lines) roads. Zipcodes (1831 in total) are color coded,  ones with 0 in their approx. CIs are ``yellow" (marked in ``A"), ones below zero are ``orange" (sizable regions marked with ``B"), and ones above are ``blue" (regions marked with ``C"). Some major cities are marked with arrows.}
	\label{fig::ne-speff-real}
\end{figure}

 \medskip
\newpage

\bibliography{spatial-tweedie0}

\end{document}